\documentstyle[11pt,arXiv_diss,amsmath,amssymb,epsfig,epic,cite]{report}

\input xy
\xyoption{all}
\def\l{\left}			
\def\r{\right}			
\def\lang{\left\langle}		
\def\rang{\right\rangle}	
\def\lPB{\left\{}		
\def\rPB{\right\}}		
\def\lpb{\left[\,}		
\def\rpb{\,\right]}		
\def\com{\,{\mathchar"213B}\,}	
\def\dcom{{\mathchar"213B}}	
\def\imi{{\rm i}}		
\def\ldef{\mathrel{\raisebox{.069ex}{:}\!\!=}}
\def\rdef{\mathrel{=\!\!\raisebox{.069ex}{:}}}
\def\pd{\partial}		
\def\fd{\delta}			
\def\grad{\nabla}		
\def\lapl{\grad^2}		
\def\invlapl{{\grad^{-2}}}	
\def\heavyside{\Theta}		
\def\pminor{\mu}		
\def\half{{\textstyle {1\over2}}}
\def\d{\,{\mathrm{d}}}		
\def\unvx{{\mathbf{\hat x}}}	

\newcommand{\Orb}{\operatorname{Orb}}

\def\etal{et al.}
\def\Alfven{{Alfv\'en}}
\def\Alfvenic{{{\Alfven}ic}}

\def\LieG{G}			
\def\LieA{{\mathfrak{g}}}	
\def\LieAx{{\mathfrak{h}}}	
\def\LieAxb{{\mathfrak{a}}}	
\def\SOthree{SO(3)}		
\def\SOtwoone{SO(2,1)}		
\def\sothree{so(3)}		
\def\vs{V}			
\def\reals{{\mathbb{R}}}	
\def\field{K}			
\def\manifold{{\mathcal{M}}}	
\def\functionalspace{{\mathcal{F}}}
\def\fv{{\xi}}			
\def\dotfv{{\skew{4}\dot\fv}}	
\def\fvz{{\varpi}}		
\def\dotfvz{{\skew{3}\dot\fvz}}	
\def\fdomain{\Omega}		
\def\magfcont{\Psi}		
\def\vort{\omega}		
\def\dotvort{{\skew{2}\dot\vort}}
\def\gvort{q}			
\def\Fgvort{{\mathcal{F}}}	
\def\fgvort{f}			
\def\Kfunc{{\mathcal{K}}}	
\def\magf{\psi}			
\def\dotmagf{{\skew{6.25}\dot\magf}}
\def\elecp{\phi}		

\def\ecurrent{J}		
\def\streamf{\phi}		
\def\pvel{v}			
\def\dotpvel{{\skew{3}\dot\pvel}}
\def\pres{p}			
\def\Telec{{T_{\mathrm{e}}}}	
\def\dotpres{{\skew{3}\dot\pres}}
\def\plasd{\chi}		
\def\vel{{\mathbf{v}}}		
\def\Bf{{\mathbf{B}}}		
\def\crmhdbeta{{\beta_{\mathrm{e}}}}
\def\equil{{\mathrm{e}}}	
\def\fveq{\fv_\equil}		
\def\fvzeq{{\fvz_\equil}}	
\def\vorteq{{\vort_\equil}}
\def\gvorteq{{\gvort_\equil}}
\def\magfeq{{\magf_\equil}}
\def\elecpeq{{\elecp_\equil}}
\def\ecurrenteq{{\ecurrent_\equil}}
\def\streamfeq{{\streamf_\equil}}
\def\pveleq{{\pvel_\equil}}
\def\preseq{{\pres_\equil}}

\def\velperpeq{{\vel_{\equil\perp}}}

\def\Bperpeq{{\Bf_{\equil\perp}}}
\def\binv{{\crmhdbeta^{-1}}}
\def\phib{\Phi}			
\def\Streamf{\Phi}		
\def\Vort{\Omega}		
\def\Elecp{\Phi}		
\def\Magf{\Psi}			
\def\Fv{\Xi}			
\def\afu{u}			
\def\Dafu{{\mathbb{D}\mskip.3\thinmuskip}}
\def\Hafu{k}			
\def\alfvenc{{c}}		
\def\alfvenspd{{v_{\mathrm{A}}}}
\def\knograd{{\kappa}}		
\def\nunograd{{\nu}}		
\def\magfvarii{\Theta}		
\def\W{W}			
\def\Wt{{\smash{\mbox{$\widetilde\W$}}\!\mskip.8\thinmuskip}}
\def\Wb{{\smash{\mbox{$\overline\W$}}\!\mskip.8\thinmuskip}}
\def\strconst{c}		
\def\ckform{{\mathfrak{K}}}	
\def\Wn{g}			
\def\Wni{{\skew{3}\bar g}}	
\def\Wcob{V}			
\def\coW{A}			
\def\Qt{Q}			
\def\Proj{P}			
\def\ww{w}			
\def\mm{m}			
\def\Nilb{N}			
\def\M{M}			
\def\Cas{C}			
\def\Casi{{\mathcal{C}}}	
\def\CasiSD{{\mathcal{C}_{\mathrm{sd}}}}
\def\Ham{H}			
\def\FrE{F}			
\def\Vpot{{\mathcal{V}}}	
\def\Vpotq{{\mathfrak{V}}}	
\def\Vpotl{{\mathfrak{v}}}	
\def\modVpot{{\mathcal{W}}}	
\def\modVpoti{{\mathcal{W}}}	
\def\rhs{{\mathcal{F}}}		
\def\fsaop{{\mathfrak{F}}}	
\def\constmot{{\mathfrak{K}}}	
\def\quadform{{\mathcal{Q}}}	
\def\dynacgen{{\mathcal{G}}}	
\def\dynac{{\mathrm{da}}}	
\def\dynacg{{\chi}}		
\def\zerorone{\theta}		
\def\ev{\Lambda}		
\def\evone{\zerorone}		
\def\xv{{\bf x}}		
\def\ad{{\rm ad}}		
\def\Ad{{\rm Ad}}		
\def\range{{\rm range\, }}	
\def\pmz{\lambda}		
\def\cycperm{{\mathrm{cyc.\ perm.}}}
\def\ag{{\mathcal{D}}}		
\def\agc{D}			
\def\ae{{\mathcal{E}}}		
\def\af{f}			
\def\afi{f}			
\def\afii{g}			
\def\afiii{h}			
\def\afiv{k}			
\def\afsd{f}			

\def\qbert{{Q*Bert}}
\def\qberttm{{\qbert${}^{\text{TM}}$}}

\newtheorem{case}{Case}

\newcommand{\eqlabel}[1]{\label{eq:#1}}
\newcommand{\chlabel}[1]{\label{ch:#1}}
\newcommand{\apxlabel}[1]{\label{apx:#1}}
\newcommand{\seclabel}[1]{\label{sec:#1}}
\newcommand{\figlabel}[1]{\label{fig:#1}}
\newcommand{\tablabel}[1]{\label{table:#1}}
\newcommand{\caselabel}[1]{\label{case:#1}}

\renewcommand{\eqref}[1]{(\ref{eq:#1})}
\newcommand{\chref}[1]{Chapter~\ref{ch:#1}}

\newcommand{\apxref}[1]{Appendix~\ref{apx:#1}}
\newcommand{\secref}[1]{Section~\ref{sec:#1}}
\newcommand{\secreftwo}[2]{Sections~\ref{sec:#1} and~\ref{sec:#2}}

\newcommand{\figref}[1]{Figure~\ref{fig:#1}}

\newcommand{\tabref}[1]{Table~\ref{table:#1}}
\newcommand{\caseref}[1]{Case~\ref{case:#1}}

\onehalfspacequote

\author{Jean-Luc Thiffeault}

\title{Classification, Casimir Invariants, and \\
	Stability of Lie--Poisson Systems}

\previousdegrees{B.S., M.A.}

\degree{DOCTOR OF PHILOSOPHY}

\degreeabbr{Ph.D.}

\graduationmonth{December}

\graduationyear{1998}

\address{}

\supervisor{Philip J. Morrison}

\makeindex

\begin{document}

\titlepage

\abstract
We classify Lie--Poisson brackets that are formed from Lie algebra extensions.
The problem is relevant because many physical systems owe their Hamiltonian
structure to such brackets.  A classification involves reducing all brackets
to a set of normal forms, independent under coordinate transformations, and is
achieved with the techniques of \emph{Lie algebra cohomology}.  For extensions
of order less than five, we find that the number of normal forms is small and
they involve no free parameters.  A special extension, known as the Leibniz
extension, is shown to be the unique ``maximal'' extension.

We derive a general method of finding Casimir invariants of Lie--Poisson
bracket extensions. The Casimir invariants of all brackets of order less than
five are explicitly computed, using the concept of \emph{coextension}.  We
obtain the Casimir invariants of Leibniz extensions of arbitrary order.  We
also offer some physical insight into the nature of the Casimir invariants of
compressible reduced magnetohydrodynamics.

We make use of the methods developed to study the stability of extensions for
given classes of Hamiltonians.  This helps to elucidate the distinction
between semidirect extensions and those involving \emph{cocycles}.  For
compressible reduced magnetohydrodynamics, we find the cocycle has a
destabilizing effect on the steady-state solutions.

\tableofcontents

\listoftables
\listoffigures


\chapter{Introduction}
\chlabel{intro}

The topic of this thesis is the classification and analysis of the properties
of Lie--Poisson brackets obtained from extensions of Lie algebras. A large
class of finite- and infinite-dimensional dynamical equations admit a
Hamiltonian formulation using noncanonical brackets of the Lie--Poisson type.
Finite-dimensional examples include the Euler equations for the rigid
body~\cite{Arnold}, the moment reduction of the Kida
vortex~\cite{Meacham1997}, and a low-order model of atmospheric
dynamics~\cite{Bokhove1996}. Infinite-dimensional examples include the Euler
equation for the ideal
fluid~\cite{Kuznetsov1980,Marsden1983,Morrison1982,Morrison1980a,Olver1982},
the quasigeostrophic equations~\cite{Holm1998,Weinstein1983}, and the Vlasov
equation~\cite{Marsden1982b,Morrison1980b}.

In mathematical terms, Lie--Poisson brackets naturally define a Poisson
structure (i.e., a symplectic structure~\cite{Weinstein1983b}) on the dual of
a Lie algebra.  For the rigid body, the Lie algebra is the one associated with
the rotation group,~$\SOthree$, while for the Kida vortex moment reduction the
underlying group is~$\SOtwoone$.  For the two-dimensional ideal fluid, the
relevant Lie algebra corresponds to the group of volume-preserving
diffeomorphisms of the fluid domain.

\index{Lie--Poisson system}
Lie--Poisson structures often occur as a result of
reduction~\cite{Marsden1974}.  Reduction is, in essence, a method of taking
advantage of the symmetries of a system to lower its order.  However in so
doing one perhaps loses the \emph{canonical} nature of the system: there are
no longer any well-defined conjugate positions and momenta.  This does not
preclude the system from being Hamiltonian, that is these conjugate variables
can exist \emph{locally}, up to some possible degeneracy in the system (the
\emph{symplectic leaves}).  The resulting Hamiltonian system (after reduction)
is often of Lie--Poisson type.  For example, the reduction of the rigid
body\index{rigid body} in Euler angle coordinates (three angles and three
canonical momenta, for a total of six coordinates) gives Euler's equations (in
terms of only the angular momenta, three coordinates), which have a
Lie--Poisson structure.

Why seek a bracket formulation of a system at all?  If we care about whether a
system is Hamiltonian or not, then for noncanonical systems it is a simple way
of showing that the equations have such a structure.  We are then free to use
the powerful machinery of Hamiltonian mechanics.  For example, we know that
the eigenvalue spectrum of the linearized system has to have four-fold
symmetry in the complex plane~\cite{Arnold1966b}.  If we are concerned with
the properties of the truncation of a hydrodynamic system, then knowing the
bracket formulation can serve as a guide for finding a finite-dimensional
representation of the system which retains the Hamiltonian
structure~\cite{McLachlan1997,Zeitlin1991}.  Also, there exists moment
reductions---finite-dimensional subalgebras of infinite-dimensional
algebras---that provide exact
closures~\cite{McLachlan1993,McLachlan1997,Meacham1997,Brad}.

\index{extension}
We will classify low-order bracket extensions and find their Casimir
invariants. An extension is simply a new Lie bracket, derived from a base
algebra (for example, $\SOthree$), and defined on~$n$-tuples of that algebra.
We are ruling out extensions where the individual brackets that appear are not
of the same form as that of the base algebra.  We are thus omitting some
brackets~\cite{Morrison1984b,Morrison1980a,Nore1997}, but the brackets we are
considering are amenable to a general classification.

The method of extension yields interesting and physically relevant algebras.
Using this method we can describe finite-dimensional systems of several
variables and infinite-dimensional systems of several fields.  For the
finite-dimensional case, an example is the two vector model of the heavy
\index{rigid body!heavy top} top~\cite{Holmes1983}, where the two vectors are
the angular momentum an the position of the center of mass.  For
infinite-dimensional systems there are examples of models of
two~\cite{Benjamin1984,McLachlan1997,Morrison1984},
three~\cite{Hazeltine1985,Kuvshinov1994,Morrison1984}, and
four~\cite{Hazeltine1987,Morrison1984b} fields.  Knowing the bracket allows
one to find the Casimir invariants of the
system~\cite{Hernandez1998,Kuroda1991,Trofimov}.  These are quantities which
commute with every functional on the Poisson manifold, and thus are conserved
by the dynamics for any Hamiltonian.  They are useful for analyzing the
constraints in the system~\cite{Thiffeault1998} and for establishing stability
criteria~\cite{Hazeltine1984,Holm1985,Morrison1987,Morrison1998,Morrison1986}.

\section{Overview}

The outline of the thesis is as follows. In \chref{LiePoisson}, we review the
general theory behind Lie--Poisson brackets. We give some examples of physical
systems of Lie--Poisson type, both finite- and infinite-dimensional. We
introduce the concept of Lie algebra extensions and derive some of their basic
properties. \chref{cohoext} is devoted to the more abstract treatment of
extensions through the theory of Lie algebra
cohomology~\cite{Chevalley1948,Azcarraga,Knapp}. We define some terminology
and special extensions such as the semidirect sum and the Leibniz
extension. In \chref{classext}, we use the cohomology techniques to treat the
specific type of extension with which we are concerned, brackets
over~$n$-tuples.  We give an explicit classification of low-order
extensions. By classifying, we mean reducing---through coordinate
changes---all possible brackets to independent normal forms. We find that the
normal forms are relatively few and that they involve no free parameters---at
least for low-order extensions.

In \chref{casinv}, we turn to the problem of finding the Casimir invariants of
the brackets, those functionals that commute with every other functional in
the algebra.  We derive some general techniques for doing this that apply to
extensions of any order.  We treat explicitly some examples, including the
Casimir invariants of a particular model of magnetohydrodynamics (MHD), which
are also given a physical interpretation. A formula for the invariants of
Leibniz extensions of any order is also derived. Then in
\secref{caslowdim} we use the classification of \secref{lowdimext} to derive
the Casimir invariants for low-order extensions.

We address general stability of Lie--Poisson systems in \chref{stability}.  We
begin by reviewing the concept of stability in \secref{genstab}, discussing
the distinctions between spectral, linearized, formal, and nonlinear
stability.  We consider the difficulties that arise for infinite-dimensional
systems.  In \secref{energycasimir} we present a review of the energy-Casimir
method for finding equilibria and establishing sufficient conditions for
stability.  We use the method on compressible reduced MHD.  In
\secref{dynaccess}, we turn to a more general method for stability analysis,
that of dynamical accessibility.  The method uses variations that have been
restricted to symplectic leaves.  We then treat several different classes of
Hamiltonian and Lie--Poisson brackets and discuss the role of cocycles in
equilibria and their stability.  Finally, in \chref{conclusion} we offer some
concluding remarks and discuss future directions for research.

\chapter{Lie--Poisson Brackets}
\chlabel{LiePoisson}

Lie--Poisson brackets\index{bracket!Lie--Poisson} define a natural Poisson
structure on duals of Lie algebras. Physically, they often arise in the
\emph{reduction}\index{reduction} of a system.  For our purposes, a
reduction is a mapping of the dynamical variables of a system to a smaller set
of variables, such that the transformed Hamiltonian and bracket depend only on
the smaller set of variables. (For a more detailed mathematical treatment, see
for example~%
\cite{AbrahamMarsden,Audin,Guillemin,Marsden1982,MarsdenRatiu,Marsden1974}\@.)
The simplest example of a reduction is the case in which a cyclic variable is
eliminated, but more generally a reduction exists as a consequence of an
underlying symmetry of the system. For instance, the Lie--Poisson bracket for
the rigid body is obtained from a reduction of the canonical Euler angle
description using the rotational symmetry of the system~\cite{Holmes1983}.
The Euler equation for the two-dimensional ideal fluid is obtained from a
reduction of the Lagrangian description of the fluid, which has a relabeling
symmetry~\cite{Bretherton1970,Morrison1998,Newcomb1967,Padhye1996a}.

Here we shall take a more abstract viewpoint: we do not assume that the
Lie--Poisson bracket is obtained from a reduction, though it is always
possible to do so by the method of Clebsch variables~\cite{Morrison1998}.
Rather we proceed directly from a given Lie algebra to build a Lie--Poisson
bracket. The choice of algebra can be guided by the symmetries of the system.

In \secref{lpbasic}, we give some definitions and review the basic theory
behind Lie--Poisson brackets.  We then give examples in \secref{lpexample}:
the free rigid body, reduced magnetohydrodynamics (RMHD), and compressible
reduced magnetohydrodynamics (CRMHD).  These last two cases are examples of
\emph{Lie algebra extensions}.  We describe general Lie algebra
extensions in \secref{theproblem}.  This introduces the problem, and
establishes the framework for the remainder of the thesis.

\section{Lie--Poisson Brackets on Duals of Lie Algebras}
\seclabel{lpbasic}

Recall that a \emph{Lie algebra}~$\LieA$\index{Lie algebra} is a vector space
on which is defined a bilinear operation~\hbox{$\lpb \com \rpb : \LieA \times
\LieA \rightarrow \LieA$}, called the Lie bracket.\index{bracket!Lie}  The Lie
bracket is \emph{antisymmetric},
\[
	\lpb\alpha\com\beta\rpb = -\lpb\beta\com\alpha\rpb,
\]
and satisfies the \emph{Jacobi identity}\index{Jacobi identity},
\[
	\lpb\alpha\com\lpb\beta\com\gamma\rpb\rpb
	+ \lpb\beta\com\lpb\gamma\com\alpha\rpb\rpb
	+ \lpb\gamma\com\lpb\alpha\com\beta\rpb\rpb = 0,
\]
for arbitrary elements~$\alpha$,~$\beta$,~$\gamma$ in~$\LieA$.  Lie algebras
are differentiable manifolds.

A real-valued \emph{functional}\index{functional} defined on a differentiable
manifold~$\manifold$ is simply a map from~$\manifold$ to~$\reals$.  (From now
on, when we say functional it will be understood that we mean a real-valued
functional.)  The vector space of all functionals on~$\manifold$ is denoted
by~$\functionalspace(\manifold)$.

The \emph{dual}~$\LieA^*$ of~$\LieA$\index{dual} is the set of all
\emph{linear} functionals on~$\LieA$.  The elements of~$\LieA^*$ are denoted by
\[
	\lang\,\fv\com\cdot\,\rang : \LieA \rightarrow \reals,
	\qquad \lang\,\fv\com\cdot\,\rang \in \LieA^*,
\]
where~$\fv$ identifies the elements of~$\LieA^*$.  It is customary, however,
to simply say~\hbox{$\fv\in\LieA^*$} and express the
\emph{pairing}\index{pairing} by~\hbox{$\lang\ \com\ \rang: \LieA^* \times
\LieA \rightarrow \reals$}.  This simplifies the procedure of
identifying~$\LieA$ and~$\LieA^*$, especially for infinite-dimensional Lie
algebras, where the pairing is typically an integral.  Note that functionals
can be defined on~$\LieA^*$, since it is a differentiable manifold.  In finite
dimensions,~$\LieA$ and~$\LieA^*$ are isomorphic as vector spaces (they have
the same dimension).  However,~$\LieA^*$ does not naturally inherit a Lie
algebra structure from~$\LieA$.  In infinite dimensions, the two spaces need
not be isomorphic.

Let~$\manifold$ be a differentiable manifold.  A \emph{Poisson
structure}\index{Poisson structure} on~$\functionalspace(\manifold)$ is a Lie
algebra on~$\functionalspace(\manifold)$ with bracket~$\lPB\com\rPB$ that
satisfies the derivation property
\[
	\lPB F\com G H\rPB = \lPB F\com G\rPB H + G \lPB F\com H\rPB,
\]
where~$F$,~$G$,~$H$~\hbox{$\in \functionalspace(\manifold)$}.  (This property
is also called the Leibniz rule.)\index{Leibniz rule}\index{derivation
property} The manifold~$\manifold$ with the bracket~$\lPB\com\rPB$ is called a
\emph{Poisson manifold}.\index{Poisson manifold}

\index{bracket!Lie--Poisson|(}
For the remainder of the thesis, we will be interested in the case
where~$\manifold$ is the dual~$\LieA^*$ of a Lie algebra~$\LieA$.  The
\emph{Lie--Poisson bracket} provides a natural Poisson structure
on~$\functionalspace(\LieA^*)$, given the Lie bracket~$\lpb\com\rpb$
in~$\LieA$.  It is defined as
\begin{equation}
	{\lPB F\com G \rPB}_\pm(\fv) = \pm\lang\fv\com
		{\lpb \frac{\fd F}{\fd\fv}\com\frac{\fd G}{\fd\fv}
		\rpb}\rang,
	\eqlabel{LPB}
\end{equation}
where~$F$ and~$G$ are real-valued functionals on~$\LieA^*$, that
is,~\hbox{$F,\ G:\LieA^*\rightarrow\reals$}, and~\hbox{$\fv \in \LieA^*$}.
The functional derivative~\hbox{$\fd F/\fd\fv \in \LieA$} is defined by
\begin{equation}
	\fd F[\,\fv;\fd\fv\,] \ldef 
	{\l.\frac{d}{d\epsilon}F[\fv + \epsilon\,
		\fd\fv]\r|}_{\epsilon=0}
	\rdef \lang\fd\fv\com\frac{\fd F}{\fd \fv}\rang .
	\eqlabel{funcder}
\end{equation}
We shall refer to the bracket~$\lpb\com\rpb$ as the \emph{inner
bracket}\index{bracket!inner} and to the bracket~$\lPB\com\rPB$ as the
Lie--Poisson bracket. The dual~$\LieA^*$ together with the Lie--Poisson
bracket is a Poisson manifold.  The sign choice in~\eqref{LPB} comes from
whether we are considering right invariant ($+$) or left invariant ($-$)
functions on the cotangent bundle of the Lie
group~\cite{MarsdenRatiu,Marsden1983}, but for our purposes we simply choose
the sign as needed.

For finite-dimensional algebras, the
Lie--Poisson bracket~\eqref{LPB} was first written down by Lie~\cite{Lie} and
was rediscovered by Berezin~\cite{Berezin1967}; it is also closely related to
work of Arnold~\cite{Arnold1966a}, Kirillov~\cite{Kirillov1962},
Kostant~\cite{Kostant1966}, and Souriau~\cite{Souriau}.
\index{bracket!Lie--Poisson|)}

Before we can describe the dynamics generated by Lie--Poisson brackets, we
need a few more definitions.  The \emph{adjoint action}\index{action}
of~$\LieA$ on itself is the same as the bracket in~$\LieA$,
\[
	\ad_\alpha\,\beta \equiv \lpb\alpha\com\beta\rpb,
\]
where~$\alpha$, $\beta \in \LieA$. From this we define the \emph{coadjoint
action}\index{action!coadjoint}~$\ad_\alpha^\dagger$ of~$\LieA$ on~$\LieA^*$
by\footnote{We are using the convention of Arnold~\cite[p.~321]{Arnold}, but
some authors define~$\ad^\dagger$ with a minus sign, so that the canonical
bracket and its coadjoint bracket have the same sign in \eqref{canicobrak}
when~$\LieA$ and~$\LieA^*$ are identified.}
\begin{equation}
	\lang \ad_\alpha^\dagger\,\fv\com \beta\,\rang \ldef
		\lang \,\fv \com\,\ad_\alpha\, \beta\,\rang ,
	\eqlabel{coadj}
\end{equation}
where~$\fv \in \LieA^*$. We also define the
\emph{coadjoint\index{bracket!coadjoint}
bracket}~\hbox{$\lpb\com\rpb^{\dag}:\LieA
\times \LieA^* \rightarrow \LieA^*$} to be
\begin{equation}
	\lpb\alpha\com\fv\,\rpb^{\dag} \ldef \ad_\alpha^\dagger\,\fv\, ,
	\eqlabel{cobrak}
\end{equation}
so that
\begin{equation}
	\lang \lpb\alpha\com\fv\rpb^\dagger\com \beta\,\rang \ldef
		\lang \,\fv \com\,\lpb\alpha \com \beta\rpb\,\rang ;
	\eqlabel{cobracket}
\end{equation}
the bracket~${\lpb\com\rpb}^\dagger$ satisfies the identity
\[
	\lang \lpb\alpha\com\fv\rpb^\dagger\com \beta\,\rang =
	-\lang \lpb\beta\com\fv\rpb^\dagger\com \alpha\,\rang.
\]

Since the inner bracket is Lie, it satisfies the Jacobi identity, and
consequently the form given by~\eqref{LPB} for the Lie--Poisson bracket will
automatically satisfy the Jacobi identity~\cite[p.~614]{AMR}\index{Jacobi
identity}. This is proved in \apxref{lpjacobi}.

We are of course interested in generating dynamics from the Lie--Poisson
bracket.  This is done in the usual manner, by inserting a Hamiltonian
functional in the bracket.  For any Poisson structure, given a Hamiltonian
functional~\hbox{$H:\manifold\rightarrow\reals$}, the equation of motion
for~$\fv \in \manifold$ is
\[
	\dotfv = \lPB\fv\com \Ham\rPB,
\]
where a dot denotes a time derivative.  For a Lie--Poisson bracket, we
have~$\manifold=\LieA^*$, and we use the definition~\eqref{LPB}
of~$\lPB\com\rPB$ to write
\[
	\dotfv = \lang\fv\com\lpb{\Delta}\com
			\frac{\fd\Ham}{\fd \fv}\rpb\rang,
\]
where~$\Delta$ is a Kronecker or Dirac delta, or a combination of both for an
infinite-dimensional system of several fields (that is, $\fv$ can be a vector
of field variables).  We then use the definition of the coadjoint
bracket~\eqref{cobracket},
\[
	\dotfv = -\Bigl\langle\lpb
		\frac{\fd \Ham}{\fd \fv}\com\,\fv\rpb^\dagger
		\com\,\Delta\Bigr\rangle,
\]
and finally use the property of the delta function to identify~$\dotfv$ with
the left slot of the pairing,\index{equations of motion}
\begin{equation}
	\dotfv = -\lpb \frac{\fd \Ham}{\fd \fv}\com\,\fv\rpb^\dagger.
	\eqlabel{motion}
\end{equation}
Thus, for Lie--Poisson brackets the dynamical evolution of~$\fv$ is generated
by the coadjoint bracket.

We close this section by commenting on the nature of the dynamics generated by
Lie--Poisson brackets.  The elements of a Lie algebra~$\LieA$ are usually
regarded as infinitesimal generators of the elements of a \emph{Lie
group}~$\LieG$ near the identity.  (We also say that the Lie algebra is the
tangent space of the Lie group at the identity.)  The \emph{coadjoint
orbit}\index{coadjoint orbit} through~\hbox{$\fv\in\LieA^*$} is defined as
\[
	\Orb(\fv) \ldef \l\{\Ad^\dagger_a\,\fv \mid a\in\LieG\r\}.
\]
(We will not rigourously define it here, but simply think
of~\hbox{$\Ad^\dagger_a:\LieA^*\rightarrow\LieA^*$} as a finite version of the
infinitesimal coadjoint
action~\hbox{$\ad^\dagger_\fv:\LieA^*\rightarrow\LieA^*$}.  See for example
Arnold~\cite[pp.~319--321]{Arnold}.)  The coadjoint orbits tell us what parts
of~$\LieA^*$ can be reached from a given element~$\fv^*$ by acting with the
group elements.  For example, the coadjoint orbits for the rotation
group~$\SOthree$\index{so3@$\SOthree$} are
spheres~\cite[p.~400]{MarsdenRatiu}, so two elements of~$\LieA^*$ belong to
the same coadjoint orbit if they lie on the same sphere (the elements can be
mapped onto each other by a rotation).

The infinitesimal generator at~$\fv$ of the coadjoint action is
\begin{equation}
	\eta_{\LieA^*}(\fv) \ldef \ad^\dagger_\eta\,\fv
	\eqlabel{coadgen}
\end{equation}
Comparing this to the equation of motion~\eqref{motion}, and recalling the
definition of the coadjoint bracket~\eqref{cobrak}, we see that~$\dotfv$ lies
along the direction of the infinitesimal
generator~$\ad^\dagger_{\fd\Ham/\fd\fv}$ at~$\fv$.

What does this all mean?  The time-evolved trajectory~\hbox{$\l\{\fv(t) \mid t
\ge 0\r\}$ of~$\fv$}\index{trajectory} must go through points in~$\LieA^*$
that can be reached by~$\Ad^\dagger_a\,\fv(0)$, where~$\fv(0)$ is an initial
condition, for some~\hbox{$a\in \LieG$}.  To put it more succinctly,
\begin{equation}
	\l\{\fv(t) \mid t \ge 0\r\} \subseteq \Orb(\fv(0)).
	\eqlabel{trajinOrb}
\end{equation}
For~\hbox{$\LieG=\SOthree$},\index{so3@$\SOthree$} since the coadjoint orbits
are spheres then the only trajectories allowed must lie on spheres.  This
makes~$\SOthree$ the natural group to describe the motion of the rigid body,
as we will see in \secref{rigidbody}.  Note that equality in~\eqref{trajinOrb}
does not usually hold, since the trajectory is one-dimensional, whereas the
coadjoint orbits are usually of higher dimension.

\section{Examples of Lie--Poisson Systems}
\seclabel{lpexample}

We will say that a physical systems can be described by a given Lie--Poisson
bracket\index{bracket!Lie--Poisson} and Hamiltonian\index{Hamiltonian} if its
equations of motion\index{equations of motion} can be written as
\eqref{motion} for some~$\Ham$; the system is then said to be
Hamiltonian\index{Hamiltonian!system} of the Lie--Poisson
type\index{Lie--Poisson system}. We give four examples: the first is
finite-dimensional (the free rigid body, \secref{rigidbody}) and the second
infinite-dimensional (Euler's equation for the ideal fluid,
\secref{twodfluid}). The third and fourth examples are also
infinite-dimensional and serve to introduce the concept of extension. They are
low--beta reduced magnetohydrodynamics (MHD) in
\secref{lowbetaRMHD} and compressible reduced MHD in \secref{CRMHD}. These
last two examples are meant to illustrate the physical relevance of Lie
algebra extensions.

\subsection{The Free Rigid Body}
\seclabel{rigidbody}

\index{rigid body}
The classic example of a Lie--Poisson bracket is obtained by taking
for~$\LieA$ the Lie algebra of the rotation
group~$\SOthree$\index{so3@$\SOthree$}. If the~$\hat {\bf e}_{(i)}$ denote a
basis of~$\LieA = \sothree$\index{so3@$\sothree$}, the Lie bracket is given by
\[
	\lpb \hat {\bf e}_{(i)}\com \hat {\bf e}_{(j)}\rpb
		= c_{ij}^k\, \hat {\bf e}_{(k)}\,,
\]
where the~$c_{ij}^k = \varepsilon_{ijk}$ are the structure
constants\index{structure constants} of the algebra, in this case the totally
antisymmetric symbol.  Using as a pairing\index{pairing} the usual contraction
between upper and lower indices, with~\eqref{LPB} we are led to the
Lie--Poisson bracket
\index{bracket!for the rigid body}
\[
	\lPB f\com g\rPB = -c_{ij}^k\, \ell_k\,\frac{\pd f}{\pd \ell_i}
		\,\frac{\pd g}{\pd \ell_j}\,,
\]
where the three-vector~$\ell$ is in~$\LieA^*$, and we have chosen the minus
sign in \eqref{LPB}. The coadjoint bracket\index{bracket!coadjoint} is
obtained using~\eqref{coadj},
\begin{equation}
	{\lpb \beta\com \ell\rpb}_i^\dagger = -c_{ij}^k\,\beta^j\,\ell_k.
	\eqlabel{rbcobrak}
\end{equation}
If we use this coadjoint bracket and insert the Hamiltonian
\index{Hamiltonian!for the rigid body}
\begin{equation}
	\Ham = \half {(I^{-1})}^{ij}\,\ell_i\,\ell_j
	\eqlabel{rbHam}
\end{equation}
in~\eqref{motion} we obtain
\index{equations of motion!for the rigid body|(}
\[
	\dot \ell_m = \lPB \ell_m\com \Ham\rPB 
		= c_{mj}^k\, {(I^{-1})}^{jp}\,\ell_k\,\ell_p\,.
\]
Notice how the moment of inertia tensor~$I$ plays the role of a
metric\index{metric}---it allows us to build a quadratic form (the
Hamiltonian) from two elements of~$\LieA^*$.  If we take~$I = {\rm
diag}(I_1,I_2,I_3)$, we recover Euler's equations for the motion of the free
rigid body
\[
	\dot \ell_1 = \l(\frac{1}{I_2} - \frac{1}{I_3}\r)\,\ell_2\,\ell_3,
\]
and cyclic permutations of 1,2,3.  \index{equations of motion!for the rigid
body|)} The~$\ell_i$ are the angular momenta\index{angular momentum} about the
axes and the~$I_i$ are the principal moments of inertia\index{moment of
inertia}. This result is naturally appealing because we expect the rigid body
equations to be invariant under the rotation group, hence the choice
of~$\SOthree$ for~$G$.

\subsection{The Two-dimensional Ideal Fluid}
\seclabel{twodfluid}

\index{two-dimensional ideal fluid}
Consider now an ideal fluid with the flow taking place over a two-dimensional
domain~$\fdomain$. Let~$\LieA$ be the infinite-dimensional Lie algebra
associated with the Lie group of volume-preserving diffeomorphisms
of~$\fdomain$. In two spatial dimensions this is the same as the group of
canonical transformations on~$\fdomain$. The bracket in~$\LieA$ is the
canonical bracket\index{bracket!canonical}
\begin{equation}
	\lpb a \com b \rpb = \frac{\pd a}{\pd x}\,\frac{\pd b}{\pd y}
		- \frac{\pd b}{\pd x}\,\frac{\pd a}{\pd y}.
	\eqlabel{canibrak}
\end{equation}
We formally identify~$\LieA$ and~$\LieA^*$ by using as the
pairing\index{pairing}~$\lang\com\rang$ the usual integral over the fluid
domain,
\[
	\lang F\com G \rang = \int_\fdomain F(\xv)\,G(\xv)\d^2x,
\]
where~$\xv \ldef (x,y)$.  For infinite-dimensional spaces, there are
functional analytic issues about whether we can make this identification, and
take~$\LieA^{**}=\LieA$. We will assume here that these relationships hold
formally. See Marsden and Weinstein~\cite{Marsden1982} for references on this
subject and Audin~\cite{Audin} for a treatment of the identification
of~$\LieA$ and~$\LieA^*$.

For simplicity, we assume that the boundary conditions are such that surface
terms vanish, and we get
\begin{equation}
	{\lpb\com\rpb}^\dagger = -\lpb\com\rpb
	\eqlabel{canicobrak}
\end{equation}
from \eqref{cobracket}. (Without this assumption the coadjoint bracket would
involve extra boundary terms.)  We take the vorticity~$\vort$\index{vorticity}
as the field variable~$\fv$ and write for the Hamiltonian
\index{Hamiltonian!for the 2-D ideal fluid}
\[
	\Ham[\vort] = -\half\lang \vort\com \invlapl\,\vort\rang ,
\]
where
\[
	(\invlapl\,\vort)(\xv)
		\ldef \int_\fdomain K(\xv|\xv')\,\vort(\xv')\d^2x',
\]
and~$K$ is Green's function\index{Green's function} for the Laplacian. The
Green's function plays the role of a metric\index{metric} since it maps an
element of~$\LieA^*$ (the vorticity~$\vort$) into an element of~$\LieA$ to be
used in the right slot of the pairing. This relationship is only weak: the
mapping~$K$ is not surjective, and thus the metric cannot formally inverted
(it is called \emph{weakly nondegenerate}). When we have identified~$\LieA$
and~$\LieA^*$ we shall often drop the comma in the pairing and write
\[
	\Ham[\vort] = -\half\lang \vort\,\streamf\rang =
		\half\lang|\grad\streamf|^2\rang,
\]
where~$\vort=\lapl\streamf$ defines the
streamfunction~$\streamf$\index{streamfunction}. We work out the evolution
equation for~$\vort$ explicitly:
\index{bracket!for the 2-D ideal fluid}
\index{equations of motion!for the 2-D ideal fluid}
\[
\begin{split}
	\dotvort(\xv) &= \lPB\vort\com \Ham\rPB 
		= \int_\fdomain \vort(\xv')
			\lpb\frac{\fd\vort(\xv)}{\fd\vort(\xv')}\com
			\frac{\fd \Ham}{\fd\vort(\xv')}\rpb\d^2x'\\
	&= \int_\fdomain\vort(\xv')
			\lpb\delta(\xv-\xv')\com {-\streamf(\xv')}\rpb\d^2x'
			\\
	&= \int_\fdomain\delta(\xv-\xv')
			\lpb\vort(\xv')\com\streamf(\xv')\rpb\d^2x'
			\\
	&= \lpb\vort(\xv)\com\streamf(\xv)\rpb\, .
\end{split}
\]
This is Euler's equation for a two-dimensional ideal fluid. We could also
have written this result down directly from \eqref{motion}
using~\hbox{${\lpb\com\rpb}^\dagger = -{\lpb\com\rpb}$}.

\subsection{Low-beta Reduced MHD}
\seclabel{lowbetaRMHD}

\index{reduced MHD}
This example will illustrate the concept of a Lie algebra
extension,\index{extension|(} the central topic of this thesis. Essentially,
the idea is to use an algebra of~$n$-tuples,\index{ntuples@$n$-tuples!algebra
of} which we call an extension, to describe a physical system with more than
one dynamical variable. As in \secref{twodfluid} we consider a flow taking
place over a two-dimensional domain~$\fdomain$. The Lie algebra~$\LieA$ is
again taken to be that of volume preserving diffeomorphisms on~$\fdomain$, but
now we consider also the vector space~$\vs$ of real-valued functions
on~$\fdomain$ (an Abelian Lie algebra under addition). The \emph{semidirect
sum}\index{semidirect sum} of~$\LieA$ and~$\vs$ is a new Lie algebra
whose elements are two-tuples~$(\alpha,v)$ with a bracket defined by
\begin{equation}
	\lpb(\alpha,v)\com(\beta,w)\rpb \ldef
	\l(\lpb\alpha\com\beta\rpb\com
	\lpb\alpha\com w\rpb - \lpb\beta\com v\rpb
	\r),
	\eqlabel{RMHDbrak}
\end{equation}
where~ $\alpha$ and~$\beta\in\LieA$, $v$ and~$w\in\vs$. This is a Lie
algebra, so we can use the prescription of \secref{lpbasic} to
build a Lie--Poisson bracket,\index{bracket!for reduced MHD}
\[
	{\lPB F\com G \rPB} = \int_\fdomain
		\l\lgroup
		\vort{\lpb \frac{\fd F}{\fd\vort}\com\frac{\fd G}{\fd\vort}
		\rpb}
	+ \magf \l({\lpb \frac{\fd F}{\fd\vort}\com\frac{\fd G}{\fd \magf}
		\rpb}
	- {\lpb \frac{\fd G}{\fd\vort}\com\frac{\fd F}{\fd \magf}
		\rpb}\r) \r\rgroup\d^2x.
\]
Let $\vort=\lapl\elecp$ be the (scalar) parallel vorticity, where $\elecp$ is
the electric potential, $\magf$ is the poloidal magnetic flux\index{magnetic
flux}, and $\ecurrent=\lapl\magf$\index{current} is the poloidal current. (We
use the same symbol for the electric field as for the streamfunction in
\secref{twodfluid} since they play a similar role.)\index{streamfunction} The
pairing\index{pairing} used is a dot product of the vectors followed by an
integral over the fluid domain (again identifying~$\LieA$ and~$\LieA^*$ as in
\secref{twodfluid}). The Hamiltonian
\index{Hamiltonian!for reduced MHD}
\[
	\Ham[\vort;\psi] = \frac{1}{2}\int_\fdomain\,\l\lgroup
		|\grad\elecp|^2+|\grad\magf|^2\r\rgroup\d^2x
\]
with the above bracket leads to the equations of motion
\index{equations of motion!for reduced MHD}
\begin{equation}
\begin{split}
	\dotvort &= \lpb\vort\com\elecp\rpb + \lpb\magf\com\ecurrent\rpb \ ,\\
	\dotmagf &= \lpb\magf\com\elecp\rpb \, .
	\eqlabel{RMHDeom}
\end{split}
\end{equation}
This is a model for low-beta reduced
MHD~\cite{Morrison1984,Strauss1976,Zeitlin1992}, obtained by an expansion in
the inverse aspect ratio~$\epsilon$ of a tokamak, with~$\epsilon$ small.  With
a strong toroidal magnetic field, the dynamics are then approximately
two-dimensional.  The model is referred to as low-beta because the electron
beta (the ratio of electron pressure to magnetic pressure,
see~\eqref{elecbeta}) is of order~$\epsilon^2$.

For high-beta reduced MHD, the electron beta is taken to be of
order~$\epsilon$.  There is then an additional advected pressure variable,
which couples to the vorticity equation, and the system still has a semidirect
sum structure~\cite{Hazeltine1985c,Strauss1977}.

Benjamin~\cite{Benjamin1984} used a system with a similar Lie--Poisson
structure, but for waves in a density-stratified fluid. Semidirect sum
structures are ubiquitous in advective systems: one variable (in this
example,~$\elecp$) ``drags'' the others along~\cite{Thiffeault1998}.

\subsection{Compressible Reduced MHD}
\seclabel{CRMHD}

\index{compressible reduced MHD}
In general there are other, more general ways to extend Lie algebras besides
the semidirect sum. The model derived by Hazeltine \emph{et
al.}~\cite{Hazeltine1987,Hazeltine1985b,Hazeltine1985c} for two-dimensional
compressible reduced MHD (CRMHD) is an example. This model has four fields,
and as for the low-beta reduced MHD system in \secref{lowbetaRMHD} it is
obtained from an expansion in the inverse aspect ratio of a tokamak. It
includes compressibility and finite ion Larmor radius effects. The Hamiltonian
is
\index{Hamiltonian!for compressible reduced MHD}
\begin{equation}
	\Ham[\vort,\pvel,\pres,\magf] =
		\frac{1}{2}\int_\fdomain
		\l\lgroup
			|\grad\elecp|^2 + \pvel^2
			+ \frac{(\pres-2\crmhdbeta\,x)^2}{\crmhdbeta}
			+ |\grad\magf|^2
		\r\rgroup \d^2x,
	\eqlabel{CRMHDHam}
\end{equation}
where $\pvel$ is the ion parallel (toroidal) velocity, $\pres$ is the electron
pressure,\footnote{The variable~$\pres$ is actually a deviation of the
pressure from a linear gradient.  The total pressure is~\hbox{$\overline\pres
= \pres-2\crmhdbeta\,x$}.} $\crmhdbeta$ is the electron beta,
\begin{equation}
	\crmhdbeta \ldef \frac{2\, \Telec}{\alfvenspd^2}\,,
	\eqlabel{elecbeta}
\end{equation}
a parameter that measures compressibility,~$\alfvenspd$ is the \Alfven\ speed,
and~$\Telec$ is the electron temperature. The other variables are as in
\secref{lowbetaRMHD}. The coordinate~$x$ points outward from the center of the
tokamak in the horizontal plane and~$y$ is the vertical coordinate.  The
motion is made two-dimensional by the strong toroidal magnetic field. The
bracket we will use is
\index{bracket!for compressible reduced MHD}
\begin{multline}
	\lPB F\com G\rPB =
		\int_\fdomain\l\lgroup\vort\lpb\frac{\fd F}{\fd \vort}
			\com\frac{\fd G}{\fd \vort}\rpb\r.
			+ \pvel\l(\lpb\frac{\fd F}{\fd \vort}
			\com\frac{\fd G}{\fd \pvel}\rpb
			+ \lpb\frac{\fd F}{\fd \pvel}
			\com\frac{\fd G}{\fd \vort}\rpb\r)\\
		+ \pres\l(\lpb\frac{\fd F}{\fd \vort}
			\com\frac{\fd G}{\fd \pres}\rpb
		+ \lpb\frac{\fd F}{\fd \pres}
			\com\frac{\fd G}{\fd \vort}\rpb\r)
		+ \magf\l(\lpb\frac{\fd F}{\fd \vort}
			\com\frac{\fd G}{\fd \magf}\rpb
		+ \lpb\frac{\fd F}{\fd \magf}
			\com\frac{\fd G}{\fd \vort}\rpb\r)\\
	- \l.\crmhdbeta\,\magf\l(
		\lpb\frac{\fd F}{\fd \pres}
		\com\frac{\fd G}{\fd \pvel}\rpb
		+ \lpb\frac{\fd F}{\fd \pvel}
		\com\frac{\fd G}{\fd \pres}\rpb\r)\r\rgroup\d^2x.
	\eqlabel{CRMHDbracket}
\end{multline}
Together this bracket and the Hamiltonian \eqref{CRMHDHam} lead to the
equations
\index{equations of motion!for compressible reduced MHD}
\begin{align*}
	\dotvort &= \lpb \vort\com\elecp\rpb + \lpb\magf\com \ecurrent\rpb 
		+ 2\lpb \pres\com x\rpb\\
	\dotpvel &= \lpb\pvel\com\elecp\rpb + \lpb\magf\com\pres \rpb
		+ 2\crmhdbeta\lpb x \com \magf \rpb\\
	\dotpres &= \lpb\pres\com\elecp\rpb
		+ \crmhdbeta\lpb\magf\com\pvel\rpb\\
	\dotmagf &= \lpb \magf\com\elecp\rpb ,
\end{align*}
which reduce to the example of \secref{lowbetaRMHD} in the limit
of~\hbox{$\pvel=\pres=\crmhdbeta=0$} (when compressibility effects are
unimportant).  In the limit of~$\crmhdbeta=0$, the parallel velocity decouples
from the other equations, and we recover the three equations of high-beta
reduced MHD for~$\vort$,~$\magf$, and~$\pres$~\cite{Hazeltine1985c}.

It is far from clear that the Jacobi identity is satisfied
for~\eqref{CRMHDbracket}. A direct verification is straightforward (if
tedious), but we shall see in \secref{theproblem} that there is an easier way.

\section{General Lie Algebra Extensions}
\seclabel{theproblem}

We wish to generalize the types of bracket used in
\secreftwo{lowbetaRMHD}{CRMHD}. We build an algebra extension by forming an~$n$-tuple of elements of a single Lie algebra~$\LieA$,
\index{ntuples@$n$-tuples}
\begin{equation}
	\alpha \ldef \l(\alpha_1,\dots,\alpha_n\r), \eqlabel{algtuple}
\end{equation}
where~\hbox{$\alpha_i \in \LieA$}. The most general bracket on this~$n$-tuple
space obtained from a linear combination of the one in~$\LieA$ has components
\index{bracket!for~$n$-tuples|(}
\begin{equation}
	{\lpb\alpha\com\beta\rpb}_\lambda = \sum_{\mu,\nu=1}^n
		{\W_\lambda}^{\mu\nu}\,
		\lpb\alpha_\mu\com\beta_\nu\rpb\,,
		\ \ \ \lambda=1,\dots,n,
	\eqlabel{extbrack}
\end{equation}
where the~${\W_\lambda}^{\mu\nu}$ are constants.  (From now on we will assume
that repeated indices are summed unless otherwise noted.) Since the bracket
in~$\LieA$ is antisymmetric the~$\W$'s must be symmetric in their upper
indices,
\begin{equation}
	{\W_\lambda}^{\mu\nu} = {\W_\lambda}^{\nu\mu}\,.
	\eqlabel{upsym}
\end{equation}
\index{Jacobi identity}
This bracket must also satisfy the Jacobi identity
\[
	{\lpb\alpha\com\lpb\beta\com\gamma\rpb\rpb}_\lambda +
	{\lpb\beta\com\lpb\gamma\com\alpha\rpb\rpb}_\lambda +
	{\lpb\gamma\com\lpb\alpha\com\beta\rpb\rpb}_\lambda = 0, \ \
	\lambda=1,\dots,n.
\]
The first term can be written
\[
	{\lpb\alpha\com\lpb\beta\com\gamma\rpb\rpb}_\lambda = 
	{\W_\lambda}^{\sigma\tau}\,{\W_\sigma}^{\mu\nu}\,
	{\lpb\alpha_\tau\com\lpb\beta_\mu\com\gamma_\nu\rpb\rpb},
\]
which when added to the other two gives
\[
	{\W_\lambda}^{\sigma\tau}\,{\W_\sigma}^{\mu\nu}\,\l(
	{\lpb\alpha_\tau\com\lpb\beta_\mu\com\gamma_\nu\rpb\rpb}
	+ {\lpb\beta_\tau\com\lpb\gamma_\mu\com\alpha_\nu\rpb\rpb}
	+ {\lpb\gamma_\tau\com\lpb\alpha_\mu\com\beta_\nu\rpb\rpb}
	\r) = 0.
\]
We cannot yet make use of the Jacobi identity in~$\LieA$: the subscripts
of~$\alpha$, $\beta$, and~$\gamma$ are different in each term so they
represent different elements of~$\LieA$.  We first relabel the sums and then
make use of the Jacobi identity in~$\LieA$ to obtain
\begin{multline*}
\l({\W_\lambda}^{\sigma\tau}\,{\W_\sigma}^{\mu\nu}
	- {\W_\lambda}^{\sigma\nu}\,{\W_\sigma}^{\tau\mu}\r)\,
	{\lpb\alpha_\tau\com\lpb\beta_\mu\com\gamma_\nu\rpb\rpb}\\
+ \l({\W_\lambda}^{\sigma\mu}\,{\W_\sigma}^{\nu\tau}
	- {\W_\lambda}^{\sigma\nu}\,{\W_\sigma}^{\tau\mu}\r)\,
	{\lpb\beta_\mu\com\lpb\gamma_\nu\com\alpha_\tau\rpb\rpb} = 0\,.
\end{multline*}
This identity is satisfied if and only if
\begin{equation}
	{\W_\lambda}^{\sigma\tau}\,{\W_\sigma}^{\mu\nu}
	= {\W_\lambda}^{\sigma\nu}\,{\W_\sigma}^{\tau\mu}\,,
	\eqlabel{Wjacob}
\end{equation}
which together with~\eqref{upsym} implies that the
quantity~${\W_\lambda}^{\sigma\tau}\,{\W_\sigma}^{\mu\nu}$ is symmetric in all
three free upper indices. If we write the~$\W$'s as~$n$\index{bracket!matrix representation}
matrices~${\W}^{(\nu)}$ with rows labeled by~$\lambda$ and columns by~$\mu$,
\index{extension!representation by matrices}
\begin{equation}
	{{\l[{\W}^{(\nu)}\r]}_\lambda}^\mu \ldef {\W_\lambda}^{\mu\nu},
	\eqlabel{Wupdef}
\end{equation}
then~\eqref{Wjacob} says that those matrices pairwise commute:
\begin{equation}
	\W^{(\nu)}\,\W^{(\sigma)} = \W^{(\sigma)}\,\W^{(\nu)}.
	\eqlabel{Wcommute}
\end{equation}
Equations~\eqref{upsym} and~\eqref{Wcommute} form a necessary and sufficient
condition: a set of~$n$ commuting matrices of size~$n\times n$ satisfying the
symmetry given by~\eqref{upsym} can be used to make a good Lie algebra
bracket. From this Lie bracket we can build a Lie--Poisson bracket using the
prescription of~\eqref{LPB} to obtain
\[
	{\lPB F\com G \rPB}_\pm(\fv) = \pm\sum_{\lambda,\mu,\nu=1}^n
		{\W_\lambda}^{\mu\nu}\lang\fv^\lambda\com
		{\lpb \frac{\fd F}{\fd\fv^\mu}\com\frac{\fd G}{\fd\fv^\nu}
		\rpb}\rang .
\]
\index{bracket!for~$n$-tuples|)}

We now return to the two extension examples of \secreftwo{lowbetaRMHD}{CRMHD}
and examine them in light of the general extension concept introduced here.

\subsection{Low-beta Reduced MHD}
\seclabel{matlowbetaRMHD}

For this example we have~$(\fv^0,\fv^1)=(\vort,\magf)$, with
\[
	\W^{(0)} = \begin{pmatrix} 1 & 0 \\ 0 & 1\end{pmatrix},\ \ \ \
	\W^{(1)} = \begin{pmatrix} 0 & 0 \\ 1 & 0\end{pmatrix}.
\]
The reason why we start labeling at~$0$ will become clearer in
\secref{semisimple}. The two~$\W^{(\mu)}$ must commute since~$\W^{(0)}=I$, the
identity. The tensor~$\W$ also satisfies the symmetry
property~\eqref{upsym}. Hence, the bracket is a good Lie algebra bracket.

\subsection{Compressible Reduced MHD}
\seclabel{matCRMHD}

We have~$n=4$ and take~$(\fv^0,\fv^1,\fv^2,\fv^3)=(\vort,\pvel,\pres,\magf)$,
so the tensor~$\W$ is given by
\begin{xalignat}{3}
	\W^{(0)} &= \begin{pmatrix}
				1 & 0 & 0 & 0\\
				0 & 1 & 0 & 0\\
				0 & 0 & 1 & 0\\
				0 & 0 & 0 & 1\\ \end{pmatrix}, \qquad
	&\W^{(1)} &= \begin{pmatrix}
				0 & 0 & 0 & 0\\
				1 & 0 & 0 & 0\\
				0 & 0 & 0 & 0\\
				0 & 0 & -\crmhdbeta & 0\\ \end{pmatrix},
		\nonumber\\[8pt]
	\W^{(2)} &= \begin{pmatrix}
				0 & 0 & 0 & 0\\
				0 & 0 & 0 & 0\\
				1 & 0 & 0 & 0\\
				0 & -\crmhdbeta & 0 & 0\\ \end{pmatrix},\qquad
	&\W^{(3)} &= \begin{pmatrix}
				0 & 0 & 0 & 0\\
				0 & 0 & 0 & 0\\
				0 & 0 & 0 & 0\\
				1 & 0 & 0 & 0\\ \end{pmatrix}.
	\eqlabel{matCRMHD}
\end{xalignat}
It is easy to verify that these matrices commute and that the tensor~$\W$
satisfies the symmetry property, so that the Lie--Poisson bracket given
by~\eqref{CRMHDbracket} satisfies the Jacobi identity\index{Jacobi
identity}. (See \secref{semisimple} for an explanation of why the labeling is
chosen to begin at zero.)\index{extension|)}

The 3-tensor~$\W$ can be represented as a cubical array of numbers, in the
same way a matrix is a square array.  In \figref{crmhdextpic} we show a
schematic representation of~$\W$ for CRMHD.  The blocks represent nonzero
elements.
\begin{figure}
\centerline{\psfig{file=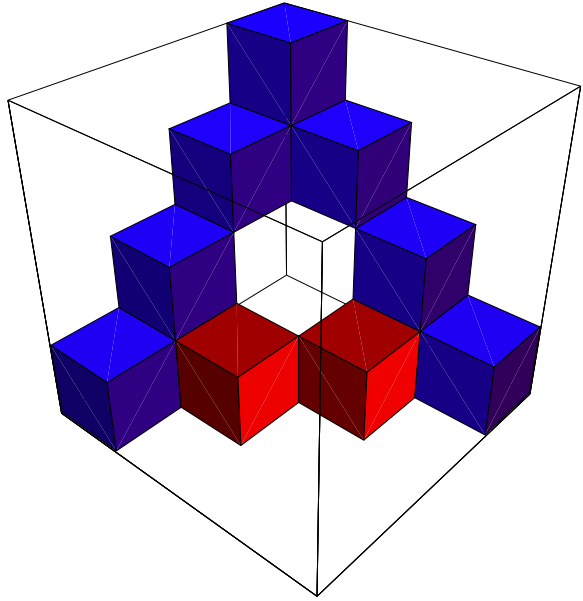,width=3in}}
\caption{Schematic representation of the 3-tensor~$\W$ for compressible
reduced MHD.  The blue cubes represent unit entries, the red cubes are equal
to~$-\crmhdbeta$, and all other entries vanish.  The vertical axis is the
lower index~$\lambda$ of~${\W_{\lambda}}^{\mu\nu}$, and the two horizontal
axes are the symmetric upper indices~$\mu$ and~$\nu$.  The origin is at the
top-rear.}
\figlabel{crmhdextpic}
\end{figure}

\chapter{Extension of a Lie Algebra}
\chlabel{cohoext}

In this chapter we review the theory of Lie algebra cohomology and its
application to extensions. This is useful for shedding light on the methods
used in \chref{classext} for classifying the extensions. However, the
mathematical details presented in this chapter can be skipped without
seriously compromising the flavor of the classification scheme of
\chref{classext}.  Most necessary mathematical concepts will be defined as
needed, but the reader wishing more extensive definitions may want to consult
books such as Azc\'arraga and Izquierdo~\cite{Azcarraga} or Choquet-Bruhat and
DeWitt-Morette~\cite{Cecile}.

\section{Cohomology of Lie Algebras}
\seclabel{cohoalgebra}

We now introduce the abstract formalism of Lie algebra cohomology.
Historically there were two different reasons for the development of this
theory. One, known as the Chevalley--Eilenberg
formulation~\cite{Chevalley1948}, was developed from de Rham cohomology\@. de
Rham cohomology concerns the relationship between exact and closed
differential forms, which is determined by the global properties (topology) of
a differentiable manifold. A Lie group is a differentiable manifold and so has
an associated de Rham cohomology. If invariant differential forms are used in
the computation, one is led to the cohomology of Lie algebras presented in
this section~\cite{Azcarraga,Cecile,Knapp}.  The second motivation is the one
that concerns us: we will show in \secref{extension} that the extension
problem---the problem of enumerating extensions of a Lie algebra---can be
related to the cohomology of Lie algebras.

\index{cohomology!of Lie algebras|(}
Let~$\LieA$ be a Lie algebra, and let the vector space~$\vs$ over the
field~$\field$ (which we take to be the real numbers later) be a
left~$\LieA$-\emph{module},\index{module}%
\footnote{When~$\vs$ is a right~$\LieA$-module, we have~\hbox{$\rho_{\lpb
\alpha\com\alpha'\rpb} = -\lpb \rho_\alpha \com \rho_{\alpha'} \rpb$}.
The results of this section can be adapted to a right action by changing the
sign every time a commutator appears.  This sign choice is for similar reasons
as that of \eqref{LPB}.}  that is, there is an operator~$\rho:
\LieA \times \vs \rightarrow \vs$ such that
\begin{align}
	\rho_\alpha\, (v + v') &= \rho_\alpha\,v + \rho_\alpha\,v',
		\nonumber\\
	\rho_{\alpha + \alpha'}\, v &= \rho_\alpha\,v + \rho_{\alpha'}\,v,
		\nonumber\\
	\rho_{\lpb \alpha\com\alpha'\rpb}v &= 
		\lpb \rho_\alpha \com \rho_{\alpha'} \rpb\,v\,,
	\eqlabel{rhomo}
\end{align}
for~$\alpha, \alpha' \in \LieA$ and~$v,v' \in \vs$. The operator~$\rho$ is
known as a left \emph{action}\index{action}.  A~$\LieA$-module gives a
representation of~$\LieA$ on~$\vs$.  The action~$\rho$ defines a Lie algebra
\emph{homomorphism} from~$\LieA$ to the algebra of linear transformations
on~$\vs$.  A Lie algebra
homomorphism\index{homomorphism}~\hbox{$f:\LieA\rightarrow\LieAxb$} is a
linear mapping between two Lie algebras~$\LieA$ and~$\LieAxb$ which preserves
the Lie algebra structure, that is
\[
	f({\lpb\alpha\com\beta\rpb}_{\LieA})
	= {\lpb f(\alpha)\com f(\beta)\rpb}_{\LieAxb},
	\qquad \alpha,\beta \in \LieA.
\]

An $n$-dimensional $\vs$-valued \emph{cochain}~$\omega_n$ for~$\LieA$, or
just~$n$-cochain\index{cochain} for short, is a skew-symmetric~$n$-linear
mapping
\[
	\omega_n :\ \stackrel{\longleftarrow n \longrightarrow}
	{\LieA \times \LieA \times \dots \times \LieA}\
	\longrightarrow \vs.
\]
Cochains are Lie algebra cohomology analogues of differential forms on a
manifold.  Addition and scalar multiplication of~$n$-cochains are defined in
the obvious manner by
\begin{align*}
	(\omega_n + \omega_n')(\alpha_1,\dots,\alpha_n)
	&\ldef \omega_n(\alpha_1,\dots,\alpha_n)
	+ \omega_n'(\alpha_1,\dots,\alpha_n),\\
	(a\,\omega_n)(\alpha_1,\dots,\alpha_n)
	&\ldef a\,\omega_n(\alpha_1,\dots,\alpha_n),
\end{align*}
where~\hbox{$\alpha_1,\dots,\alpha_n \in \LieA$} and~\hbox{$a \in \field$}.
The set of all~$n$-cochains thus forms a vector space over the field~$\field$
and is denoted by~$C^n(\LieA,\vs)$\index{Cn@$C^n(\LieA,\vs)$}.  The
$0$-cochains are defined to be just elements of~$\vs$, so
that~\hbox{$C^0(\LieA,\vs) = \vs$}.

The \emph{coboundary operator}\index{coboundary!operator} is the map between
cochains,
\[
	s_n : C^n(\LieA,\vs) \longrightarrow C^{n+1}(\LieA,\vs),
\]
defined by
\begin{multline*}
	(s_n\,\omega_n)(\alpha_1,\dots,\alpha_{n+1}) \ldef
	\sum_{i=1}^{n+1}(-)^{i+1}\rho_{\alpha_i}\omega_n(\alpha_1,\dots,
	\hat\alpha_i,\dots,\alpha_{n+1})\\
	+ \sum_{\scriptscriptstyle{j,k=1}\atop\scriptscriptstyle{j<k}}^{n+1}
	(-)^{j+k}\omega_n(\lpb\alpha_j\com\alpha_k\rpb,\alpha_1,\dots,
	\hat\alpha_j,\dots,\hat\alpha_k,\dots,\alpha_{n+1}),
\end{multline*}
where the caret means an argument is omitted.  We shall often drop the~$n$
subscript on~$s_n$, deducing it from the dimension of the cochain on which~$s$
acts.

We shall make use mostly of the first few cases,
\begin{align}
	(s\,\omega_0)(\alpha_1) &= \rho_{\alpha_1}\,\omega_0,
	\eqlabel{1cobound}\\
	(s\,\omega_1)(\alpha_1,\alpha_2) &=
		\rho_{\alpha_1}\,\omega_1(\alpha_2)
		- \rho_{\alpha_2}\,\omega_1(\alpha_1)
		- \omega_1(\lpb\alpha_1\com\alpha_2\rpb),
	\eqlabel{2cobound}\\
	(s\,\omega_2)(\alpha_1,\alpha_2,\alpha_3) &=
		\rho_{\alpha_1}\,\omega_2(\alpha_2,\alpha_3)
		+ \rho_{\alpha_2}\,\omega_2(\alpha_3,\alpha_1)
		+ \rho_{\alpha_3}\,\omega_2(\alpha_1,\alpha_2)\nonumber\\
	& \mbox{}
		- \omega_2(\lpb\alpha_1\com\alpha_2\rpb,\alpha_3)
		- \omega_2(\lpb\alpha_2\com\alpha_3\rpb,\alpha_1) 
		- \omega_2(\lpb\alpha_3\com\alpha_1\rpb,\alpha_2)\,.
	\eqlabel{3cobound}
\end{align}
It is easy to verify that~$s\,\omega_n$ defines an $(n+1)$-cochain, and it is
straightforward (if tedious) to show that~\hbox{$s_{n+1}s_n = s^2=0$}.  For
this to be true, the homomorphism property~\eqref{rhomo} of~$\rho$ is crucial.

An~$n$-\emph{cocycle}\index{cocycle} is an element~$\omega_n$
of~$C^n(\LieA,\vs)$ such that~$s_n\,\omega_n = 0$.
An~$n$-\emph{coboundary}\index{coboundary} $\omega_{\rm cob}$ is an element
of~$C^n(\LieA,\vs)$ for which there exists an element~$\omega_{n-1}$
of~$C^{n-1}(\LieA,\vs)$ such that~\hbox{$\omega_{\rm cob} = s\omega_{n-1}$}.
Note that all coboundaries are cocycles, but not vice-versa.

Let
\[
	Z^{n}_\rho(\LieA,\vs) = \ker s_n
\]
\index{Zn@$Z^n_\rho(\LieA,\vs)$}
be the vector subspace of all~$n$-cocycles,~\hbox{$Z^{n}_\rho(\LieA,\vs)
\subset C^n(\LieA,\vs)$}, and let
\[
	B^{n}_\rho(\LieA,\vs) = \range s_{n-1}
\]
\index{Bn@$B^n_\rho(\LieA,\vs)$}
be the vector subspace of all~$n$-coboundaries,~\hbox{$B^{n}_\rho(\LieA,\vs)
\subset C^n(\LieA,\vs)$}.  The~$n$th \emph{cohomology
group}\index{cohomology!group} of~$\LieA$ with coefficients in~$\vs$ is
defined to be the quotient vector space
\begin{equation}
	H^n_\rho(\LieA,\vs) \ldef Z^{n}_\rho(\LieA,\vs)/B^{n}_\rho(\LieA,\vs).
\end{equation}
\index{Hn@$H^n_\rho(\LieA,\vs)$|see{cohomology group}}%
Note that for~\hbox{$n > \dim \LieA$}, we have~\hbox{$H^n_\rho(\LieA,\vs) =
Z^{n}_\rho(\LieA,\vs) = B^{n}_\rho(\LieA,\vs) = 0$}.  This is because one
cannot build a nonvanishing antisymmetric quantity with more indices than the
dimension of the space (at least two of the indices would always be equal,
which implies that the quantity is zero).
\index{cohomology!of Lie algebras|)}

\section{Application of Cohomology to Extensions}
\seclabel{extension}

\index{cohomology!application to extensions}
In \secref{theproblem} we gave a definition of extension that is specific to
our problem, in terms of the tensors~$\W$.  We will now define extensions in a
more abstract manner.  We then show how the cohomology of Lie algebras of
\secref{cohoalgebra} is related to the problem of classifying extensions.  In
\chref{classext} we will return to the more concrete concept of extension, of
the form given in
\secref{theproblem}.

Let~$f_i: \LieA_i \rightarrow \LieA_{i+1}$ be a collection of Lie algebra
homomorphisms,
\[
\xymatrix@M=4pt{
	\dots \ar[r] & {\LieA_i} \ar[r]^-{f_i} &
	{\LieA_{i+1}} \ar[r]^-{f_{i+1}} & {\LieA_{i+2}} \ar[r] & \dots\
} .
\]
By the homomorphism property\index{homomorphism} of~$f_i$, we have
\[
	f_i({\lpb\alpha\com\beta\rpb}_{\LieA_i})
	= {\lpb f_i(\alpha)\com f_i(\beta)\rpb}_{\LieA_{i+1}},
	\qquad \alpha,\beta \in \LieA_i.
\]
The subscript on the brackets denotes the algebra to which it belongs.

The sequence~$f_i$ is called an \emph{exact sequence}\index{sequence!exact} of
Lie algebra homomorphisms if
\[
	\range f_i = \ker f_{i+1}\,.
\]

Let~$\LieA$, $\LieAx$, and~$\LieAxb$ be Lie algebras. The algebra~$\LieAx$ is
said to be an~\emph{extension}\index{extension!abstract} of~$\LieA$
by~$\LieAxb$ if there is a short exact sequence of Lie algebra homomorphisms
\begin{equation}\xymatrix@M=4pt{
	0 \ar[r] & {\LieAxb} \ar[r]^{i}
	& {\LieAx} \ar@_{->}@<-.5ex>[r]_{\pi}
	& {\LieA} \ar@_{-->}@<-.5ex>[l]_{\tau} \ar[r] & 0
	}.
	\eqlabel{extdef}
\end{equation}
The homomorphism~$i$ is an insertion (injection), and~$\pi$ is a projection
(surjection). We shall distinguish brackets in the different algebras by
appropriate subscripts.  We also define~\hbox{$\tau:\LieA\rightarrow\LieAx$}
to be a linear mapping such that~\hbox{$\pi\circ\tau = 1_{|\LieA}$} (the
identity mapping in~$\LieA$).  Note that~$\tau$ is not unique, since the
kernel of~$\pi$ is not trivial.  Let~$\beta \in
\LieAx$,~$\eta \in \LieAxb$; then
\[
	\pi{\lpb\beta\com i\,\eta\rpb}_\LieAx 
		= {\lpb\pi\,\beta\com\pi\,i\,\eta\rpb}_\LieA
		= 0,
\]
using the homomorphism property of~$\pi$ and~\hbox{$\pi\circ i=0$}, a
consequence of the exactness of the sequence. Thus~\hbox{${\lpb\beta\com
i\,\eta\rpb}_\LieAx \in
\ker \pi = \range i$}, and~$i\,\LieAxb$ is an ideal in~$\LieAx$
since~\hbox{$\lpb\beta\com i\eta\rpb \in i \LieAxb$}.
Hence, we can form the quotient algebra~$\LieAx/\LieAxb$, with equivalence
classes denoted by~\hbox{$\beta + \LieAxb$}. By exactness~\hbox{$\pi(\beta +
\LieAxb) = \pi\,\beta$}, so~$\LieA$ is isomorphic to~$\LieAx/\LieAxb$ and we
write~\hbox{$\LieA = \LieAx/\LieAxb$}.

Though~$i\,\LieAxb$ is a subalgebra of~$\LieAx$,~$\tau\,\LieA$ is not
necessarily a subalgebra of~$\LieAx$, for in general
\[
	{\lpb\tau\,\alpha\com\tau\,\beta\rpb}_\LieAx
	\ne \tau\,{\lpb\alpha\com\beta\rpb}_\LieA,
\]
for~\hbox{$\alpha,\beta \in \LieA$}; that is,~$\tau$ is not necessarily a
homomorphism. The classification problem\index{classification!abstract}
essentially resides in the determination of how much~$\tau$ differs from a
homomorphism. The cohomology machinery of \secref{cohoalgebra} is the key to
quantifying this difference, and we proceed to show this.

To this end, we use the algebra~$\LieAxb$ as the vector space~$\vs$ of
\secref{cohoalgebra}, so that~$\LieAxb$ will be a left $\LieA$-module. We
define the left action\index{action} as
\begin{equation}
	\rho_\alpha\,\eta \ldef
		i^{-1}{\lpb \tau\,\alpha\com i\,\eta\rpb}_\LieAx
	\eqlabel{rhodef}
\end{equation}
for~\hbox{$\alpha \in \LieA$} and~\hbox{$\eta \in \LieAxb$}. For~$\LieAxb$ to
be a left~$\LieA$-module, we need~$\rho$ to be a homomorphism, i.e.,~$\rho$
must satisfy~\eqref{rhomo}.  Therefore consider
\[
\begin{split}
	{\lpb\rho_\alpha\com\rho_\beta\rpb}\,\eta &=
		(\rho_\alpha\rho_\beta - \rho_\beta\rho_\alpha)\,\eta
		\\
	&= \rho_\alpha\,i^{-1}{\lpb\tau\,\beta\com i\,\eta\rpb}_\LieAx
		- \rho_\beta\,i^{-1}{\lpb\tau\,\alpha\com i\,\eta\rpb}_\LieAx
		\\
	&= i^{-1}{\lpb\tau\,\alpha\com
		{\lpb\tau\,\beta\com i\,\eta\rpb}_\LieAx\rpb}_\LieAx
		- i^{-1}{\lpb\tau\,\beta\com
		{\lpb\tau\,\alpha\com i\,\eta\rpb}_\LieAx\rpb}_\LieAx,
\end{split}
\]
which upon using the Jacobi identity in~$\LieAx$ becomes
\begin{equation}
\begin{split}
	{\lpb\rho_\alpha\com\rho_\beta\rpb}\,\eta &=
		i^{-1}{\lpb{\lpb\tau\,\alpha\com \tau\,\beta\rpb}_\LieAx\com
		i\,\eta\rpb}_\LieAx\\
	&= i^{-1}{\lpb\tau\,{\lpb\alpha\com \beta\rpb}_\LieA\com
		i\,\eta\rpb}_\LieAx
		+ i^{-1}{\lpb\l(
		{\lpb\tau\,\alpha\com \tau\,\beta\rpb}_\LieAx
		- \tau\,{\lpb\alpha\com \beta\rpb}_\LieA\r)\com
		i\,\eta\rpb}_\LieAx\\
	&= \rho_{{\lpb\alpha\com\beta\rpb}_\LieA}\,\eta
		+ i^{-1}{\lpb\l(
		{\lpb\tau\,\alpha\com \tau\,\beta\rpb}_\LieAx
		- \tau\,{\lpb\alpha\com \beta\rpb}_\LieA\r)\com
		i\,\eta\rpb}_\LieAx.
	\eqlabel{mismatch}
\end{split}
\end{equation}
By applying~$\pi$ on the expression in parentheses of the last term
of~\eqref{mismatch}, we see that it vanishes and so is in~$\ker \pi$, and by
exactness it is also in~$i\,\LieAxb$. Thus the $\LieAx$ commutator above
involves two elements of~$i\,\LieAxb$. We define~\hbox{$\omega:
\LieA \times \LieA \rightarrow \LieAxb$} by
\begin{equation}
	\omega(\alpha,\beta) \ldef i^{-1}\l(
	{\lpb\tau\,\alpha\com\tau\,\beta\rpb}_\LieAx
	- \tau\,{\lpb\alpha\com\beta\rpb}_\LieA\r).
	\eqlabel{cocycle}
\end{equation}
The mapping~$i^{-1}$ is well defined on~$i\,\LieAxb$. Equation
\eqref{mismatch} becomes
\begin{equation}
	{\lpb\rho_\alpha\com\rho_\beta\rpb}\,\eta =
		\rho_{{\lpb\alpha\com\beta\rpb}_\LieA}\,\eta + {\lpb
		\omega(\alpha,\beta) \com
		\eta\rpb}_\LieAxb.
	\eqlabel{rhohomo}
\end{equation}
Therefore,~$\rho$ satisfies the homomorphism property if either of the
following is true:
\begin{enumerate}
\item[(i)] $\LieAxb$ is Abelian,
\item[(ii)] $\tau$ is a homomorphism,
\end{enumerate}
Condition~(i) implies~$\lpb\com\rpb_\LieAxb=0$, while condition~(ii)
means
\[
	{\lpb\tau\,\alpha\com \tau\,\beta\rpb}_\LieAx =
		\tau\,{\lpb\alpha\com\beta\rpb}_\LieA,
\]
which implies~$\omega\equiv 0$. If either of these conditions is
satisfied,~$\LieAxb$ with the action~$\rho$ is a left~$\LieA$-module.  We
treat these two cases separately in \secreftwo{abelianext}{semidext},
respectively.

\section{Extension by an Abelian Lie Algebra}
\seclabel{abelianext}

In this section we assume that the homomorphism condition~(i) at the end of
\secref{extension} is met.  Therefore~$\LieAxb$ is a left~$\LieA$-module, and
we can define~$\LieAxb$-valued cochains on~$\LieA$.  In particular,~$\omega$
defined by~\eqref{cocycle} is a 2-cochain,~\hbox{$\omega
\in C^2(\LieA,\LieAxb)$}, that measures the ``failure''
of~$\tau$ to be a homomorphism. We now show, moreover, that~$\omega$ is a
2-cocycle,~\hbox{$\omega \in Z^2_\rho(\LieA,\LieAxb)$}.  By
using~\eqref{3cobound},
\begin{align*}
	(s\,\omega)(\alpha,\beta,\gamma) &=
		\rho_{\alpha}\,\omega(\beta,\gamma)
		+ \rho_{\beta}\,\omega(\gamma,\alpha)
		+ \rho_{\gamma}\,\omega(\alpha,\beta)\\
	&\phantom{=} - \omega({\lpb\alpha\com\beta\rpb}_\LieA,\gamma)
		- \omega({\lpb\beta\com\gamma\rpb}_\LieA,\alpha) 
		- \omega({\lpb\gamma\com\alpha\rpb}_\LieA,\beta)\,,
		\\
	&= i^{-1}{\lpb\tau\,\alpha\com i\,\omega(\beta,\gamma)\rpb}_\LieAx
		- \omega({\lpb\alpha\com\beta\rpb}_\LieA,\gamma)
		+ \cycperm,
\end{align*}
where we have written ``$\cycperm$'' to mean cyclic permutations
of~$\alpha$,~$\beta$, and~$\gamma$.  Using the definition~\eqref{cocycle}
of~$\omega$, we have
\begin{align*}
	(s\,\omega)(\alpha,\beta,\gamma) &= i^{-1}{\lpb\tau\,\alpha\com
		{\lpb\tau\,\beta\com\tau\,\gamma\rpb}_\LieAx
		- \tau\,{\lpb\beta\com\gamma\rpb}_\LieA\rpb}_\LieAx\\
	&\phantom{=} - i^{-1}\l({\lpb\tau\,{\lpb\alpha\com\beta\rpb}_\LieA\com
		\tau\,\gamma\rpb}_\LieAx 
		- \tau\,{\lpb{\lpb
		\alpha\com\beta\rpb}_\LieA\com\gamma\rpb}_\LieA\r)
		+ \cycperm,\\
	&= i^{-1}\l({\lpb\tau\,\alpha\com
		{\lpb\tau\,\beta\com\tau\,\gamma\rpb}_\LieAx\rpb}_\LieAx
		+ \cycperm\r)\\
	&\phantom{=} + i^{-1}\tau\l({\lpb{\lpb
		\alpha\com\beta\rpb}_\LieA\com\gamma\rpb}_\LieA
		+ \cycperm\r) = 0.
\end{align*}
The first parenthesis vanishes by the Jacobi identity in~$\LieAx$, the second
by the Jacobi identity in~$\LieA$, and the other terms were canceled in
pairs. Hence~$\omega$ is a 2-cocycle.

Two extensions~$\LieAx$ and~$\LieAx'$ are
equivalent\index{extension!equivalence of|(}
\index{equivalent extensions|(} if there exists a Lie algebra
isomorphism~$\sigma$ such that the diagram
\begin{equation}
	\xymatrix@M=4pt{
	& & {\LieAx} \ar[dr]^{\pi} \ar[dd]^{\sigma} & & \\
	0 \ar[r] & {\LieAxb} \ar[ur]^i \ar[dr]_{i'} & & {\LieA} \ar[r] & 0 \\
	& & {\LieAx'} \ar[ur]_{\pi'} & &
	}
	\eqlabel{equivext}
\end{equation}
is commutative, that is if~\hbox{$\sigma\circ i=i'$ and~$\pi =
\pi'\circ\sigma$}.

There will be an injection~$\tau$ associated with~$\pi$ and a~$\tau'$
associated with~$\pi'$, such that~\hbox{$\pi\circ\tau = 1_{|\LieA} =
\pi'\circ\tau'$.} The linear map~\hbox{$\nu = \sigma^{-1}\tau' - \tau$} must
be from~$\LieA$ to~$i\,\LieAxb$, so~\hbox{$i^{-1}\nu \in C^1(\LieA,\LieAxb)$}.
Consider~$\rho$ and~$\rho'$ respectively defined using~$\tau,i$ and~$\tau',i'$
by~\eqref{rhodef}. Then
\begin{equation}
\begin{split}
	(\rho_\alpha - {\rho'}_\alpha)\,\eta 
		&= i^{-1}{\lpb\tau\,\alpha\com i\,\eta\rpb}_\LieAx -
		{i'}^{-1}{\lpb\tau'\,\alpha\com i'\,\eta\rpb}_{\LieAx'}
		\\
	&= i^{-1}{\lpb\tau\,\alpha\com i\,\eta\rpb}_\LieAx -
		i^{-1}\sigma^{-1}{\lpb\sigma(\nu + \tau)\,\alpha\com \sigma
		i\,\eta\rpb}_{\LieAx'}\\
	&= i^{-1}{\lpb\tau\,\alpha\com i\,\eta\rpb}_\LieAx -
		i^{-1}{\lpb(\nu + \tau)\,\alpha\com
		i\,\eta\rpb}_\LieAx\\
	&= -i^{-1}{\lpb\nu\,\alpha\com i\,\eta\rpb}_\LieAx = 0,
	\eqlabel{samerho}
\end{split}
\end{equation}
since~$\LieAxb$ is Abelian. Hence, $\tau$ and~$\tau'$ define the same~$\rho$.
Now consider the 2-cocycles~$\omega$ and~$\omega'$ defined from~$\tau$
and~$\tau'$ by~\eqref{cocycle}. We have
\begin{align*}
	\omega'(\alpha,\beta) - \omega(\alpha,\beta)
		&= {i'}^{-1}
		\l({\lpb\tau'\,\alpha\com\tau'\,\beta\rpb}_{\LieAx'}
		- \tau'\,{\lpb\alpha\com\beta\rpb}_\LieA\r)\\
	&\phantom{=} - i^{-1}\l({\lpb\tau\,\alpha\com\tau\,\beta\rpb}_\LieAx
		- \tau\,{\lpb\alpha\com\beta\rpb}_\LieA\r)\\
	&= i^{-1}\sigma^{-1}\l({\lpb\sigma(\nu+\tau)\,\alpha\com
		\sigma(\nu+\tau)\,\beta\rpb}_{\LieAx'} 
		- \sigma(\nu+\tau)\,{\lpb\alpha\com\beta\rpb}_\LieA\r)
		\\
	&\phantom{=} - i^{-1}\l({\lpb\tau\,\alpha\com\tau\,\beta\rpb}_\LieAx
		- \tau\,{\lpb\alpha\com\beta\rpb}_\LieA\r)\\
	&= i^{-1}\l({\lpb(\nu+\tau)\,\alpha\com(\nu+\tau)\,\beta\rpb}_\LieAx
		- \nu\,{\lpb\alpha\com\beta\rpb}_\LieA -
		{\lpb\tau\,\alpha\com\tau\,\beta\rpb}_\LieAx\r)\\
	&= i^{-1}\l({\lpb\tau\,\alpha\com\nu\,\beta\rpb}_\LieAx
		+ {\lpb\nu\,\alpha\com\tau\,\beta\rpb}_\LieAx
		- \nu\,{\lpb\alpha\com\beta\rpb}_\LieA\r)\\
	&= \rho_\alpha\,(i^{-1}\nu\,\beta) - \rho_\beta\,(i^{-1}\nu\,\alpha)
		- i^{-1}\nu\,{\lpb\alpha\com\beta\rpb}_\LieA.
\end{align*}
Comparing this with~\eqref{2cobound}, we see that
\begin{equation}
	\omega' - \omega = s\,(i^{-1}\nu),
	\eqlabel{cobdiff}
\end{equation}
so~$\omega$ and~$\omega'$ differ by a coboundary. Hence, they represent the
same element in~$H^2_\rho(\LieA,\LieAxb)$. Equivalent extensions uniquely
define an element of the second cohomology group~$H^2_\rho(\LieA,\LieAxb)$.
Note that this is true in particular for~$\LieAx=\LieAx'$,~$\sigma=1$, so that
the element of~$H^2_\rho(\LieA,\LieAxb)$ is independent of the choice
of~$\tau$.equivalent\index{extension!equivalence of|)}
\index{equivalent extensions|)}

We are now ready to write down explicitly the bracket in~$\LieAx$. We can
represent an element~\hbox{$\alpha \in \LieAx$} as a two-tuple:~\hbox{$\alpha
= (\alpha_1, \alpha_2)$} where~\hbox{$\alpha_1 \in \LieA$} and~\hbox{$\alpha_2
\in \LieAxb$} (\hbox{$\LieAx = \LieA \oplus \LieAxb$} as a vector space). The
injection~$i$ is then~\hbox{$i\,\alpha_2 = (0,\alpha_2)$}, the
projection~$\pi$ is~\hbox{$\pi\,(\alpha_1,\alpha_2) = \alpha_1$}, and since
the extension is independent of the choice of~$\tau$ we
take~\hbox{$\tau\,\alpha_1 = (\alpha_1,0)$}. By linearity,
\begin{multline*}
	{\lpb\alpha,\beta\rpb}_\LieAx =
		{\lpb(\alpha_1,0),(\beta_1,0)\rpb}_\LieAx
		+ {\lpb(0,\alpha_2),(0,\beta_2)\rpb}_\LieAx\\
	+ {\lpb(\alpha_1,0),(0,\beta_2)\rpb}_\LieAx
		+ {\lpb(0,\alpha_2),(\beta_1,0)\rpb}_\LieAx.
\end{multline*}
We know that~${\lpb(0,\alpha_2),(0,\beta_2)\rpb}_\LieAx = 0$ since~$\LieAxb$
is Abelian. By definition of the cocycle~$\omega$, Eq.~\eqref{cocycle}, we
have
\begin{align*}
	{\lpb(\alpha_1,0),(\beta_1,0)\rpb}_\LieAx &=
		{\lpb\tau\,\alpha_1\com\tau\,\beta_1\rpb}_\LieAx \\
	&= i\,\omega(\alpha_1,\beta_1) +
		\tau\,{\lpb\alpha_1\com\beta_1\rpb}_\LieA\\
	&= ({\lpb\alpha_1\com\beta_1\rpb}_\LieA\,,\,\omega(\alpha_1,\beta_1)).
\end{align*}
Finally, by the definition of~$\rho$, Eq.~\eqref{rhodef}, 
\begin{equation}
	{\lpb(\alpha_1,0),(0,\beta_2)\rpb}_\LieAx
		= {\lpb\tau\,\alpha_1,i\,\beta_2\rpb}_\LieAx
		= \rho_{\alpha_1}\,\beta_2,
	\eqlabel{a1b2com}
\end{equation}
and similarly for~${\lpb(0,\alpha_2),(\beta_1,0)\rpb}_\LieAx$, with opposite
sign. So the bracket is\index{bracket!for abstract extension}
\begin{equation}
	{\lpb\alpha,\beta\rpb}_\LieAx =
		\Bigl({\lpb\alpha_1\com\beta_1\rpb}_\LieA\, , \,
		\rho_{\alpha_1}\,\beta_2 - \rho_{\beta_1}\,\alpha_2
		+ \omega(\alpha_1,\beta_1)\Bigr).
	\eqlabel{abelianbracket}
\end{equation}
As a check we work out the Jacobi identity in~$\LieAx$:
\begin{align*}
	{\lpb\alpha\com{\lpb\beta\com\gamma\rpb}_\LieAx\rpb}_\LieAx &=
		\l({\lpb\alpha_1\com{\lpb\beta\com\gamma\rpb}_1\rpb}_\LieA
		\com \rho_{\alpha_1}\,{\lpb\beta\com\gamma\rpb}_2
		- \rho_{{\lpb\beta\com\gamma\rpb}_1}\,\alpha_2
		+ \omega(\alpha_1,{\lpb\beta\com\gamma\rpb}_1)\r)\\
	&= \Bigl({\lpb\alpha_1\com{\lpb\beta_1
		\com\gamma_1\rpb}_\LieA\rpb}_\LieA
		\com \rho_{\alpha_1}(\rho_{\beta_1}\,\gamma_2
		- \rho_{\gamma_1}\,\beta_2 + \omega(\beta_1,\gamma_1))
		\\
	&\phantom{=\Bigl({\lpb\alpha_1\com{\lpb\beta_1
		\com\gamma_1\rpb}_\LieA\rpb}_\LieA\com}
		- \rho_{{\lpb\beta_1\com\gamma_1\rpb}_\LieA}\,\alpha_2
		+ \omega(\alpha_1,{\lpb\beta_1\com\gamma_1\rpb}_\LieA)\Bigr).
\end{align*}
Upon adding permutations, the first component will vanish by the Jacobi
identity in~$\LieA$.  We are left with
\begin{multline*}
	{\lpb\alpha\com{\lpb\beta\com\gamma\rpb}_\LieAx\rpb}_\LieAx 
		+ \cycperm = \Bigl(0\com
	\l(\rho_{\alpha_1}\rho_{\beta_1} - \rho_{\beta_1}\rho_{\alpha_1}
		- \rho_{{\lpb\alpha_1\com\beta_1\rpb}_\LieA}\r)
		\gamma_2\\
		+ \rho_{\alpha_1}\,\omega(\beta_1,\gamma_1)
		- \omega({\lpb\alpha_1\com\beta_1\rpb}_\LieA,\gamma_1)\Bigr)
		+ \cycperm,
\end{multline*}
which vanishes by the the homomorphism property of~$\rho$ and the fact
that~$\omega$ is a 2-cocycle, Eq.~\eqref{3cobound}.

Equation~\eqref{abelianbracket} is the most general form of the Lie bracket
for extension by an Abelian Lie algebra. It turns out that the theory of
extension by a non-Abelian algebra can be reduced to the study of extension by
the center of~$\LieAxb$, which is Abelian~\cite{Azcarraga}. We will not need
this fact here, as the only extensions by non-Abelian algebras we will deal
with are of the simpler type of \secref{semidext}.

We have thus shown that equivalent extensions are enumerated by the second
cohomology group~$H^2_\rho(\LieA,\LieAxb)$.  The coordinate
transformation~$\sigma$ used in~\eqref{equivext} to define equivalence of
extensions preserves the form of~$\LieA$ and~$\LieAxb$ as subsets of~$\LieAx$.
However, we have the freedom to choose coordinate transformations which do
transform these subsets. All we require is that the isomorphism~$\sigma$
between~$\LieAx$ and~$\LieAx'$ be a Lie algebra homomorphism.  We can
represent this by the diagram
\begin{equation}\xymatrix@M=4pt{
	0 \ar[r] & {\LieAxb} \ar[r]^{i} & {\LieAx} \ar[r]^{\pi}
		\ar[d]^\sigma
		& {\LieA} \ar[r] & 0 \\
	0 \ar[r] & {\LieAxb'} \ar[r]^{i} & {\LieAx'} \ar[r]^{\pi}
		& {\LieA'} \ar[r] & 0. \\
	}
	\eqlabel{equivext2}
\end{equation}
The primed and the unprimed extensions are not equivalent, but they are
isomorphic~\cite[p.~199]{Weiss}.  Cohomology for us is not the whole story,
since we are interested in isomorphic extensions, but it will guide our
classification scheme.\index{classification!abstract} We discuss this point
further in \secref{furthertrans}.

Diagrams~\eqref{equivext} and~\eqref{equivext2} are related to the ``Short
Five Lemma,''\index{short five lemma} which states that if the diagram of Lie
algebra homomorphisms
\[
	\xymatrix@M=4pt{
	0 \ar[r] & {\LieAxb} \ar[r]^{i} \ar[d]^\gamma & {\LieAx} \ar[r]^{\pi}
		\ar[d]^\sigma
		& {\LieA} \ar[r] \ar[d]^\delta & 0 \\
	0 \ar[r] & {\LieAxb'} \ar[r]^{i} & {\LieAx'} \ar[r]^{\pi}
		& {\LieA'} \ar[r] & 0  \\
	}
\]
is commutative, with the top and bottom rows exact, then
\begin{alignat*}{3}
\text{(i)}&\ \ &\gamma,\ \delta\ \ &\text{monomorphisms}\ \ &\Longrightarrow
			\ \sigma\ \
			&\text{monomorphism};\\
\text{(ii)}&\ \ &\gamma,\ \delta\ \ &\text{epimorphisms}\ \ &\Longrightarrow
			\ \sigma\ \
			&\text{epimorphism};\\
\text{(iii)}&\ \ &\gamma,\ \delta\ \ &\text{isomorphisms}\ \ &\Longrightarrow
			\ \sigma\ \
			&\text{isomorphism}.
\end{alignat*}
A monomorphism is injective, an epimorphism is surjective, and an isomorphism
is bijective.  The important point is that the converse of the Lemma is not
true: if~$\sigma$ is an isomorphism then it says nothing about the properties
(or even the existence) of~$\gamma$ and~$\delta$.  Note that~(iii) follows
immediately from~(i) and~(ii).  The proof can be found in Mac Lane and
Birkhoff~\cite{MacLane} or Hungerford~\cite{Hungerford}, for example.

\section{Semidirect and Direct Extensions}
\seclabel{semidext}


Assume now that~$\omega$ defined by~\eqref{cocycle} is a coboundary.
By~\eqref{cobdiff} there exists an equivalent extension with~\hbox{$\omega
\equiv 0$}.  For that equivalent extension~$\tau$ is a homomorphism and
condition~(ii) at the end of \secref{extension} is satisfied. Thus the
sequence
\begin{equation}
\xymatrix@M=4pt{
	{\LieAx} & {\LieA} \ar[l]_{\tau} & 0 \ar[l]
	}
	\eqlabel{sdpdef}
\end{equation}
is an exact sequence of Lie algebra homomorphisms, as well as the sequence
given by~\eqref{extdef}. We then say that the extension is a semidirect
extension (or a semidirect sum of algebras)\index{semidirect sum|(} by analogy
with the group case. More generally, we say that~$\LieAx$ splits if it is
isomorphic to a semidirect sum, which corresponds to~$\omega$ being a
coboundary, not necessarily zero. If~$\LieAxb$ is not Abelian,
then~\eqref{samerho} is not satisfied and two equivalent extensions (or two
different choices of~$\tau$) do not necessarily lead to the same~$\rho$.

Representing elements of~$\LieAx$ as 2-tuples, as in \secref{abelianext}, we
can derive the bracket in~$\LieAx$ for a semidirect sum.  The difference
is that~$\tau$ is a homomorphism so that
\[
	{\lpb(\alpha_1,0),(\beta_1,0)\rpb}_\LieAx =
		{\lpb\tau\,\alpha_1\com\tau\,\beta_1\rpb}_\LieAx
	= \tau\,{\lpb\alpha_1\com\beta_1\rpb}_\LieA
	= ({\lpb\alpha_1\com\beta_1\rpb}_\LieA\com 0),
\]
and~$\LieAxb$ is not assumed Abelian,
\[
	{\lpb(0,\alpha_2),(0,\beta_2)\rpb}_\LieAx =
		{\lpb i\,\alpha_2\com i\,\beta_2\rpb}_\LieAx
	= i\,{\lpb\alpha_2\com\beta_2\rpb}_\LieAxb
	= (0\com {\lpb\alpha_2\com\beta_2\rpb}_\LieAxb),
\]
which together with~\eqref{a1b2com} gives
\begin{equation}
	{\lpb\alpha,\beta\rpb}_\LieAx =
		\Bigl({\lpb\alpha_1\com\beta_1\rpb}_\LieA\, , \,
		\rho_{\alpha_1}\,\beta_2 - \rho_{\beta_1}\,\alpha_2
		+ {\lpb\alpha_2\com\beta_2\rpb}_\LieAxb\Bigr),
	\eqlabel{sdbracket}
\end{equation}
Verifying Jacobi for~\eqref{sdbracket} we find the~$\rho$ must also satisfy
\[
	\rho_{\alpha_1}\,{\lpb\beta_2\com\gamma_2\rpb}_\LieAxb
		= {\lpb\rho_{\alpha_1}\,\beta_2\com\gamma_2\rpb}_\LieAxb
		+ {\lpb\beta_2\com\rho_{\alpha_1}\,\gamma_2\rpb}_\LieAxb\, ,
\]
which is trivially satisfied if~$\LieAxb$ is Abelian, but in general this
condition states that~$\rho_\alpha$ is a derivation on~$\LieAxb$.

Now consider the case where~$i^{-1}$ is a homomorphism and
\[
	\ker i^{-1} = \range \tau.
\]
Then the sequence
\[\xymatrix@M=4pt{
	0 \ar@_{->}@<-.5ex>[r]
	& {\LieAxb} \ar@_{->}@<-.5ex>[l] \ar@_{->}@<-.5ex>[r]_{i}
	& {\LieAx} \ar@_{->}@<-.5ex>[l]_{i^{-1}} \ar@_{->}@<-.5ex>[r]_{\pi}
	& {\LieA} \ar@_{->}@<-.5ex>[l]_{\tau} \ar@_{->}@<-.5ex>[r]
	& 0 \ar@_{->}@<-.5ex>[l]
	}
\]
is exact in both directions and, hence, both~$i$ and~\hbox{$\pi=\tau^{-1}$}
are bijections.  The action of~$\LieA$ on~$\LieAxb$ is
\[
	\rho_{\alpha}\,\eta = i^{-1}{\lpb\tau\,\alpha\com i\eta\rpb}_\LieAx
		 = {\lpb i^{-1}\tau\,\alpha\com \eta\rpb}_\LieAxb = 0
\]
since by exactness~\hbox{$i^{-1}\circ\tau = 0$}. This is called a direct
sum. Note that in this case the role of~$\LieA$ and~$\LieAxb$ is
interchangeable and they are both ideals in~$\LieAx$. The bracket in~$\LieAx$
is easily obtained from~\eqref{sdbracket} by letting~\hbox{$\rho=0$},
\begin{equation}
	{\lpb\alpha,\beta\rpb}_\LieAx =
		\Bigl({\lpb\alpha_1\com\beta_1\rpb}_\LieA\com
		{\lpb\alpha_2\com\beta_2\rpb}_\LieAxb\Bigr).
	\eqlabel{dbracket}
\end{equation}
Semidirect and direct extensions play an important role in physics. A simple
example of a semidirect sum structure is when $\LieA$ is the Lie
algebra~$\sothree$ associated with the rotation group $\SOthree$ and $\LieAxb$
is $\reals^3$. Their semidirect sum is the algebra of the six parameter
Euclidean group of rotations and translations\index{semidirect sum!and
Euclidean group}. This algebra can be used in a Lie--Poisson bracket to
describe the dynamics of the heavy top\index{rigid body!heavy top} (see for
example~\cite{Holmes1983,Marsden1984,Vinogradov1977}). We have already
discussed the semidirect sum in \secref{lowbetaRMHD}. The bracket
\eqref{RMHDbrak} is a semidirect sum, with~$\LieA$ the algebra of the group of
volume-preserving diffeomorphisms and~$\LieAxb$ the Abelian Lie algebra of
functions on~$\reals^2$. The action is just the adjoint\index{action!adjoint}
action~\hbox{$\rho_\alpha\,v \ldef
\lpb\alpha\com v\rpb$} obtained by identifying~$\LieA$ and~$\LieAxb$.

In general, semidirect Lie--Poisson structures appear in systems where the
field variables are in some sense ``slaved'' to the base variable (the one
associated with~$\LieAx$)~\cite{Marsden1982,Thiffeault1998}.  Here, the
advected quantities are forced to move on the coadjoint orbits\index{coadjoint
orbit} of the Lie group~$\LieG$.  This is seen directly from the equations of
motion~\eqref{motion}, since, for a semidirect sum,
\[
	\dotfv^\mu = -\lpb \frac{\fd H}{\fd \fv^0}\com\,\fv^\mu\rpb^\dagger
	= -\ad^\dagger_{{\fd H}/{\fd \fv^0}}\,\fv^\mu,
\]
which is by definition the infinitesimal generator of the coadjoint orbits of
the Lie group~\cite{MarsdenRatiu} (see \secref{lpbasic}).  For example, the
coadjoint orbits of~$\SOthree$ are spheres, so the semidirect product%
\footnote{Semidirect \emph{product} is the term used for groups, semidirect
\emph{sum} for algebras.}
\index{semidirect product}
of~$SO(3)$ and~$\reals^3$ leads to a physical system where the dynamics are
confined to spheres, which naturally describes rigid body motion.  In other
words, the coadjoint orbits of the semidirect product of~$\LieG$
and~$\reals^3$ are isomorphic to the coadjoint orbits of~$\LieG$.  We shall
have more to say on this in~\secref{SDPstab}.

A Lie--Poisson bracket built from a direct sum is just a sum of the separate
brackets. The dynamical interaction between the variables can only come from
the Hamiltonian or from constitutive equations.  For example, in the
baroclinic instability model of two superimposed two fluid layers with
different potential vorticities, the two layers are coupled through the
potential vorticity relation~\cite{McLachlan1997}.  A very similar model
with a direct sum structure exists in MHD for studying magnetic
reconnection~\cite{Cafaro1998}.
\index{semidirect sum|)}

\subsection{Classification of Splitting Extensions}

\index{classification!of splitting extensions|(}
\index{semidirect sum!classification}
We now briefly mention the connection between the first cohomology group and
splitting extensions.  This will not be used directly in the classification
scheme of~\chref{classext}, but we include it for completeness.  We assume in
this section that~$\LieAxb$ is Abelian.  In~\eqref{sdpdef} we had chosen the
canonical~$\tau$,~\hbox{$\tau(\alpha) = (\alpha,0)$}.  Now suppose we use
instead
\begin{equation}
	\tau'(\alpha) = \l(\alpha\com\nu(\alpha)\r).
\end{equation}
Here~$\nu$ is a linear map from~$\LieA$ to~$\LieAxb$ and is thus an element
of~\hbox{$C^1(\LieA,\LieAxb)$}, a 1-cochain.  If~$\tau'$ is a Lie algebra
homomorphism,
\begin{equation}
	\tau'({\lpb\alpha\com\beta\rpb}_\LieA) =
		\l({\lpb\alpha\com\beta\rpb}_\LieA\com
		\nu({\lpb\alpha\com\beta\rpb}_\LieA)\r)
	\eqlabel{taupdef}
\end{equation}
must be equal to
\begin{equation}
	{\lpb\tau'(\alpha)\com\tau'(\beta)\rpb}_\LieAx =
		\l({\lpb\alpha\com\beta\rpb}_\LieA\com
		\rho_\alpha\,\nu(\beta) - \rho_\beta\,\nu(\alpha)
		\r)
	\eqlabel{tauphomo}
\end{equation}
 subtracting~\eqref{taupdef}
and~\eqref{tauphomo} gives
\begin{equation}
	\rho_\alpha\,\nu(\beta) - \rho_\beta\,\nu(\alpha)
		- \nu({\lpb\alpha\com\beta\rpb}_\LieA)
	= s\,\nu(\alpha,\beta) = 0,
\end{equation}
from~\eqref{2cobound}.  Hence~$\nu$ is a cocycle, with coboundaries given
by
\begin{equation}
	\nu(\alpha) = \rho_\alpha\,\eta_0,\qquad \eta_0 \in \LieAxb,
	\eqlabel{splitcobound}
\end{equation}
The first cohomology group~$H^1_\rho(\LieA,\LieAxb)$ classifies splitting
extensions of~$\LieAx$ by~$\LieAxb$ modulo those given in terms of the
coboundaries~\eqref{splitcobound}.

\index{classification!of splitting extensions|)}

\chapter{Classification of Extensions of a Lie Algebra}
\chlabel{classext}

In this chapter we return to the main problem introduced in
\secref{theproblem}: the classification of algebra
extensions built by forming~$n$-tuples of elements of a single Lie
algebra~$\LieA$.  The elements of this Lie algebra~$\LieAx$ are
written\index{ntuples@$n$-tuples} as~\hbox{$\alpha \ldef
\l(\alpha_1,\dots,\alpha_n\r)$},~\hbox{$\alpha_i \in \LieA$}, with a
bracket defined by\index{bracket!for~$n$-tuples}
\begin{equation}
	{\lpb\alpha\com\beta\rpb}_\lambda = {\W_\lambda}^{\mu\nu}\,
		\lpb\alpha_\mu\com\beta_\nu\rpb,
	\tag{\ref{eq:extbrack}}
\end{equation}
where~${\W_\lambda}^{\mu\nu}$ are constants. We will call~$n$ the \emph{order}
of the extension\index{extension!order of}.  Recall (see
\secref{theproblem}) that the~$\W$'s are symmetric in their upper indices,
\begin{equation}
	{\W_\lambda}^{\mu\nu} = {\W_\lambda}^{\nu\mu}\,,
	\tag{\ref{eq:upsym}}
\end{equation}
and commute,
\begin{equation}
	\W^{(\nu)}\,\W^{(\sigma)} = \W^{(\sigma)}\,\W^{(\nu)},
	\tag{\ref{eq:Wcommute}}
\end{equation}
where the~\hbox{$n\times n$} matrices~$\W^{(\nu)}$ are defined
by~${{[\W^{(\nu)}]}_\lambda}^{\mu} := {\W_\lambda}^{\nu\mu}$.  Since
the~$\W$'s are 3-tensors we can also represent their elements by matrices
obtained by fixing the lower index,\index{extension!representation by
matrices}
\begin{equation}
	\W_{(\lambda)}\ :\ {\l[\W_{(\lambda)}\r]}^{\mu\nu} :=
	{\W_{\lambda}}^{\mu\nu},
	\eqlabel{lowindex}
\end{equation}
which are symmetric but do not commute.  Either collection of
matrices,~\eqref{Wupdef} or~\eqref{lowindex}, completely describes the Lie
bracket, and which one we use will be understood by whether the parenthesized
index is up or down.

\index{classification}
What do we mean by a classification?  A classification is achieved if we
obtain a set of normal forms for the extensions which are independent, that is
not related by linear transformations.  We use linear transformations because
they preserve the Lie--Poisson structure---they amount to transformations of
the~$\W$ tensor.  We thus begin by assuming the most general~$\W$ possible.

We first show in \secref{directprod} how an extension can be broken down into
a direct sum of degenerate subblocks (degenerate in the sense that the
eigenvalues have multiplicity greater than unity). The classification scheme
is thus reduced to the study of a single degenerate subblock. In
\secref{classcoho} we couch our particular extension problem in terms of the
Lie algebra cohomology language of \secref{extension} and apply the techniques
therein. The limitations of this cohomology approach are investigated in
\secref{furthertrans}, and we look at other coordinate transformations
that do not necessarily preserve the extension structure of the algebra, as
expressed in diagram~\eqref{equivext2}.  In \secref{Leibniz} we introduce a
particular type of extension, called the Leibniz extension, that is in a
sense the ``maximal'' extension. Finally, in \secref{lowdimext} we give an
explicit classification of solvable extensions up to order four.

\section{Direct Sum Structure}
\seclabel{directprod}

\index{extension!direct sum|(}
A set of commuting matrices can be put into simultaneous block-diagonal form
by a coordinate transformation, each block corresponding to a degenerate
eigenvalue~\cite{Suprunenko}. Let us denote the change of basis by a
matrix~${\M_{\beta}}^{\bar\alpha}$, with
inverse~${\l(M^{-1}\r)_{\bar\alpha}}^{\beta}$, such that the
matrix~\hbox{${\Wt}^{(\nu)}$}, whose components are given by
\[
	{\Wt_{\bar\beta}}\,{}^{\bar\alpha\nu} =
		{(M^{-1})_{\bar\beta}}^{\lambda} \,
		{\W_{\lambda}}^{\mu\nu} \, {\M_{\mu}}^{\bar\alpha}\ ,
\]
is in block-diagonal form for all~$\nu$~\cite{Suprunenko}.
However,~${\W_\lambda}^{\mu\nu}$ is a 3-tensor and so the third index is
also subject to the coordinate change:
\[
	{\Wb_{\bar\beta}}^{\bar\alpha\bar\gamma} =
	{\Wt_{\bar\beta}}\,{}^{\bar\alpha\nu}
		{\M_{\nu}}^{\bar\gamma}\, .
\]
This last step only adds linear combinations of the~$\Wt^{(\nu)}$'s
together, so the~$\Wt^{(\nu)}$'s and
the~$\Wb^{(\bar\gamma)}$'s have the same block-diagonal structure.
Note that the~${\Wb_{\bar\beta}}^{\bar\alpha\bar\gamma}$ are still
symmetric in their upper indices, since this property is preserved by a change
of basis:
\begin{align*}
	{\Wb_{\bar\beta}}^{\bar\alpha\bar\gamma} &=
		{(M^{-1})_{\bar\beta}}^{\lambda} \,
		{\W_{\lambda}}^{\mu\nu} \, {\M_{\mu}}^{\bar\alpha}\,
		{\M_{\nu}}^{\bar\gamma}\,  \\
	&= {(M^{-1})_{\bar\beta}}^{\lambda} \,
		{\W_{\lambda}}^{\nu\mu} \, {\M_{\nu}}^{\bar\alpha}\,
		{\M_{\mu}}^{\bar\gamma}\,
		\qquad\text{(Relabeling~$\mu$ and~$\nu$)}\\
	&= {(M^{-1})_{\bar\beta}}^{\lambda} \,
		{\W_{\lambda}}^{\mu\nu} \, {\M_{\mu}}^{\bar\gamma}\,
		{\M_{\nu}}^{\bar\alpha} \\
	&= {\Wb_{\bar\beta}}^{\bar\gamma\bar\alpha}\ .
\end{align*}
So from now on we just assume that we are working in a basis where
the~$\W^{(\nu)}$'s are block-diagonal and symmetric in their upper indices;
this symmetry means that if we look at a~$\W$ as a cube, then in the
block-diagonal basis it consists of smaller cubes along the main
diagonal. This is the 3-tensor equivalent of a block-diagonal matrix, as
illustrated in \figref{extcubes}, a pictorial representation of a direct sum
of extensions.
\begin{figure}
\centerline{\psfig{file=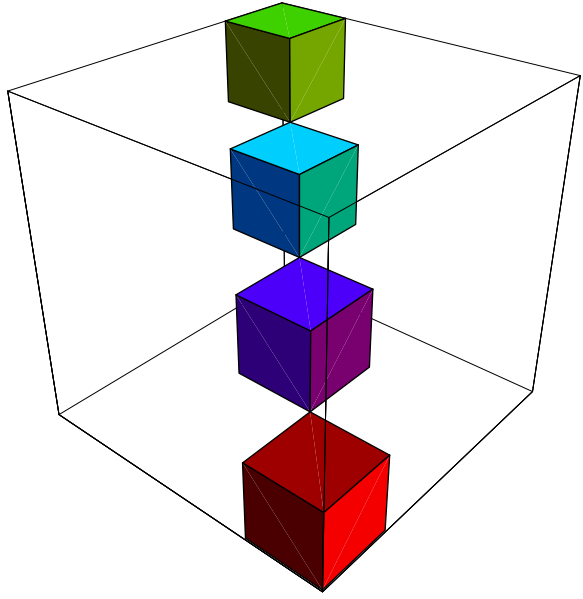,width=3in}}
\caption{Schematic representation of the 3-tensor~$\W$ for a direct sum of
extensions.  The cubes represent potentially nonzero elements.}
\figlabel{extcubes}
\end{figure}

\subsection{Example: three-field model of MHD}

We consider as an example of a direct sum structure a three-field model of MHD
due to Hazeltine~\cite{Hazeltine1983b,Hazeltine1985}.  In addition to the
vorticity~$\vort$ and the magnetic flux~$\magf$ (see \secref{lowbetaRMHD}),
the model also includes a field~$\plasd$ which measures plasma density
perturbations.  The model includes as limits the RMHD system
of~\secref{lowbetaRMHD} and the Charney--Hasegawa--Mima
equation~\cite{Horton1994}.  We thus have~\hbox{$\fv = (\vort,\magf,\plasd)$},
with the Hamiltonian
\begin{equation}
	\Ham = \half \lang |\grad\elecp|^2 + |\grad\magf|^2
		+ \alpha\,\plasd^2\rang,
	\eqlabel{tfRMHDHam}
\end{equation}
and bracket represented by the matrices
\[
	\W^{(1)} = \begin{pmatrix}
		1 & 0 & 0 \\
		0 & 1 & 0 \\
		0 & 0 & 1
	\end{pmatrix},\quad
	\W^{(2)} = \begin{pmatrix}
		0 & 0 & 0 \\
		1 & 0 & 1 \\
		0 & 0 & 0
	\end{pmatrix},\quad
	\W^{(3)} = \begin{pmatrix}
		0 & 0 & 0 \\
		0 & 1 & 0 \\
		1 & 0 & 1
	\end{pmatrix}.\quad
\]
The matrices commute and obey the symmetry~\eqref{upsym}, so they form a good
bracket.  As in~\secref{lowbetaRMHD}, the electric potential is denoted
by~$\elecp$ and the electric current by~$\ecurrent$.  The equations of motion
are given by
\begin{equation}
\begin{split}
	\dotvort &= \lpb\vort\com\elecp\rpb + \lpb\magf\com\ecurrent\rpb,\\
	\dotmagf &= \lpb\magf\com\elecp\rpb
		+ \alpha\,\lpb\plasd\com\magf\rpb,\\
	\dot\plasd &= \lpb\plasd\com\elecp\rpb + \lpb\magf\com\ecurrent\rpb.
	\eqlabel{tfRMHDeom}
\end{split}
\end{equation}
The~$\W^{(\mu)}$'s are not in block triangular form, and since~$\W^{(3)}$ has
eigenvalues which are not threefold degenerate we know the extension can be
blocked-up further.  Indeed, the coordinate
transformation~\hbox{$\eta^{\bar\mu} = \fv^\nu\,{\M_\nu}^{\bar\mu}$}, with
\begin{equation}
	\M = \begin{pmatrix}
		0 & 0 & 1 \\
		0 & 1 & 0 \\
		1 & 0 & -1
	\end{pmatrix},
	\eqlabel{mtrans}
\end{equation}
will transform the extension to
\[
	\Wb^{(1)} = \l(\begin{array}{cc|c}
		1 & 0 & 0 \\
		0 & 1 & 0 \\ \hline
		0 & 0 & 0
	\end{array}\r),\quad
	\Wb^{(2)} = \l(\begin{array}{cc|c}
		0 & 0 & 0 \\
		1 & 0 & 0 \\ \hline
		0 & 0 & 0
	\end{array}\r),\quad
	\Wb^{(3)} = \l(\begin{array}{cc|c}
		0 & 0 & 0 \\
		0 & 0 & 0 \\ \hline
		0 & 0 & 1
	\end{array}\r),\quad
\]
where we have explicitly indicated the blocks.  The extension is also
block-diagonal in the alternate, lower-indexed representation,
\[
	\Wb_{(\bar 1)} = \l(\begin{array}{cc|c}
		1 & 0 & 0 \\
		0 & 0 & 0 \\ \hline
		0 & 0 & 0
	\end{array}\r),\quad
	\Wb_{(\bar 2)} = \l(\begin{array}{cc|c}
		0 & 1 & 0 \\
		1 & 0 & 0 \\ \hline
		0 & 0 & 0
	\end{array}\r),\quad
	\Wb_{(\bar 3)} = \l(\begin{array}{cc|c}
		0 & 0 & 0 \\
		0 & 0 & 0 \\ \hline
		0 & 0 & 1
	\end{array}\r).\quad
\]
This is what was meant by ``cubes'' at the end of the previous section.

At the bracket level the variables~${\eta}^{\bar 1}$ and~${\eta}^{\bar 2}$ are
decoupled from~${\eta}^{\bar 3}$.  But under the transformation~\eqref{mtrans}
the Hamiltonian~\eqref{tfRMHDHam} becomes
\[
	\bar\Ham = \half \lang |\grad({\eta}^{\bar 1}
		+ {\eta}^{\bar 3})|^2
		+ |\grad{\eta}^{\bar 2}|^2
		+ \alpha\,|{\eta}^{\bar 1}|^2\rang.
\]
The new equations of motion are thus
\[
\begin{split}
	\dot{\eta}^{\bar 1} &= \bigl[{\eta}^{\bar 1}\com\bar\elecp\bigr]
		+ \bigl[{\eta}^{\bar 2}\com\bar\ecurrent\,\bigr],\\
	\dot{\eta}^{\bar 2} &= \bigl[{\eta}^{\bar 2}
		\com\bar\elecp - \alpha\,{\eta}^{\bar 1}\,\bigr],\\
	\dot{\eta}^{\bar 3} &= \bigl[{\eta}^{\bar 3}\com\bar\elecp\,\bigr].
\end{split}
\]
with~\hbox{$\lapl\bar\elecp \ldef {\eta}^{\bar 1} + {\eta}^{\bar 3}$}
and~\hbox{$\bar\ecurrent \ldef \lapl{\eta}^{\bar 2}$}.  The
variable~$\eta^{\bar 3}$ is still coupled to the other variables through the
defining relation for~$\bar\elecp$.
\index{extension!direct sum|)}

\subsection{Lower-triangular Structure}
\seclabel{lowertri}

\index{extension!lower-triangular structure|(}
Block-diagonalization is the first step in the classification: each subblock
of~$\W$ is associated with an ideal (hence, a subalgebra) in the
full~$n$-tuple algebra~$\LieA$.  (A subset~$\LieAxb\subseteq\LieAx$ is an
ideal\index{ideal} in the Lie algebra~$\LieAx$
if~\hbox{$\lpb\LieAx\com\LieAxb\rpb\subseteq\LieAxb$}.  Ideals are
subalgebras.)  Hence, by the definition of
\secref{semidext}, the algebra~$\LieA$ is a direct sum of the algebra denoted
by each subblock. Each of these algebras can be studied independently, that is
we can focus our attention on a \emph{single subblock}.  So from now on we
assume that we have~$n$ commuting matrices, each with~$n$-fold degenerate
eigenvalues. The eigenvalues can, however, be different for each matrix.

Such a set of commuting matrices can be put into lower-triangular form by a
coordinate change, and again the transformation of the third index preserves
this structure (though it can change the eigenvalue of each matrix).  The
eigenvalue of each matrix lies on the diagonal; we denote the eigenvalue
of~$\W^{(\mu)}$ by~$\ev^{(\mu)}$. We write the quantity~${\W_1}^{\mu\nu}$ as
the matrix
\[
	{\W_{(1)}} = \begin{pmatrix}
		\ev^{(1)} & 0 & 0 & \cdots & 0 \\
		\ev^{(2)} & 0 & 0 & \cdots & 0 \\
		\vdots  &   &   &        & \vdots \\
		\ev^{(n)} & 0 & 0 & \cdots & 0
		\end{pmatrix},
\]
which consists of the first row of the lower-triangular matrices~$\W^{(\mu)}$
as prescribed by \eqref{lowindex}.  Evidently, the symmetry of~${\W_{(1)}}$
requires
\[
	\ev^{(\nu)} = \evone\,{\delta_1}^\nu\, ,
\]
that is, all the matrices~$\W^{(\mu)}$ are nilpotent (their eigenvalues
vanish) except for~$\W^{(1)}$ when~\hbox{$\evone\ne 0$}.  If this first
eigenvalue is nonzero then it can be scaled to~$\evone=1$ by the coordinate
transformation~${\M_{\nu}}^{\bar\alpha} =
\evone^{-1}~{\delta_\nu}^{\bar\alpha}$. We will use the
symbol~$\zerorone$ to mean a variable which can take the value 0 or 1.
\figref{solvextpic} shows the structure, with~$\evone=0$, of a degenerate
extension, after lower-triangularity and symmetry of the upper indices
of~$\W$ are taken into account.
\begin{figure}
\centerline{\psfig{file=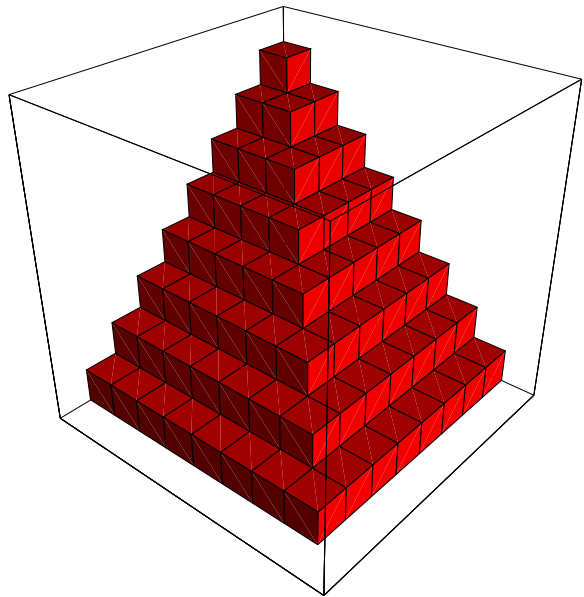,width=3in}}
\caption{Schematic representation of the 3-tensor~$\W$ for a solvable
extension.  The cubes represent potentially nonzero elements.  The vertical
axis is the lower index~$\lambda$ of~${\W_{\lambda}}^{\mu\nu}$, and the two
horizontal axes are the symmetric upper indices~$\mu$ and~$\nu$.  The origin
is at the top-rear.  The pyramid-like structure is a consequence of the
symmetry of~$\W$ and of its lower-triangular structure in this basis.}
\figlabel{solvextpic}
\end{figure}
\index{extension!lower-triangular structure|)}

\section{Connection to Cohomology}
\seclabel{classcoho}

\index{classification!and cohomology}
\index{extension!connection to cohomology}
We now bring together the abstract notions of \chref{cohoext} with
the~$n$-tuple extensions of \secref{theproblem}.  It is shown in
\secref{prelimsplit} that we need only classify the case
of~\hbox{$\zerorone=0$}.  This case will be seen to correspond to solvable
extensions, which we classify in \secref{solvext}.

\subsection{Preliminary Splitting}
\seclabel{prelimsplit}

Assume we are in the basis described at the end of \secref{directprod} and,
for now, suppose~$\evone = 1$.  To place the structure of~$\W$ in the context
of Lie algebras, we first give some definitions.  The \emph{derived
series}~$\LieA^{(k)}$\index{derived series} of~$\LieA$ has terms
\begin{align}
	\LieA^{(0)} &= \LieA\nonumber \\
	\LieA^{(1)} &= \lpb\LieA\com\LieA\rpb\nonumber \\
	\LieA^{(2)} &= \lpb\LieA^{(1)}\com\LieA^{(1)}\rpb\nonumber \\
	&\ \, \vdots\nonumber\\
	\LieA^{(k)} &= \lpb\LieA^{(k-1)}\com\LieA^{(k-1)}\rpb,
\end{align}
where by~$\lpb\LieA\com\LieA\rpb$ we mean the set obtained by taking all the
possible Lie brackets of elements of~$\LieA$.  The \emph{lower
central}\index{lower central series}
series~$\LieA^k$ has terms defined by
\begin{align}
	\LieA^{0} &= \LieA\nonumber \\
	\LieA^{1} &= \lpb\LieA\com\LieA\rpb\nonumber \\
	\LieA^{2} &= \lpb\LieA\com\LieA^{1}\rpb\nonumber \\
	&\ \, \vdots\nonumber\\
	\LieA^{k} &= \lpb\LieA\com\LieA^{(k-1)}\rpb.
	\eqlabel{lowercent}
\end{align}
An algebra~$\LieA$ is said to be \emph{solvable}\index{solvable algebra} if
its derived series terminates,~$\LieA^{(k)}=0$, for some~$k$.  An
algebra~$\LieA$ is said to be \emph{nilpotent}\index{nilpotent algebra} if its
lower central series terminates,~$\LieA^{k}=0$, for some~$k$.  Note that a
nilpotent algebra is solvable, but not vice-versa~\cite{Jacobson}.

The set of elements of the form~$\beta =
\l(0,\beta_2,\dots,\beta_n\r)$ is a nilpotent ideal in~$\LieAx$ that we
denote by~$\LieAxb$ ($\LieAxb$ is thus a solvable subalgebra).  To see this,
observe that~\eqref{lowercent} involves nested brackets, so that the
elements~$\LieAxb^{k}$ of the lower central series will involve~$k$th powers
of the~$\W^{(\mu)}$.  But since the~$\W^{(\mu)}$ with~\hbox{$\mu>1$} are
lower-triangular with zeros along the diagonal, we
have~$(\W^{(\mu)})^{n-1}=0$, and the lower central series must eventually
vanish.

Because~$\LieAxb$ is an ideal, we can construct the algebra~$\LieA =
\LieAx/\LieAxb$, so that~$\LieAx$ is an extension of~$\LieA$ by~$\LieAxb$.
If~$\LieA$ is semisimple, then~$\LieAxb$ is the radical of~$\LieAx$ (the
maximal solvable ideal).\index{semisimple algebra} It is easy to see that the
elements of~$\LieA$ embedded in~$\LieAx$ are of the form~$\alpha =
\l(\alpha_1,0,\dots,0\r)$.  We will now show that~$\LieAx$ splits; that is,
there exist coordinates in which~$\LieAx$ is manifestly the semidirect sum
of~$\LieA$ and the (in general non-Abelian) algebra~$\LieAxb$.

In~\apxref{woneident} we give a lower-triangular coordinate transformation
that makes~$\W^{(1)}=I$, the identity matrix.  Assuming we have effected this
transformation, the mappings~$i$,~$\pi$, and~$\tau$ of \secref{extension} are
given by
\begin{alignat*}{3}
	i & :\LieAxb\ &\longrightarrow \LieAx,\ \ \ 
		&i(\alpha_2,\dots,\alpha_n) = (0,\alpha_2,\dots,\alpha_n),
		\nonumber\\
	\pi & :\LieAx\ &\longrightarrow \LieA,\ \ \  
		&\pi(\alpha_1,\alpha_2,\dots,\alpha_n) 
		= \alpha_1,\\
	\tau & :\LieA\ &\longrightarrow \LieAx,\ \ \  
		&\tau(\alpha_1) 
		= (\alpha_1,0,\dots,0),\nonumber
\end{alignat*}
and the cocycle\index{cocycle} of Eq.~\eqref{cocycle} is
\[
\begin{split}
	i\,\omega(\alpha,\beta) &=
	{\lpb\tau\,\alpha\com\tau\,\beta\rpb}_\LieAx
	- \tau\,{\lpb\alpha\com\beta\rpb}_\LieA\\
	&= {\lpb(\alpha_1,0,\dots,0)\com(\beta_1,0,\dots,0)\rpb}_\LieAx
	- (\lpb\alpha_1\com\beta_1\rpb,0,\dots,0)\\
	&= \l({\W_{1}}^{11}\,\lpb\alpha_1\com\beta_1\rpb,0,\dots,0\r)
	- (\lpb\alpha_1\com\beta_1\rpb,0,\dots,0)\\
	&= 0,
\end{split}
\]
since~\hbox{${\W_{1}}^{11}=1$}.  Hence, the extension is a semidirect sum.
The coordinate transformation that made~$\W^{(1)}=I$ removed a coboundary,
making the above cocycle vanish identically.  For the case where~$\LieA$ is
finite-dimensional and semisimple,\index{semisimple algebra} we have an
explicit demonstration of the Levi decomposition theorem: any
finite-dimensional\footnote{The inner bracket can be infinite dimensional, but
the order of the extension is finite.} Lie algebra~$\LieAx$ (of characteristic
zero) with radical~$\LieAxb$ is the semidirect sum of a semisimple Lie
algebra~$\LieA$ and~$\LieAxb$~\cite{Jacobson}.

\subsection{Solvable Extensions}
\seclabel{solvext}

\index{extension!solvable}
Above we assumed the eigenvalue~$\evone$ of the first matrix was unity;
however, if this eigenvalue vanishes, then we have a solvable algebra
of~$n$-tuples to begin with.  Since~$n$ is arbitrary we can study these two
solvable cases together.

Thus, we now suppose~$\LieAx$ is a solvable Lie algebra of~$n$-tuples (we
reuse the symbols~$\LieAx$,~$\LieA$, and~$\LieAxb$ to parallel the notation of
\secref{cohoalgebra}), where all of the the~$\W^{(\mu)}$'s are
lower-triangular with zeros along the diagonal.  Note
that~$\W^{(n)}=0$, so the set of elements of the form~$\alpha =
(0,\dots,0,\alpha_n)$ forms an Abelian subalgebra of~$\LieAx$.  In fact, this
subalgebra is an ideal.  Now assume~$\LieAx$ contains an Abelian ideal of
order~\hbox{$n-m$} (the order of this ideal is at least~$1$), which we denote
by~$\LieAxb$.  The elements of~$\LieAxb$ can always be cast in the form
\[
	\alpha = (0,\dots,0,\alpha_{m+1},\dots,\alpha_n)
\]
via a coordinate transformation that preserves the lower-triangular,
nilpotent form of the~${\W}^{(\mu)}$.

We also denote by~$\LieA$ the algebra of~$m$-tuples with bracket
\[
	{{\lpb(\alpha_1,\dots,\alpha_{m})
	\com(\beta_1,\dots,\beta_{m})\rpb}_\LieA}_\lambda
	= \sum_{\mu,\nu = 1}^{m} {\W_\lambda}^{\mu\nu}\,
	\lpb\alpha_\mu\com\beta_\nu\rpb\,,\ \ \lambda=1,\dots,m.
\]
It is trivial to show that~$\LieA = \LieAx/\LieAxb$, so that~$\LieAx$ is an
extension of~$\LieA$ by~$\LieAxb$.  Since~$\LieAxb$ is Abelian we can use the
formalism of \secref{cohoalgebra} (the other case we used above was
for~$\LieAxb$ non-Abelian but where the extension was semidirect).  The
injection and projection maps are given by
\begin{alignat*}{3}
	i& : \LieAxb\ &\longrightarrow \LieAx,\ \ \ 
		&i(\alpha_{m+1},\dots,\alpha_n)
		= (0,\dots,0,\alpha_{m+1},\dots,\alpha_n),\nonumber\\
	\pi& : \LieAx\ &\longrightarrow \LieA,\ \ \  
		&\pi(\alpha_1,\alpha_2,\dots,\alpha_n) 
		= (\alpha_1,\dots,\alpha_{m}),\\
	\tau & :\LieA\ &\longrightarrow \LieAx,\ \ \  
		&\tau(\alpha_1,\dots,\alpha_{m}) 
		= (\alpha_1,\dots,\alpha_{m},0,\dots,0).\nonumber
\end{alignat*}
From the definition of the action\index{action}, Eq.~\eqref{rhodef}, we have
for~\hbox{$\alpha \in \LieA$} and~\hbox{$\eta \in \LieAxb$},
\begin{equation}
\begin{split}
i\,\rho_\alpha\,\eta &= {\lpb \tau\,\alpha\com i\,\eta\rpb}_\LieAx\\
	&= {\lpb (\alpha_1,\dots,\alpha_{m},0,\dots,0)
		\com (0,\dots,0,\eta_{m+1},\dots,\eta_n)\rpb}_\LieAx
		\\
	&= \sum_{\mu=1}^{m}\,\sum_{\nu=m+1}^{n-1}(0,\dots,0,
		\W_{m+2}^{\,\,\mu\nu}{\lpb\alpha_\mu\com\eta_\nu\rpb}
		,\dots,
		{\W_n}^{\mu\nu}{\lpb\alpha_\mu\com\eta_\nu\rpb}).
	\eqlabel{solvaction}
\end{split}
\end{equation}
In addition to the action, the solvable extension is also characterized by the
cocycle defined in Eq.~\eqref{cocycle},
\index{cocycle}
\begin{align}
	i\,\omega(\alpha,\beta) &=
	{\lpb\tau\,\alpha\com\tau\,\beta\rpb}_\LieAx
	- \tau\,{\lpb\alpha\com\beta\rpb}_\LieA\nonumber\\
	&= {\lpb(\alpha_1,\dots,\alpha_{m},0,\dots,0)
		\com(\beta_1,\dots,\beta_{m},0,\dots,0)\rpb}_\LieAx
		\nonumber\\
	&\phantom{=} - \tau\,{\lpb(\alpha_1,\dots,\alpha_{m})
		\com(\beta_1,\dots,\beta_{m})\rpb}_\LieA\nonumber\\
	&= \sum_{\mu,\nu=1}^{m}(0,\dots,0,
		\W_{m+1}^{\!\!\mu\nu}{\lpb\alpha_\mu\com\beta_\nu\rpb}
		,\dots,{\W_n}^{\mu\nu}{\lpb\alpha_\mu\com\beta_\nu\rpb}).
	\eqlabel{solvcocycle}
\end{align}
We can illustrate which parts of the~$\W$'s contribute to the action and
which to the cocycle by writing
\begin{equation}
	\W_{(\lambda)} = \l(\begin{array}{c|c}
		{\bf\ww}_\lambda\ & \ \rule[0em]{0cm}{1em}{\bf r}_\lambda \\
		\hline
		\rule[0em]{0cm}{1em}{\bf r}_\lambda^T\   & {\bf 0}
	\end{array}\r),\ \ \lambda=m+1,\dots,n,
	\eqlabel{Wform}
\end{equation}
where the~${\bf\ww}_\lambda$'s are~\hbox{$m\times m$} symmetric matrices that
determine the cocycle~$\omega$ and the~${\bf r}_\lambda$'s are~\hbox{$m\times
(n-m)$} matrices that determine the action~$\rho$.  The zero matrix of
size~\hbox{$(n-m)\times(n-m)$} on the bottom right of the~$\W_{(\lambda)}$'s
appears as a consequence of~$\LieAxb$ being Abelian.

The algebra~$\LieA$ is completely characterized by the~$\W_{(\lambda)}$,
$\lambda = 1,\dots,m$.  Hence, we can look for the maximal Abelian ideal
of~$\LieA$ and repeat the procedure we used for the full~$\LieAx$.  It is
straightforward to show that although coordinate transformations of~$\LieA$
might change the cocycle~$\omega$ and the action~$\rho$, they will not alter
the \emph{form} of~\eqref{Wform}.

Recall that in \secref{cohoalgebra} we defined 2-coboundaries as 2-cocycles
obtained from 1-cochains by the coboundary operator,~$s$.  The 2-coboundaries
turned out to be removable obstructions to a semidirect sum structure.  Here
the coboundaries are associated with the parts of the~$\W_{(\lambda)}$ that
can be removed by (a restricted class of) coordinate transformations, as shown
below.

\index{coboundary|(}
Let us explore the connection between 1-cochains and coboundaries in the
present context.  Since a 1-cochain is just a linear mapping from~$\LieA$
to~$\LieAxb$, for~\hbox{$\alpha = (\alpha_1,\dots,\alpha_{m}) \in \LieA$} we
can write this as
\begin{equation}
	\omega^{(1)}_\mu(\alpha) =
		-\sum_{\lambda=1}^{m}{k_\mu}^\lambda\, \alpha_\lambda\,,
		\ \ \mu=m+1,\dots,n,
	\eqlabel{thecobound}
\end{equation}
where the~${k_\mu}^\lambda$ are arbitrary constants.  To find the form of a
2-coboundary we act on the 1-cochain~\eqref{thecobound} with the coboundary
operator; using~\eqref{2cobound} and~\eqref{solvaction} we obtain
\begin{align}
	\omega^{\rm cob}_\lambda(\alpha,\beta)
		&= (s\,\omega^{(1)})(\alpha,\beta)
		\nonumber\\
	&= \rho_\alpha\omega^{(1)}(\beta) + \rho_\beta\omega^{(1)}(\alpha)
		- \omega^{(1)}({\lpb\alpha\com\beta\rpb}_\LieA)
		\nonumber\\
	&= \sum_{\mu=1}^{m}\,\sum_{\nu=m+1}^{n}\,{\W_\lambda}^{\mu\nu}
		\lpb\alpha_\mu\com\omega^{(1)}_\nu(\beta)\rpb
		- \sum_{\mu=1}^{m}\,\sum_{\nu=m+1}^{n}\,{\W_\lambda}^{\mu\nu}
		\lpb\beta_\mu\com\omega^{(1)}_\nu(\alpha)\rpb\nonumber\\
	&\phantom{=} + \sum_{\mu,\nu,\sigma=1}^{m}{k_\lambda}^\sigma\,
		{\W_\sigma}^{\mu\nu}
		\lpb\alpha_\mu\com\beta_\nu\rpb.
	\eqlabel{cob1}
\end{align}
After inserting~\eqref{thecobound} into~\eqref{cob1} and relabeling, we obtain
the general form of a 2-coboundary
\[
	\omega^{\rm cob}_\lambda(\alpha,\beta) = \sum_{\mu,\nu=1}^{m}\,
		{\Wcob_\lambda}^{\mu\nu}
		\lpb\alpha_\mu\com\beta_\nu\rpb,
		\ \ \ \lambda=m+1,\dots,n,
\]
where
\begin{equation}
	{\Wcob_\lambda}^{\mu\nu} \ldef
		\sum_{\tau=1}^{m}\,
		{k_\lambda}^\tau\,{\W_\tau}^{\mu\nu}
		- \sum_{\sigma=m+1}^{n}\,
		\l(
			{k_\sigma}^\mu\,{\W_\lambda}^{\nu\sigma}
			+ {k_\sigma}^\nu\,{\W_\lambda}^{\mu\sigma}
		\r).
	\eqlabel{cob}
\end{equation}

\index{coboundary!removing}
To see how coboundaries are removed, consider the lower-triangular coordinate
transformation
\[
	\l[{\M_\sigma}^{\bar \tau}\r] = \l(\begin{array}{c|c}
		{\bf I}\ \  & \,\ {\bf 0} \\ \hline
		\rule[0em]{0cm}{1em}{\bf k}\ \ & \ c\,{\bf I}
	\end{array}\r),
\]
where~$\sigma$ labels rows.  This transformation subtracts~$\Wcob_{(\lambda)}$
from~$\W_{(\lambda)}$ for~\hbox{$\lambda>m$} and leaves the first~$m$ of the
$\W_{(\lambda)}$'s unchanged.  In other words, if~${\Wb}$ is the
transformed~$\W$,
\begin{equation}
	{\Wb_{(\lambda)}} =
	\begin{cases}
		\,\W_{(\lambda)}&\lambda = 1,\dots,m;\\[8pt]
		\l(\begin{array}{c|c}
			c^{-1}\,({\bf w}_\lambda - {\bf \Wcob}_\lambda)\ 
				& \ {\bf r}_\lambda \\ \hline
			{\bf r}_\lambda^T\   & {\bf 0}
		\end{array}\r)&\lambda=m+1,\dots,n.
	\end{cases}
	\eqlabel{newW}
\end{equation}
We have also included in this transformation an arbitrary scale factor~$c$.
Since by~\eqref{solvcocycle} the block in the upper-left characterizes the
cocycle, we see that the transformed cocycle is the cocycle characterized
by~${\bf\ww}_\lambda$ minus the coboundary characterized
by~${\bf\Wcob}_\lambda$.

The special case we will encounter most often is when the maximal Abelian
ideal of~$\LieAx$ simply consists of elements of the
form~\hbox{$(0,\dots,0,\alpha_n)$}.  For this case~$m=n-1$, and the action
vanishes since~\hbox{${\W_n}^{\mu n}=0$} (the extension is central).  The
cocycle~$\omega$ is entirely determined by~$\W_{(n)}$.  The form of the
coboundary is reduced to
\begin{equation}
	{\Wcob_n}^{\mu\nu} =
		\sum_{\tau=1}^{n-1}\,
		{k_n}^\tau\,{\W_\tau}^{\mu\nu},
	\eqlabel{cobnoaction}
\end{equation}
that is, a linear combinations of the first~$(n-1)$ matrices. Thus it is easy
to see at a glance which parts of the cocycle characterized $\W_{(n)}$ can be
removed by lower-triangular coordinate transformations.
\index{coboundary|)}

\section{Further Coordinate Transformations}
\seclabel{furthertrans}

In the previous section we restricted ourselves to lower-triangular coordinate
transformations, which in general preserve the lower-triangular structure of
the~$\W^{(\mu)}$.  But when the~$\W^{(\mu)}$ matrices are relatively sparse,
there exist non-lower-triangular coordinate transformations that nonetheless
preserve the lower-triangular structure.  As alluded to in
\secref{abelianext}, these transformations are outside the scope of cohomology
theory, which is restricted to transformations that preserve the exact form of
the action and the algebras~$\LieA$ and~$\LieAxb$, as shown by~\eqref{newW}.
In other words, cohomology theory classifies extensions \emph{given} $\LieA$,
$\LieAxb$, and~$\rho$.  We need not obey this restriction.  We can allow
non-lower-triangular coordinate transformations as long as they preserve the
lower-triangular structure of the~$\W^{(\mu)}$'s.

We now discuss a particular class of such transformations that will be useful
in \secref{lowdimext}.  Consider the case where both the algebra of
$(n-1)$-tuples~$\LieA$ and that of $1$-tuples~$\LieAxb$ are Abelian.  Then the
possible (solvable) extensions, in lower triangular form, are characterized
by~$\W_{(\lambda)}=0$, $\lambda=1,\dots,n-1$, with $\W_{(n)}$ arbitrary
(except for~${\W_n}^{\mu n}=0$).  Let us apply a coordinate change of the form
\[
	\M = \l(\begin{array}{c|c} {\bf \mm} \ & {\bf 0} \\ \hline
		{\bf 0} & \ c\,
	\end{array}\r),
\]
where~${\bf \mm}$ is an~$(n-1)\times (n-1)$ nonsingular matrix and~$c$ is
again a nonzero scale factor.  Denoting by~$\Wb$ the transformed~$\W$,
we have
\begin{equation}
	{\Wb_{(\lambda)}} =
	\begin{cases}
		\,0 & \lambda = 1,\dots,n-1;\\[8pt]
		\l(\begin{array}{c|c}
			c^{-1}\,{\bf m}^T\,{\bf \ww}_\lambda\,{\bf m}\ 
				& \ {\bf 0} \\ \hline
			{\bf 0}\   & \ \ 0\
		\end{array}\r) & \lambda=n.
	\end{cases}
	\eqlabel{newW2}
\end{equation}

This transformation does not change the lower-triangular form of the
extension, even if~${\bf\mm}$ is not lower-triangular.  The manner in
which~${\bf\ww}_{n}$ is transformed by~$\M$ is very similar to that of a
(possibly singular) metric tensor: it can be diagonalized and rescaled such
that all its eigenvalues are~$0$ or~$\pm 1$.  We can also change the overall
sign of the eigenvalues using~$c$ (something that cannot be done for a metric
tensor).  Hence, we shall order the eigenvalues such that the~$+1$'s come
first, followed by the~$-1$'s, and finally by the~$0$'s.  We will show in
\secref{lowdimext} how the negative eigenvalues can be eliminated to
harmonize the notation.

\section{Appending a Semisimple Part}
\seclabel{semisimple}

\index{semisimple algebra!appending to solvable|(}
In \secref{classcoho} we showed that because of the Levi decomposition theorem
we only needed to classify the solvable part of the extension for a given
degenerate block. Most physical applications have a semisimple part
($\evone=1$); when this is so, we shall label the matrices
by~$\W^{(0)},\W^{(1)},\dots,\W^{(n)}$, where they are now of\index{semidirect
sum!labeling of} size~\hbox{$n+1$} and~$\W^{(0)}$ is the identity.%
\footnote{The term semisimple is not quite precise: if the base algebra is not
semisimple then neither is the extension. However we will use the term to
distinguish the different cases.} Thus the matrices labeled
by~$\W^{(1)},\dots,\W^{(n)}$ will always form a solvable subalgebra. This
explains the labeling in \secreftwo{matlowbetaRMHD}{matCRMHD}.

If the extension has a semisimple part ($\zerorone=1$, or
equivalently~$\W^{(0)}=I$), we shall refer to it as
\emph{semidirect}\index{extension!semidirect}.  This was the case treated in
\secref{prelimsplit}.  A pictorial representation of an
arbitrary semidirect extension with nonvanishing cocycle is shown in
\figref{fsdpextpic}.
\begin{figure}
\centerline{\psfig{file=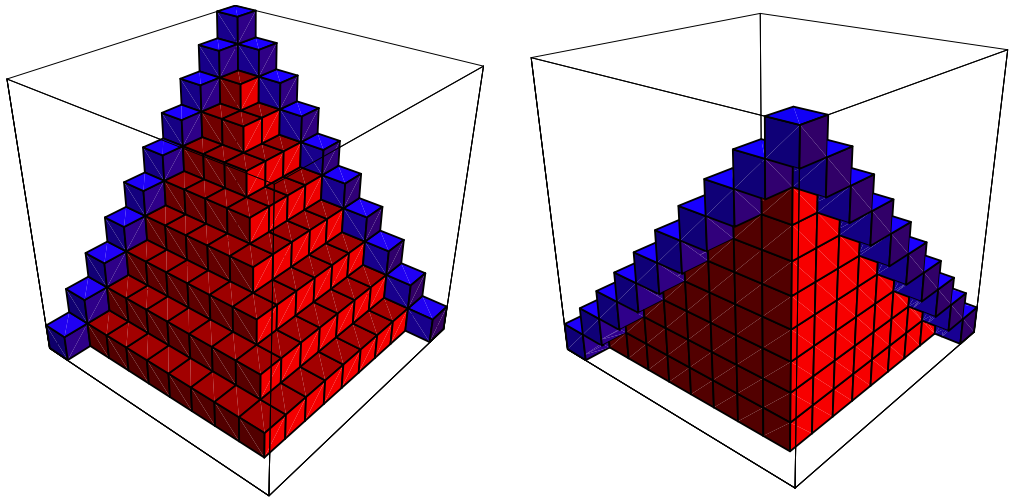,width=5.5in}}
\caption{Front and rear views of a schematic representation of the
3-tensor~$\W$ for an arbitrary semidirect extension with cocycle.  The
solvable part is in red. The semisimple part is in blue and consists of unit
entries.  The axes are as in \figref{solvextpic}.  An extension with
\emph{all} these elements nonzero cannot actually occur.}
\figlabel{fsdpextpic}
\end{figure}
If the extension is not semidirect, then it is solvable (and contains~$n$
matrices instead of~$n+1$).  This is the extension represented in
\figref{solvextpic}.

Given a solvable algebra of~$n$-tuples we can carry out in some sense the
inverse of the Levi decomposition and append a semisimple part to the
extension.  Effectively, this means that the~\hbox{$n\times n$}
matrices~$\W^{(1)},\dots,\W^{(n)}$ are made~\hbox{$n+1 \times n+1$} by adding
a row and column of zeros.  Then we simply append the matrix~$\W^{(0)}=I$ to
the extension.  In this manner we construct a semisimple extension from a
solvable one.  This is useful since we will be classifying solvable
extensions, and afterwards we will want to recover their semidirect
counterpart.

The extension obtained by appending a semisimple part to the completely
Abelian algebra of~$n$-tuples will be called \emph{pure
semidirect}\index{extension!pure semidirect}.  It is characterized
by~$\W^{(0)}=I$, and~${\W_\lambda}^{\mu\nu}=0$ for~$\mu,\nu>0$.  This is shown
schematically in \figref{psemiextpic}.
\begin{figure}
\centerline{\psfig{file=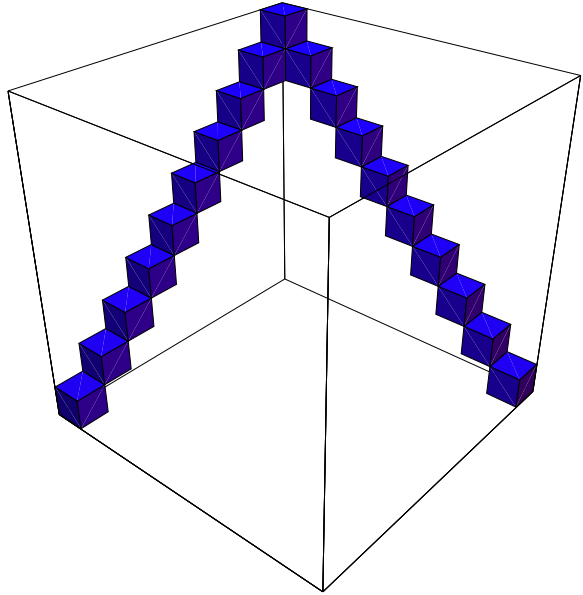,width=3in}}
\caption{Schematic representation of the 3-tensor~$\W$ for a pure semidirect
extension.  The axes are as in \figref{solvextpic}.}
\figlabel{psemiextpic}
\end{figure}
\index{semisimple algebra!appending to solvable|)}

\section{Leibniz Extension}
\seclabel{Leibniz}

\index{extension!Leibniz|(}
A particular extension that we shall consider is called the Leibniz
extension~\cite{Parthasarathy1976}.  For the solvable case this extension has
the form
\begin{equation}
	{\W}^{(1)} \rdef \Nilb = \begin{pmatrix}
	0 & & & & \\
	1 & 0 & & & \\
	  & 1 & 0 & & \\
	  &   & & \ddots & \\
	  &   & & 1 & 0
	\end{pmatrix}
	\eqlabel{Nilb}
\end{equation}
or~${\W_\lambda}^{\mu\,1} = {\delta_{\lambda-1}}^{\mu}$,~$\lambda>1$.  The
first matrix is an~\hbox{$n \times n$} Jordan block.  In this case the other
matrices, in order to commute with~$\W^{(1)}$, must be in striped
lower-triangular form~\cite{Suprunenko},
\begin{equation}
	{\W}^{(\nu)} = \begin{pmatrix}
	0 &   &   &   &   & \\
	a & 0 &   &   &   & \\
	b & a & 0 &   &   & \\
	c & b & a & 0 &   & \\
	d & c & b & a & 0 & \\
	\vdots & & & & & \ddots
	\end{pmatrix}.
	\eqlabel{striated}
\end{equation}
But by symmetry of the upper indices the first column of matrix~$\W^{(\nu)}$
must be~\hbox{${\W_\lambda}^{1(\nu)} = {\delta_\lambda}^{\nu}$}, so that
\begin{equation}
	\W^{(\nu)} = (\Nilb)^\nu,
	\eqlabel{Nilbmu}
\end{equation}
where on the right-hand side the~$\nu$ denotes an exponent, not a superscript.
An equivalent way of characterizing the Leibniz extension is
\begin{equation}
	{\W_\lambda}^{\mu\nu} = {\delta_\lambda}^{\mu+\nu}\,,
	\ \ \ \mu,\nu, \lambda = 1,\dots,n.
	\eqlabel{sLeib}
\end{equation}
The tensor~$\delta$ is an ordinary Kronecker delta.  Note that
neither~\eqref{Nilbmu} nor~\eqref{sLeib} are covariant expressions, reflecting
the coordinate-dependent nature of the Leibniz extension.

The Leibniz extension is in some sense a ``maximal'' extension: it is the only
extension that has~\hbox{$\W_{(\lambda)} \ne 0$} for
\emph{all}~\hbox{$\lambda>1$} (up to coordinate transformations).  Its
uniqueness will become clear in \secref{lowdimext}, and is proved in
\secref{maxiLeib}.  We show two schematic views of the extension in
\figref{leibextpic}.%
\begin{figure}
\centerline{\psfig{file=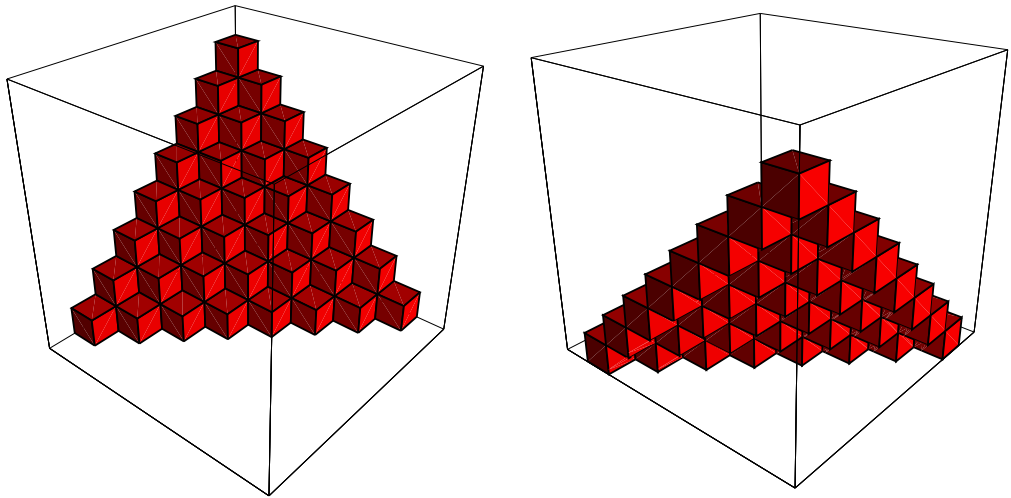,width=5.5in}}
\caption{Two views of the 3-tensor~$\W$ for a solvable
Leibniz extension, where each cube denotes a~$1$.  The axes are as in
\figref{solvextpic}.  The Leibniz extension is ``hollow.''}
\figlabel{leibextpic}
\end{figure}
Fans of 1980's arcade games will understand why the author is suggesting the
alternate name \qbert{} extension,\footnote{\qberttm{} is a trademark of the
Sony Corporation.} since Leibniz has no dearth of things named after him (see
\figref{qbertscr}).
\begin{figure}
\centerline{\psfig{file=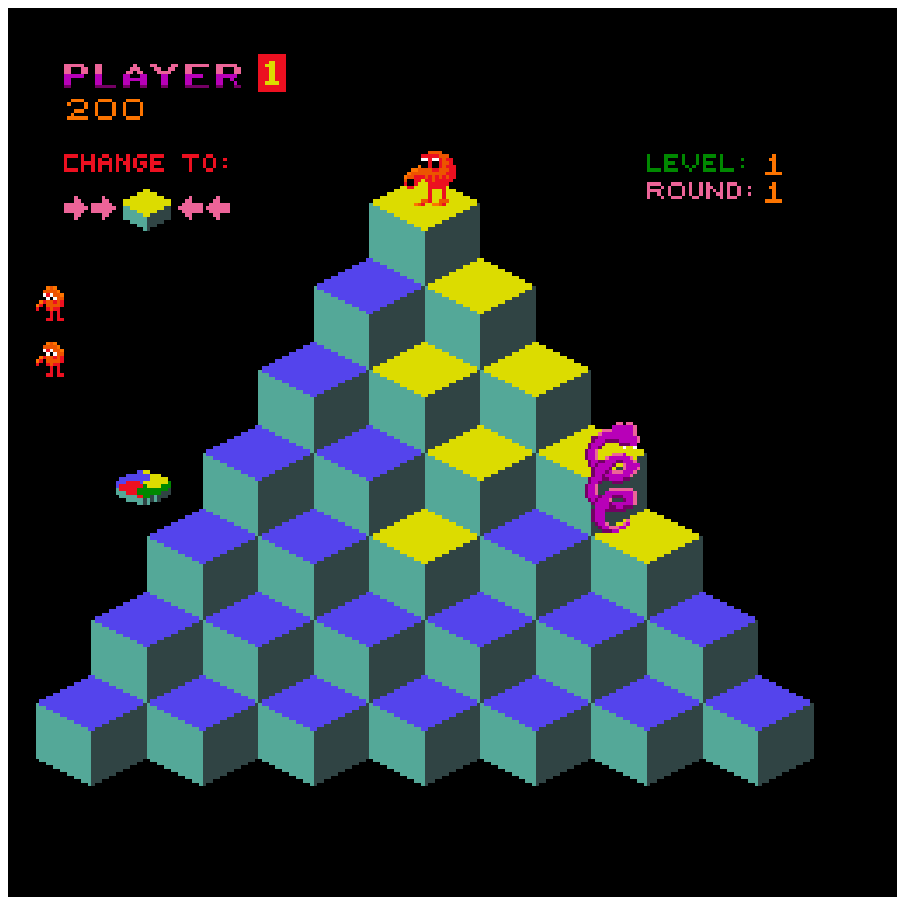,width=3in}}
\caption{Screenshot of the \qbert{} game.  Compare with \figref{leibextpic}!}
\figlabel{qbertscr}
\end{figure}

To construct the semidirect Leibniz extension, we append~$\W^{(0)}=I$, a
square matrix of size~$n+1$, to the solvable Leibniz extension above, as
described in \secref{semisimple}.  The characterization given by
Eq.~\eqref{sLeib} can be used for the semidirect Leibniz extension by simply
letting the indices run from~$0$ to~$n$.

\index{extension!Leibniz|)}

\section{Low-order Extensions}
\seclabel{lowdimext}

\index{extension!low-order}
\index{classification!of low-order extensions|(}
We now classify algebra extensions of low order. As demonstrated in
\secref{classcoho} we only need to classify solvable algebras, which
means that~$\W^{(n)}=0$ for all cases. We will do the classification up to
order~$n=4$.  For each case we first write down the most general set of
lower-triangular matrices~$\W^{(\nu)}$ (we have already used the fact that a
set of commuting matrices can be lower-triangularized) with the
symmetry~${\W_\lambda}^{\mu\nu}={\W_\lambda}^{\nu\mu}$ built in. Then we look
at what sort of restrictions the commutativity of the matrices places on the
elements. Finally, we eliminate coboundaries for each case by the methods of
\secreftwo{classcoho}{furthertrans}. This requires coordinate
transformations, but we usually will not bother using new symbols and just
assume the transformation were effected.

Note that, due to the lower-triangular structure of the extensions, the
classification found for an~$m$-tuple algebra applies to the first~$m$
elements of an~$n$-tuple algebra,~\hbox{$n>m$}.  Thus,~$\W_{(n)}$ is the
cocycle that contains all of the new information not included in the
previous~\hbox{$m=n-1$} classification.  These comments will become clearer as
we proceed.

There are three generic cases that we will encounter for any order:

\begin{enumerate}
\item The Leibniz extension, discussed in \secref{Leibniz}.
\item An extension with~\hbox{$\W_{(\lambda)} \equiv 0$},
\hbox{$\lambda=1,\dots,n-1$}, and~$\W_{(n)}$ arbitrary (and symmetric).
This extension automatically satisfies the commutativity requirement, because
the product of any two~$\W^{(\mu)}$ vanishes.  It can be further classified by
the methods of \secref{furthertrans}.  Later we will refer to this case as
having a \emph{vanishing coextension} (see \secref{cassoln} and
\figref{vanishcoextpic}).
\item The Abelian extension,  which  vanishes
identically:~\hbox{$\W_{(\lambda)} \equiv 0$},
\hbox{$\lambda=1,\dots,n$}.  This is a special case of 2, above.  When
appended to a semidirect part (as explained in \secref{semisimple}), the
Abelian extension generates the pure semidirect extension.
\end{enumerate}

We shall call an order~$n$ extension \emph{trivial} if~\hbox{$\W_{(n)} \equiv
0$}, so that the cocycle appended to the order~$n-1$ extension contributes
nothing to the bracket.

We now proceed with the classification for orders~$n=1$ to~$4$.

\subsection{n=1}
\seclabel{neqone}

This case is Abelian, with the only possible element~${W_1}^{11}=0$.

\subsection{n=2}
\seclabel{neqtwo}

The most general lower-triangular form for the matrices is
\[
\W^{(1)} = \begin{pmatrix} 0 & 0 \\ {\W_2}^{11} & 0 \end{pmatrix},
\ \ \ \
\W^{(2)} = \begin{pmatrix} 0 & 0 \\ 0 & 0 \end{pmatrix}.
\]
If~${\W_2}^{11} \ne 0$, then we can rescale it to unity.  Hence, we
let~${\W_2}^{11} \ldef \zerorone_1$, where~$\zerorone_1 = 0$ or $1$.  The
case~$\zerorone_1 = 0$ is the Abelian case, while for~$\zerorone=1$ we have
the~$n=2$ Leibniz extension (\secref{Leibniz}).  Thus for~$n=2$ there are only
two possible algebras.  The cocycle which we have added at this stage is
characterized by~$\zerorone_1$.

\subsection{n=3}
\seclabel{neqthree}

Using the result of \secref{neqtwo}, the most general lower-triangular form is
\[
\W^{(1)} = \begin{pmatrix}
	0 & 0 & 0 \\
	\zerorone_1 & 0 & 0 \\
	{\W_3}^{11} & {\W_3}^{21} & 0 \end{pmatrix},
\ \ \ \
\W^{(2)} = \begin{pmatrix}
	0 & 0 & 0 \\
	0 & 0 & 0 \\
	{\W_3}^{21} & {\W_3}^{22} & 0 \end{pmatrix},
\]
and~$\W^{(3)} = 0$.  These satisfy the symmetry condition~\eqref{upsym}, and
the requirement that the matrices commute leads to the condition
\[
	\zerorone_1\,{\W_3}^{22} = 0.
\]
The symmetric matrix representing the cocycle is
\begin{equation}
W_{(3)} = \begin{pmatrix}
	{\W_3}^{11} & {\W_3}^{21} & 0 \\
	{\W_3}^{21} & {\W_3}^{22} & 0 \\ 
	0 & 0 & 0 \end{pmatrix}.
	\eqlabel{Wiii}
\end{equation}
If~$\zerorone_1 = 1$, then~${\W_3}^{22}$ must vanish.  Then,
by~\eqref{cobnoaction} we can remove from~$\W_{(3)}$ a multiple of~$\W_{(2)}$,
and therefore we may assume~${\W_3}^{11}$ vanishes.  A suitable rescaling
allows us to write~${\W_3}^{21}=\zerorone_2$, where~$\zerorone_2 = 0$ or $1$.
The cocycle for the case~$\zerorone_1=1$ is thus
\[
W_{(3)} = \begin{pmatrix}
	0 & \zerorone_2 & 0 \\
	\zerorone_2 & 0 & 0 \\ 
	0 & 0 & 0 \end{pmatrix}.
\]
For~$\zerorone_2=1$ we have the Leibniz extension (\secref{Leibniz}).

If~$\zerorone_1 = 0$, we have the case discussed in \secref{furthertrans}.
For this case we can diagonalize and rescale~$\W_{(3)}$ such that
\[
W_{(3)} = \begin{pmatrix}
	\pmz_1 & 0 & 0 \\
	0 & \pmz_2 & 0 \\ 
	0 & 0 & 0 \end{pmatrix},
\]
where~$(\pmz_1,\pmz_2)$ can be~$(1,1)$, $(1,0)$, $(0,0)$, or $(1,-1)$.  This
last case, as alluded to at the end of \secref{furthertrans}, can be
transformed so that it corresponds to~$\zerorone_1=0$, $\zerorone_2=1$.  The
choice~$(1,0)$ can be transformed to the~$\zerorone_1=1$, $\zerorone_2=0$
case.  Finally for~$(\pmz_1,\pmz_2)=(1,1)$ we can use the complex
transformation
\[
	\fv^1\rightarrow\frac{1}{\sqrt{2}}(\fv^1+\fv^2),\ \ \
	\fv^2\rightarrow-\frac{\imi}{\sqrt{2}}(\fv^1-\fv^2),\ \ \
	\fv^3\rightarrow\fv^3,
\]
to transform to the~$\zerorone_1=0$, $\zerorone_2=1$ case.

\index{classification!and complex transformations}
We allow complex transformations in our classification because we are chiefly
interested in finding Casimir invariants for Lie--Poisson brackets.  If we
disallowed complex transformations, the final classification would contain a
few more members.  The use of complex transformations will be noted as we
proceed.


There are thus four independent extensions for~$n=3$, corresponding to
\[
	(\zerorone_1\com\zerorone_2)
	\in \l\{(0\com 0)\com(0 \com 1)\com(1 \com 0)\com(1 \com 1)\r\}.
\]
These will be referred to as Cases~$1$--$4$, respectively.
Cases~\ref{case:n4-00} and~\ref{case:n4-10} have~$\zerorone_2=0$, and so are
trivial ($\W_{(3)}=0$). \caseref{n4-01} is the solvable part of the
compressible reduced MHD bracket (\secref{matCRMHD}).  \caseref{n4-11} is the
solvable Leibniz extension.

\subsection{n=4}
\seclabel{neqfour}

Proceeding as before and using the result of \secreftwo{neqtwo}{neqthree}, we
now know that we need only write
\begin{equation}
\W_{(4)} = \begin{pmatrix}
	{\W_4}^{11} &  {\W_4}^{21} & {\W_4}^{31} & 0 \\
	{\W_4}^{21} &  {\W_4}^{22} & {\W_4}^{32} & 0 \\
	{\W_4}^{31} &  {\W_4}^{32} & {\W_4}^{33} & 0 \\
	0 & 0 & 0 & 0
	\end{pmatrix}.
	\eqlabel{Wfour}
\end{equation}
The matrices~$\W_{(1)}$, $\W_{(2)}$, and~$\W_{(3)}$ are given by their~$n=3$
analogues padded with an extra row and column of zeros (owing to the
lower-triangular form of the matrices).  The requirement that the
matrices~$\W^{(1)}\dots\W^{(4)}$ commute leads to the conditions%
\begin{equation}
\begin{split}
	\zerorone_2\,{\W_4}^{33} &= 0, \\
	\zerorone_2\,{\W_4}^{31} &= \zerorone_1\,{\W_4}^{22}, \\
	\zerorone_2\,{\W_4}^{32} &= 0, \\
	\zerorone_1\,{\W_4}^{32} &= 0.
	\eqlabel{commrel4}
\end{split}
\end{equation}
There are four cases to look at, corresponding to the possible values
of~$\zerorone_1$ and~$\zerorone_2$.

\begin{case}
$\zerorone_1=0$, $\zerorone_2=0$.
\caselabel{n4-00}
\end{case}

This is the unconstrained case discussed in \secref{furthertrans}, that is,
all the commutation relations \eqref{commrel4} are automatically satisfied.
We can diagonalize to give
\[
\W_{(4)} = \begin{pmatrix}
	\pmz_1' &  0  & 0 & 0 \\
	0 &  \pmz_2' & 0 & 0 \\
	0 & 0 & \pmz_3' & 0 \\
	0 & 0 & 0 & 0
	\end{pmatrix},
\]
where
\[
(\pmz_1',\pmz_2',\pmz_3')
	\in \l\{(1,1,1),(1,1,0),(1,0,0),(0,0,0),(1,1,-1),(1,-1,0)\r\},
\]
so there are six distinct cases.  The exact form of the transformation is
unimportant, but the~$(1,1,0)$ extension can be mapped to \caseref{n4-01} (the
transformation is complex),~$(1,0,0)$ can be mapped to
\caseref{n4-10}a, and~$(1,-1,0)$ can be mapped to
\caseref{n4-01}.  Finally the~$(1,1,1)$ extension can be mapped to
the~$(1,1,-1)$ case by a complex transformation.

After transforming that~$(1,1,-1)$ case, we are left with
\[
\W_{(4)} = \begin{pmatrix}
	0 & 0 & 0 & 0 \\
	0 & 0 & 0 & 0 \\
	0 & 0 & 0 & 0 \\
	0 & 0 & 0 & 0\end{pmatrix},
\begin{pmatrix}
	0 & 0 & 1 & 0 \\
	0 & 1 & 0 & 0 \\
	1 & 0 & 0 & 0 \\
	0 & 0 & 0 & 0
\end{pmatrix}.
\]
These will be called Cases~\ref{case:n4-00}a and~\ref{case:n4-00}b.

\begin{case}
$\zerorone_1=0$, $\zerorone_2=1$.
\caselabel{n4-01}
\end{case}

The commutation relations~\eqref{commrel4} reduce to ${\W_{4}}^{31} =
{\W_{4}}^{32} = {\W_{4}}^{33} = 0$, and we have
\[
\W_{(4)} = \begin{pmatrix}
	{\W_4}^{11} &  {\W_4}^{21}  & 0 & 0 \\
	{\W_4}^{21} &  {\W_4}^{22} & 0 & 0 \\
	0 & 0 & 0 & 0 \\
	0 & 0 & 0 & 0
	\eqlabel{fff}
	\end{pmatrix}.
\]
We can remove~${\W_4}^{21}$ because it is a coboundary (in this case a
multiple of~${\W_{(3)}}$).  We can also rescale appropriately to obtain the
four possible extensions
\[
\W_{(4)} = \begin{pmatrix}
	0 & 0 & 0 & 0 \\
	0 & 0 & 0 & 0 \\
	0 & 0 & 0 & 0 \\
	0 & 0 & 0 & 0
	\end{pmatrix},
\begin{pmatrix}
	1 & 0 & 0 & 0 \\
	0 & 0 & 0 & 0 \\
	0 & 0 & 0 & 0 \\
	0 & 0 & 0 & 0
\end{pmatrix},
\begin{pmatrix}
	1 & 0 & 0 & 0 \\
	0 & 1 & 0 & 0 \\
	0 & 0 & 0 & 0 \\
	0 & 0 & 0 & 0
\end{pmatrix},
\begin{pmatrix}
	1 & 0 & 0 & 0 \\
	0 & -1 & 0 & 0 \\
	0 & 0 & 0 & 0 \\
	0 & 0 & 0 & 0
\end{pmatrix}.
\]
Again, the form of the transformation is unimportant, but it turns out that
the second extension can be mapped to \caseref{n4-10}c, and the third and
fourth to \caseref{n4-10}b.  This last transformation is complex.  Thus there
is only one independent possibility, the trivial extension~$\W_{(4)}=0$.

\begin{case}
$\zerorone_1=1,\zerorone_2=0$.
\caselabel{n4-10}
\end{case}

We can remove~~${\W_4}^{11}$ using a coordinate transformation.  From the
commutation requirement \eqref{commrel4} we obtain~${\W_4}^{22} =
{\W_4}^{32} = 0$.  We are left with~$\W_{(3)}=0$ and
\[
\W_{(4)} = \begin{pmatrix}
	0           & {\W_4}^{21} & {\W_4}^{31} & 0 \\
	{\W_4}^{21} & 0           & 0           & 0 \\
	{\W_4}^{31} & 0           & {\W_4}^{33} & 0 \\
	0           & 0           & 0           & 0
\end{pmatrix}.
\]
Using the fact that elements of the form~$(0,\alpha_2,0,\alpha_4)$ are an
Abelian ideal of this bracket, we find that~${\W_4}^{33}{\W_4}^{31}=0$. Using
an upper-triangular transformation we can also
make~${\W_4}^{21}{\W_4}^{31}=0$.  After suitable rescaling we find there are
five cases: the trivial extension~$\W_{(4)}=0$, and
\[
\W_{(4)} =
\begin{pmatrix}
	0 & 0 & 0 & 0 \\
	0 & 0 & 0 & 0 \\
	0 & 0 & 1 & 0 \\
	0 & 0 & 0 & 0
\end{pmatrix},
\begin{pmatrix}
	0 & 0 & 1 & 0 \\
	0 & 0 & 0 & 0 \\
	1 & 0 & 0 & 0 \\
	0 & 0 & 0 & 0
\end{pmatrix},
\begin{pmatrix}
	0 & 1 & 0 & 0 \\
	1 & 0 & 0 & 0 \\
	0 & 0 & 1 & 0 \\
	0 & 0 & 0 & 0
\end{pmatrix},
\begin{pmatrix}
	0 & 1 & 0 & 0 \\
	1 & 0 & 0 & 0 \\
	0 & 0 & 0 & 0 \\
	0 & 0 & 0 & 0
\end{pmatrix}.
\]
However the last of these may be mapped to \caseref{n4-11} (below)
with~$\zerorone_3=0$.  We will refer to the trivial extension as
Case~\ref{case:n4-10}a and to the remaining three extensions as
Cases~\ref{case:n4-10}b--d, respectively.

\begin{case}
$\zerorone_1=1$, $\zerorone_2=1$.
\caselabel{n4-11}
\end{case}

The elements~${\W_4}^{11}$ and~${\W_4}^{21}$ are coboundaries that can be
removed by a coordinate transformation.  From \eqref{commrel4} we
have~${\W_4}^{33} = {\W_4}^{32} = 0, {\W_4}^{22} = {\W_4}^{31}
\rdef \zerorone_3$, so that
\[
\W_{(4)} = \begin{pmatrix}
	0 & 0 & \zerorone_3 & 0 \\
	0 & \zerorone_3 & 0 & 0 \\
	\zerorone_3 & 0 & 0 & 0 \\
	0 & 0 & 0 & 0
\end{pmatrix}.
\]
For~$\zerorone_3=1$ we have the Leibniz extension.  The two cases will be
referred to as \caseref{n4-11}a for~$\zerorone_3=0$
and~\ref{case:n4-11}b for~$\zerorone_3=1$.

\tabref{n=4extensions} summarizes the results.  There are are total
of nine independent~$n=4$ extensions, four of which are trivial
($\W_{(4)}=0$). As noted in \secref{Leibniz} only the Leibniz extension,
\caseref{n4-11}b, has nonvanishing~$\W_{(i)}$ for all~\hbox{$1<i\le n$}.

\begin{table}
\begin{center}

\begin{tabular}{lccc@{}c@{}c@{}c} \hline

Case & $\W_{(2)}$ & $\W_{(3)}$ & \multicolumn{4}{c}{$\W_{(4)}$} \\
 &  &  & a & b & c & d \\[3pt] \hline

%
%
1 &
$\begin{pmatrix} 0 \end{pmatrix}$ &
${\begin{pmatrix} 0 & 0 \\ 0 & 0 \end{pmatrix}}$ &
${\begin{pmatrix} 0 & 0 & 0 \\ 0 & 0 & 0 \\ 0 & 0 & 0 \end{pmatrix}}$ &
${\begin{pmatrix} 0 & 0 & 1 \\ 0 & 1 & 0 \\ 1 & 0 & 0 \end{pmatrix}}$& &
\rule[-2em]{0cm}{5em}
\\

%
%
2 &
$\begin{pmatrix} 0 \end{pmatrix}$ &
${\begin{pmatrix} 0 & 1 \\ 1 & 0 \end{pmatrix}}$ &
${\begin{pmatrix} 0 & 0 & 0 \\ 0 & 0 & 0 \\ 0 & 0 & 0 \end{pmatrix}}$ &
&&
\rule[-2em]{0cm}{5em}
\\

%
%
3 &
$\begin{pmatrix} 1 \end{pmatrix}$ &
${\begin{pmatrix} 0 & 0 \\ 0 & 0 \end{pmatrix}}$ &
${\begin{pmatrix} 0 & 0 & 0 \\ 0 & 0 & 0 \\ 0 & 0 & 0 \end{pmatrix}}$ &
${\begin{pmatrix} 0 & 0 & 0 \\ 0 & 0 & 0 \\ 0 & 0 & 1 \end{pmatrix}}$&
${\begin{pmatrix} 0 & 0 & 1 \\ 0 & 0 & 0 \\ 1 & 0 & 0 \end{pmatrix}}$&
${\begin{pmatrix} 0 & 1 & 0 \\ 1 & 0 & 0 \\ 0 & 0 & 1 \end{pmatrix}}$
\rule[-2em]{0cm}{5em}
\\

%
%
4 &
$\begin{pmatrix}1\end{pmatrix}$ &
${\begin{pmatrix} 0 & 1 \\ 1 & 0 \end{pmatrix}}$ &
${\begin{pmatrix} 0 & 0 & 0 \\ 0 & 0 & 0 \\ 0 & 0 & 0\end{pmatrix}}$ &
${\begin{pmatrix} 0 & 0 & 1 \\ 0 & 1 & 0 \\ 1 & 0 & 0 \end{pmatrix}}$&&
\rule[-2.4em]{0cm}{5.4em}
\\ \hline

\end{tabular}

\end{center}
\caption{Enumeration of the independent extensions up to~$n=4$.  We
have~$\W_{(1)}=0$ for all the cases, and we have left out a row and a column
of zeros at the end of each matrix.}
\tablabel{n=4extensions}

\end{table}

The surprising fact is that even to order four the normal forms of the
extensions involve no free parameters: all entries in the coefficients of the
bracket are either zero or one. There is no obvious reason this should hold
true if we try to classify extensions of order~$n>4$. It would be interesting
to find out, but the classification scheme used here becomes prohibitive at
such high order. The problem is that some of the transformations used to
relate extensions cannot be systematically derived and were obtained by
educated guessing.

\index{classification!of low-order extensions|)}

\section{Leibniz as the Maximal Extension}
\seclabel{maxiLeib}

\index{extension!Leibniz|(}
We mentioned in \secref{Leibniz} that the Leibniz extension is maximal: it is
the only extension that has~\hbox{$\W_{(\lambda)} \ne 0$} for
\emph{all}~\hbox{$\lambda>1$}.  Having seen the classification process at work
in \secref{lowdimext}, we are now in a position to show why the Leibniz
extension has this property.  We will demonstrate that the only way to extend
a Leibniz extension nontrivially (i.e., with a nonvanishing cocycle) is to
append a cocycle such that the new extension is again Leibniz.

Consider a solvable Leibniz extension of order~$n-1$, denoted by the
3-tensor~$\Wt$.  We increase the order of~$\Wt$ by one by
appending the most general cocycle possible (as was done in
\secref{lowdimext}) to obtain an extension of order~$n$ denoted by the
tensor~$\W$.  The form of the matrices~$\W^{(\mu)}$ of the new extension is
\begin{equation}
	{\W}^{(\mu)} = \l(\begin{array}{c|c}
	{\Wt}^{(\mu)} & \\ \hline
	\rule{0cm}{.5cm}
	{\W_n}^{(\mu)} & 0
	\end{array}\r),\quad \mu=1,\ldots,n-1,
	\eqlabel{leibnp1}
\end{equation}
and~\hbox{$\W^{(n)} \equiv 0$}.  The quantity~${\W_n}^{(\mu)}$ is a row vector
defined in the obvious manner as~\hbox{$[{\W_n}^{(\mu)}]^\nu =
{\W_n}^{\mu\nu}$}.

In particular, the first matrix of the $n$th order extension is
\begin{equation}
	{\W}^{(1)} = \l(\begin{array}{ccccc|c}
	0 & & & & & \\
	1 & 0 & & & & \\
	  & 1 & 0 & & & \\
	  &   & \cdots & \cdots & & \\
	  &   & & 1 & 0 & \\ \hline
	\rule{0cm}{.5cm}
	{\W_n}^{11} & {\W_n}^{12} & \cdots & {\W_n}^{1,n-2}
		& {\W_n}^{1,n-1} & 0
	\end{array}\r),
\end{equation}
where the~${\W_n}^{\mu\nu}$ represent the appended cocycle, and we have
explicitly delimited the order~$n-1$ Leibniz extension.  It is not difficult
to show that the~${\W_n}^{1\nu}$,~\hbox{$\nu=1,\dots,n-2$}, are coboundaries
and so can be removed by a coordinate transformation.  We thus assume
that~\hbox{${\W_n}^{1\nu}=0$},~\hbox{$\nu=1,\dots,n-2$}.  The only potentially
nonzero element of that row is~${\W_n}^{1,n-1}$.

Taking the commutator of two matrices of the form~\eqref{leibnp1} gives the
conditions
\[
	\sum_{\sigma=1}^{n-1}{\W_n}^{\mu\sigma}\,{\Wt_\sigma}^{\nu\tau}
	= \sum_{\sigma=1}^{n-1}{\W_n}^{\nu\sigma}\,{\Wt_\sigma}^{\mu\tau},
	\quad \mu,\nu,\tau=1,\dots,n-1.
\]
Substituting the form of the Leibniz extension~\eqref{sLeib} for~$\Wt$, this
becomes
\[
	\sum_{\sigma=1}^{n-1}{\W_n}^{\mu\sigma}\,{\delta_\sigma}^{\nu+\tau}
	= \sum_{\sigma=1}^{n-1}{\W_n}^{\nu\sigma}\,{\delta_\sigma}^{\mu+\tau},
\]
or
\begin{alignat*}{2}
	{\W_n}^{\mu,\nu+\tau} =
	\begin{cases}
		{\W_n}^{\mu+\tau,\nu}& \mu+\tau < n,\\
		0& \mu+\tau\ge n,
	\end{cases}
\end{alignat*}
where~\hbox{$\nu+\tau < n$}.  For~$\tau=1$, this is
\begin{alignat*}{2}
	{\W_n}^{\mu,\nu+1} =
	\begin{cases}
		{\W_n}^{\mu+1,\nu}& \mu < n-1,\\
		0& \mu = n-1,
	\end{cases}
\end{alignat*}
for~\hbox{$\nu=1,\dots,n-2$}, which says that~$\W_{(n)}$ has a banded
structure.  Because we have
that~\hbox{${\W_n}^{1\nu}=0$},~\hbox{$\nu=1,\dots,n-2$}, it must be that
\[
	\W_{(n)} = \begin{pmatrix}
			0 & 0 & \dots & 0 & {\W_n}^{1,n-1} & 0 \\
			0 & 0 & \dots & {\W_n}^{1,n-1} & 0 & 0 \\
			\hdotsfor{6} \\
			0 & {\W_n}^{1,n-1} & \dots & 0 & 0 & 0 \\
			{\W_n}^{1,n-1} & 0 & \dots & 0 & 0 & 0 \\
			0 & 0 & \dots & 0 & 0 & 0 \\
		   \end{pmatrix}
\]
So either~${\W_n}^{1,n-1}=0$ (the extension is trivial), or~${\W_n}^{1,n-1}$
can be rescaled to unity (the extension is of the Leibniz type).

Thus, if one has a Leibniz extension of size~$n-1$ then the only way to
nontrivially extend it is to make it the Leibniz extension of size~$n$.  But
since the Leibniz extension is the only nontrivial extension of order~$2$ (see
\secref{neqtwo}), we have shown the uniqueness of the maximal extension, up to
a change of coordinates.

\index{extension!Leibniz|)}

\chapter{Casimir Invariants for Extensions}
\chlabel{casinv}

In this chapter we will use the bracket extensions of
\chref{classext} to make Lie--Poisson brackets, following the
prescription of \chref{LiePoisson}. In \secref{cascond} we write down the
general form of the Casimir condition (the condition under which a functional
is a Casimir invariant) for a general class of inner brackets. Then in
\secref{Casdirprod} we see how the Casimirs separate for a direct sum
of algebras, the case discussed in \secref{directprod}. \secref{localcas}
discusses the particular properties of Casimirs of solvable extensions. In
\secref{cassoln} we give a general solution to the Casimir problem and
introduce the concept of \emph{coextension}. Finally, in \secref{Casex} we
work out the Casimir invariants for some specific examples, including CRMHD
and the Leibniz extension.

\section{Casimir Condition}
\seclabel{cascond}

A generalized Casimir invariant\index{Casimir} (or Casimir for
short) is a function~$\Cas:\LieA^* \rightarrow \reals$ for which
\[
	\lPB F \com\, \Cas \rPB \equiv 0,
\]
for all~$F:\LieA^* \rightarrow \reals$.  Using~\eqref{LPB}
and~\eqref{cobracket}, we can write this as
\[
	\lang \,\fv \com\,\lpb\frac{\delta F}{\delta \fv} \com
		\frac{\delta \Cas}{\delta \fv}\rpb\,\rang
	= -\lang \lpb\frac{\delta \Cas}{\delta \fv}\com\,\fv\rpb^\dagger \com\,
		\frac{\delta F}{\delta \fv}\,\rang.
\]
Since this vanishes for all~$F$ we conclude
\begin{equation}
	\lpb\frac{\delta \Cas}{\delta \fv}\com\,\fv\rpb^\dagger = 0.
	\eqlabel{cascond}
\end{equation}
To figure out the coadjoint bracket\index{bracket!coadjoint} corresponding
to~\eqref{extbrack}, we write
\[
	\lang \,\fv \com\,\lpb\alpha \com \beta\rpb\,\rang = 
	\lang \,\fv^\lambda \com\,{\W_\lambda}^{\mu\nu}
		{\lpb\alpha_\mu \com \beta_\nu\rpb}\,\rang,
\]
which after using the coadjoint bracket in~$\LieA$ becomes
\[
	\lang \lpb\beta\com\fv\rpb^\dagger\com \alpha\,\rang =
	\lang {\W_\lambda}^{\mu\nu}\lpb\beta_\nu\com\fv^\lambda\rpb^\dagger
		\com \alpha_\mu\,\rang
\]
so that
\begin{equation}
	\lpb\beta\com\fv\rpb^{\dagger\,\nu} = {\W_\lambda}^{\mu\nu}
		\lpb\beta_\mu\com\fv^\lambda\rpb^\dagger.
	\eqlabel{cobrakext}
\end{equation}
We can now write the Casimir\index{Casimir!condition}
condition~\eqref{cascond} for the bracket extension as
\begin{equation}
	{\W_\lambda}^{\mu\nu} \lpb\frac{\delta \Cas}{\delta\fv^\mu}
		\com\fv^\lambda\rpb^\dagger = 0,\ \ \ \ \nu=0,\dots,n.
	\eqlabel{cascond2}
\end{equation}

We now specialize the bracket to the case of most interested to us, where the
inner bracket is of canonical form~\eqref{canibrak}\index{bracket!canonical}.
(We will touch briefly on the finite-dimensional case in \secref{findimcas},
but the remainder of the thesis will deal with a canonical inner bracket
unless otherwise noted.)  As we saw in \chref{LiePoisson}, this is the bracket
for 2-D fluid flows.  Further, we assume that the form of the Casimir
invariants is
\begin{equation}
	\Cas[\fv] = \int_\fdomain \Casi(\fv(\xv))\d^2x,
	\eqlabel{formcas}
\end{equation}
and thus, since~$\Casi$ does not contain derivatives of~$\fv$, functional
derivatives of~$\Cas$ can be written as ordinary partial derivatives
of~$\Casi$.  We can then rewrite~\eqref{cascond2} as
\begin{equation}
	{\W_\lambda}^{\mu\nu}
		\frac{\partial^2\Casi}{\partial\fv^\mu\partial\fv^\sigma}
		\lpb\fv^\sigma\com\fv^\lambda\rpb
	= 0,\ \ \ \ \nu=0,\dots,n.
	\eqlabel{cascond5}
\end{equation}
In the canonical case where the inner bracket is like~\eqref{canibrak}
the~$\lpb\fv^\sigma\com\fv^\lambda\rpb$ are independent and antisymmetric
in~$\lambda$ and~$\sigma$.  Thus a necessary and sufficient condition for the
Casimir condition\index{Casimir!condition} to be satisfied is
\begin{equation}
	{\W_\lambda}^{\mu\nu}
		\frac{\partial^2\Casi}{\partial\fv^\mu\partial\fv^\sigma}
	= {\W_\sigma}^{\mu\nu}
		\frac{\partial^2\Casi}{\partial\fv^\mu\partial\fv^\lambda}\ ,
	\eqlabel{cascond3}
\end{equation}
for~$\lambda,\sigma,\nu=0,\dots,n$.  Sometimes we shall abbreviate this as
\begin{equation}
	{\W_\lambda}^{\mu\nu} \Casi_{\dcom\mu\sigma}
	= {\W_\sigma}^{\mu\nu} \Casi_{\dcom\mu\lambda}\ ,
	\eqlabel{cascond4}
\end{equation}
that is, any subscript~$\mu$ on~$\Casi$ following a comma indicates
differentiation with respect to~$\fv^\mu$.  Equation \eqref{cascond4} is
trivially satisfied when~$\Casi$ is a linear function of the~$\fv$'s.  That
solution usually follows from special cases of more general solutions, and we
shall only mention it in \secref{singWn} where it is the only solution.

An important result is immediate from \eqref{cascond4} for a semidirect
extension.  Whenever the extension is semidirect\index{semidirect sum!labeling
of} we shall label the variables~$\fv^0,\fv^1,\dots,\fv^n$, because the
subset~$\fv^1,\dots,\fv^n$ then forms a solvable subalgebra (see
\secref{semisimple} for terminology).  For a semidirect extension,~$\W^{(0)}$
is the identity matrix, and thus~\eqref{cascond4} gives
\begin{align*}
	{\delta_\lambda}^{\mu} \Casi_{\dcom\mu\sigma}
		&= {\delta_\sigma}^{\mu} \Casi_{\dcom\mu\lambda}\ ,\\
	\Casi_{\dcom\lambda\sigma} &= \Casi_{\dcom\sigma\lambda}\ ,
\end{align*}
which is satisfied because we can interchange the order of differentiation.
Hence, $\nu=0$ does not lead to any conditions on the Casimir. However, the
variables~$\mu,\lambda,\sigma$ still take values from~$0$ to~$n$ in
\eqref{cascond4}.


\subsection{Finite-dimensional Casimirs}
\seclabel{findimcas}

\index{Casimir!finite-dimensional|(}
For completeness, we briefly outline the derivation of
condition~\eqref{cascond4} for a finite-dimensional algebra, though we shall
be concerned with the canonical inner bracket for the remainder of the thesis.
The Lie--Poisson bracket can be written
\begin{equation}
	\lPB f \com g \rPB = {\W_\lambda}^{\mu\nu}\,\strconst_{ij}^k\,
		\fv^\lambda_k\,\frac{\pd f}{\pd\fv^\mu_i}
		\,\frac{\pd g}{\pd\fv^\nu_j},
\end{equation}
where the~$\strconst_{ij}^k$ are the structure constants of the
algebra~$\LieA$.  The roman indices denote the components of each~$\fv^\mu$,
in the same manner as the rigid body example of \secref{rigidbody}, and~$f$
and~$g$ are ordinary functions of the~$\fv^\mu_i$.  The Casimir
condition~\eqref{cascond2} is thus
\begin{equation}
	{\W_\lambda}^{\mu\nu}\,\strconst_{ij}^k\,\frac{\pd \Cas}{\pd\fv^\mu_i}
		\,\fv^\lambda_k = 0,
	\eqlabel{findimCascond}
\end{equation}
where both~$\nu$ and~$j$ are free indices.  From the structure constants we
can construct the \emph{Cartan--Killing form}\index{Cartan--Killing
form}~\cite{Hamermesh,Jacobson},
\begin{equation}
	\ckform_{ij} \ldef \strconst_{is}^t\,\strconst_{jt}^s.
\end{equation}
The Cartan--Killing form is symmetric, and is nondegenerate for a semisimple
algebra.  We assume this is the case for~$\LieA$, and denote the inverse
of~$\ckform_{ij}$ by~$\ckform^{ij}$.

For definiteness we take a Casimir of the form
\begin{equation}
	\Cas = \half\,\ckform^{ij}\,\Casi_{\mu\nu}\,\fv^\mu_i\,\fv^\nu_j,
	\eqlabel{finitedimCas}
\end{equation}
where~$\Casi_{\mu\nu}$ is a symmetric tensor.  Inserting this
into~\eqref{findimCascond}, we get
\begin{equation}
	{\W_\lambda}^{\mu\nu}\,\strconst_{ij}^k\,
		\ckform^{is}\,\Casi_{\mu\sigma}\,\fv^\sigma_s
		\,\fv^\lambda_k = 0.
	\eqlabel{findimCascond2}
\end{equation}
The symbol~\hbox{$\strconst^{sk}_j \ldef \ckform^{si}\,\strconst_{ij}^k$} can
be shown to be antisymmetric in its upper indices.  (We use the
Cartan--Killing form as a metric to raise and lower indices.)  We can then
define the
\index{bracket!on the dual}
bracket~\hbox{$\lpb\com\rpb^*:\LieA^*\times\LieA^*\rightarrow\LieA^*$} by
\begin{equation}
	{\lpb\fv\com\eta\rpb}^*_k \ldef \strconst^{ij}_k\,\fv_i\,\eta_j,
\end{equation}
which is a Lie bracket on~$\LieA^*$ induced by the Cartan--Killing
form~$\ckform$.  The Casimir condition~\eqref{findimCascond2} can be rewritten
neatly in terms of the bracket~$\lpb\com\rpb^*$ as
\begin{equation}
	{\W_\lambda}^{\mu\nu}\,\Casi_{\mu\sigma}\,
		{\lpb\fv^\sigma\com\fv^\lambda\rpb}^* = 0.
	\eqlabel{findimCascond3}
\end{equation}
This should be compared with condition~\eqref{cascond5}, for the
infinite-dimensional case, where the bracket~$\lpb\com\rpb^*$ is obtained from
the identification of~$\LieA$ and~$\LieA^*$.  The Casimir~\eqref{finitedimCas}
is thus the finite-dimensional analogue of~\eqref{formcas}.  Since
condition~\eqref{findimCascond2} has to be true for any value of the~$\fv$, it
follows that we must have
\begin{equation}
	{\W_\lambda}^{\mu\nu}\,\Casi_{\mu\sigma}
		= {\W_\sigma}^{\mu\nu}\,\Casi_{\mu\lambda},
\end{equation}
the same condition as~\eqref{cascond4}.  We conclude that, even thought we
shall be concerned with the canonical bracket case, many of the subsequent
results of this chapter apply to finite-dimensional brackets.
\index{Casimir!finite-dimensional|)}

\section{Direct Sum}
\seclabel{Casdirprod}

\index{Casimir!for a direct sum}
For the direct sum we found in \secref{directprod} that if we look at the
3-tensor~$\W$ as a cube, then it ``blocks out'' into smaller cubes, or
subblocks, along its main diagonal, each subblock representing a subalgebra.
We denote each subblock of~${\W_\lambda}^{\mu\nu}$
by~${{\W_i}_{\lambda}}^{\mu\nu}$, \hbox{$i=1,\dots,r$}, where~$r$ is the
number of subblocks.  We can rewrite~\eqref{LPB} as
\[
\begin{split}
	\lPB A\com B\rPB &= \sum_{i=1}^r \lang\fv_i^{\lambda}\com
		{{\W_i}_\lambda}^{\mu\nu}\,
		\lpb {\frac{\fd A}{\fd\fv_i^{\mu}}}\com
		{\frac{\fd B}{\fd\fv_i^{\nu}}}\rpb\rang \\
	&\rdef \sum_{i=1}^r {\lPB A\com B\rPB}_i\, ,
\end{split}
\]
where~$i$ labels the different subblocks and the greek indices run over the
size of the~$i$th subblock.  Each of the subbrackets~\hbox{${\lPB \com
\rPB}_i$} depends on different fields.  In particular, if the
functional~$\Cas$ is a Casimir, then, for any functional~$F$
\[
	\lPB F\com \Cas\rPB = \sum_{i=1}^r {\lPB F\com \Cas\rPB}_i = 0 
		\ \ \ \Longrightarrow
	\ \ \ {\lPB F\com \Cas\rPB}_i = 0,\ \ i=1,\dots,r\, .
\]
The solution for this is
\[
	\Cas[\fv] = \Cas_1[\fv_1] + \cdots
		+ \Cas_r[\fv_r]\, ,\ \ \
	{\rm where}\ {\lPB F\com \Cas_i\rPB}_i =0,\ i=1,\dots,r\, ,
\]
that is, the Casimir is just the sum of the Casimir for each subbracket.
Hence, the question of finding the Casimirs can be treated separately for each
component of the direct sum.  We thus assume we are working on a single
degenerate subblock, as we did for the classification in
\chref{classext}, and henceforth we drop the subscript~$i$.

\index{Casimir!and null eigenvectors}
There is a complication when a single (degenerate) subblock has more that one
simultaneous eigenvector.  By this we mean~$k$ vectors~$u^{(a)}$,
$a=1,\dots,k$, such that
\[
	{\W_\lambda}^{\mu(\nu)}\,u^{(a)}_\mu = \ev^{(\nu)}\,\,u^{(a)}_\lambda.
\]
Note that lower-triangular matrices always have at least the
eigenvector given by~\hbox{$u_\mu={\delta_\mu}^n$}.  Let~$\eta^{(a)} \ldef
u^{(a)}_\rho\xi^\rho$, and consider a
function~$\Casi(\eta^{(1)},\dots,\eta^{(k)})$.  Then
\begin{align*}
	{\W_\lambda}^{\mu(\nu)}
		\frac{\partial^2\Casi}{\partial\fv^\mu\partial\fv^\sigma}
	&= {\W_\lambda}^{\mu(\nu)} \sum_{a,b=1}^k
		u^{(a)}_\mu u^{(b)}_\sigma
		\frac{\partial^2\Casi}{\partial\eta^{(a)}\partial\eta^{(b)}}\,,
		\nonumber\\
	&= \ev^{(\nu)} \sum_{a,b=1}^k
		u^{(a)}_\lambda u^{(b)}_\sigma
		\frac{\partial^2\Casi}{\partial\eta^{(a)}\partial\eta^{(b)}}.
\end{align*}
Because the eigenvalue~$\ev^{(\nu)}$ does not depend on~$a$ (the block was
assumed to have degenerate eigenvalues), the above expression is symmetric
in~$\lambda$ and~$\sigma$.  Hence, the Casimir condition~\eqref{cascond3}
is satisfied.

The reason this is introduced here is that if a degenerate block splits into a
direct sum, then it will have several simultaneous eigenvectors.  The Casimir
invariants~$\Casi^{(a)}(\eta^{(a)})$ and~$\Casi^{(b)}(\eta^{(b)})$
corresponding to each eigenvector, instead of adding
as~$\Casi^{(a)}(\eta^{(a)}) + \Casi^{(b)}(\eta^{(b)})$, will combine into one
function to give~$\Casi{(\eta^{(a)},\eta^{(b)})}$, a more general functional
dependence. However, these situations with more than one eigenvector are not
limited to direct sums.  For instance, they occur in semidirect sums.  In
\secref{caslowdim} we will see examples of both cases.

\section{Local Casimirs for Solvable Extensions}
\seclabel{localcas}

\index{Casimir!for solvable extensions}
In the solvable case, when all the~$\W^{(\mu)}$'s are lower-triangular with
vanishing eigenvalues, a special situation occurs.  If we consider the Casimir
condition \eqref{cascond5}, we notice that derivatives with respect to~$\fv^n$
do not occur at all, since~$\W^{(n)}=0$.  Hence, the functional
\[
	\Cas[\fv] = \int_\fdomain\,\fv^n(\xv')\,\delta(\xv-\xv')\d^2x'
		= \fv^n(\xv)
\]
is conserved.  The variable~$\fv^n(\xv)$ is \emph{locally} conserved.  It
cannot have any dynamics associated with it.  This holds true for any other
simultaneous null eigenvectors the extension happens to have, but for the
solvable case~$\fv^n$ is always such a vector (provided the matrices have been
put in lower-triangular form, of course).

Hence, there are at most~$n-1$ dynamical variables in an order~$n$ solvable
extension.  An interesting special case occurs when the only
nonvanishing~$\W_{(\mu)}$ is for~$\mu=n$.  Then the Lie--Poisson bracket is
\[
	\lPB F\com G\rPB = \sum_{\mu,\nu=1}^{n-1}{\W_n}^{\mu\nu}
		\int_\fdomain\,\fv^n(\xv)
		\,\lpb \frac{\fd F}{\fd \fv^\mu(\xv)}\com
		\frac{\fd G}{\fd \fv^\nu(\xv)}\rpb\d^2x,
\]
where~$\fv^n(\xv)$ is some function of our choosing.  This bracket is not what
we would normally call Lie--Poisson because~$\fv^n(\xv)$ is not dynamical.
It gives equations of motion of the form
\[
	\dotfv^\nu = {\W_n}^{\nu\mu}\,\lpb \frac{\fd H}{\fd \fv^\mu}
		\com \fv^n \rpb,
\]
which can be used to model, for example, advection of scalars in a specified
flow given by~$\fv^n(\xv)$.  This bracket occurs naturally when a Lie--Poisson
bracket is linearized~\cite{MarsdenRatiu,Morrison1998}%
\index{bracket!linearized}.

\section{Solution of the Casimir Problem}
\seclabel{cassoln}

We now proceed to find the solution to~\eqref{cascond5}.  We assume that all
the~$\W^{(\mu)}$, $\mu=0,\dots,n$, are in lower-triangular form, and that the
matrix~$\W^{(0)}$ is the identity matrix (which we see saw can always be
done).  Though this is the semidirect form of the extension, we will see that
we can also recover the Casimir invariants of the solvable part.  We
assume~$\nu>0$ in~\eqref{cascond5}, since~$\nu=0$ does not lead to a condition
on the Casimir (\secref{cascond}).  Therefore~${\W_{\lambda}}^{n\nu}=0$.
Thus, we separate the Casimir condition into a part involving indices ranging
from~$0,\dots,n-1$ and a part that involves only~$n$. The condition
\[
\sum_{\mu,\sigma,\lambda=0}^{n} {\W_\lambda}^{\mu\nu} \Casi_{\dcom\mu\sigma}
	\lpb\fv^\lambda\com\fv^\sigma\rpb = 0, \ \ \ \nu > 0,
\]
becomes
\[
	\sum_{\lambda=0}^{n} \l\lgroup
	\sum_{\mu,\sigma=0}^{n-1}{\W_\lambda}^{\mu\nu} \Casi_{\dcom\mu\sigma}
	\lpb\fv^\lambda\com\fv^\sigma\rpb
	+ \sum_{\mu=0}^{n-1}{\W_\lambda}^{\mu\nu} \Casi_{\dcom\mu n}
	\lpb\fv^\lambda\com\fv^n\rpb
	\r\rgroup = 0,
\]
where we have used~${\W_{\lambda}}^{n\nu}=0$ to limit the sum on~$\mu$.
Separating the sum in~$\lambda$,
\begin{multline*}
	\sum_{\lambda=0}^{n-1} \l\lgroup
	\sum_{\mu,\sigma=0}^{n-1}{\W_\lambda}^{\mu\nu} \Casi_{\dcom\mu\sigma}
	\lpb\fv^\lambda\com\fv^\sigma\rpb
	+ \sum_{\mu=0}^{n-1}{\W_\lambda}^{\mu\nu} \Casi_{\dcom\mu n}
	\lpb\fv^\lambda\com\fv^n\rpb
	\r\rgroup\\
	+ \sum_{\mu,\sigma=0}^{n-1}{\W_n}^{\mu\nu} \Casi_{\dcom\mu\sigma}
	\lpb\fv^n\com\fv^\sigma\rpb
	+ \sum_{\mu=0}^{n-1}{\W_n}^{\mu\nu} \Casi_{\dcom\mu n}
	\lpb\fv^n\com\fv^n\rpb
	= 0.
\end{multline*}
The last sum vanishes because~$\lpb\fv^n\com\fv^n\rpb=0$. Now we separate the
condition into semisimple and solvable parts,
\begin{multline*}
	\sum_{\mu=1}^{n-1} \Biggl\lgroup
	\sum_{\lambda,\sigma=0}^{n-1}{\W_\lambda}^{\mu\nu}
		\Casi_{\dcom\mu\sigma}
	\lpb\fv^\lambda\com\fv^\sigma\rpb
	- \sum_{\sigma=0}^{n-1}{\W_\sigma}^{\mu\nu} \Casi_{\dcom\mu n}
	\lpb\fv^n\com\fv^\sigma\rpb
	\\
	\mbox{} + 
	\sum_{\sigma=0}^{n-1}{\W_n}^{\mu\nu} \Casi_{\dcom\mu\sigma}
	\lpb\fv^n\com\fv^\sigma\rpb
	\Biggr\rgroup
	+ \sum_{\lambda,\sigma=0}^{n-1}{\W_\lambda}^{0\nu}
		\Casi_{\dcom 0\sigma}
	\lpb\fv^\lambda\com\fv^\sigma\rpb\\
	- \sum_{\sigma=0}^{n-1}{\W_\sigma}^{0\nu} \Casi_{\dcom 0 n}
	\lpb\fv^n\com\fv^\sigma\rpb
	+ \sum_{\sigma=0}^{n-1}{\W_n}^{0\nu} \Casi_{\dcom 0\sigma}
	\lpb\fv^n\com\fv^\sigma\rpb
	= 0.
\end{multline*}
Using~${\W_\sigma}^{0\nu} = {\delta_\sigma}^\nu$, we can separate the
conditions into a part for~$\nu=n$ and one for~\hbox{$0<\nu<n$}.  For~$\nu=n$,
the only term that survives is the last sum
\[
	\sum_{\sigma=0}^{n-1} \Casi_{\dcom 0\sigma}
	\lpb\fv^n\com\fv^\sigma\rpb = 0.
\]
Since the commutators are independent, we have the conditions,
\begin{equation}
	\Casi_{\dcom 0\sigma} = 0, \ \ \ \sigma=0,\dots,n-1.
	\eqlabel{C0seq0}
\end{equation}
and for~\hbox{$0<\nu<n$},
\begin{multline*}
	\sum_{\mu=1}^{n-1} \Biggl\lgroup
	\sum_{\lambda,\sigma=1}^{n-1}{\W_\lambda}^{\mu\nu}
		\Casi_{\dcom\mu\sigma}
	\lpb\fv^\lambda\com\fv^\sigma\rpb
	- \sum_{\sigma=1}^{n-1}{\W_\sigma}^{\mu\nu} \Casi_{\dcom\mu n}
	\lpb\fv^n\com\fv^\sigma\rpb
	\\
	\mbox{} + 
	\sum_{\sigma=1}^{n-1}{\W_n}^{\mu\nu} \Casi_{\dcom\mu\sigma}
	\lpb\fv^n\com\fv^\sigma\rpb
	\Biggr\rgroup
	- \Casi_{\dcom 0 n} \lpb\fv^n\com\fv^\nu\rpb
	= 0,
\end{multline*}
where we have used~\eqref{C0seq0}.  Using independence of the inner brackets
gives
\begin{align}
	{\Wt_\lambda}^{\mu\nu}
		\Casi_{\dcom\mu\sigma} &=
	{\Wt_\sigma}^{\mu\nu}
		\Casi_{\dcom\mu\lambda},
	\eqlabel{cascondsubext} \\
	{\Wn}^{\nu\mu} \Casi_{\dcom\mu\sigma} &=
		{\Wt_\sigma}^{\nu\mu} \Casi_{\dcom\mu n}
		+ {\delta^\nu}_\sigma\, \Casi_{\dcom 0 n},
	\eqlabel{Axeqb}
\end{align}
for~$0< \sigma,\lambda,\nu,\mu < n$.  From now on in this section repeated
indices are summed, and all greek indices run from~$1$ to~$n-1$ unless
otherwise noted.  We have written a tilde over the~$\W$'s to stress the fact
that the indices run from~$1$ to~$n-1$, so that the~$\Wt$ represent a solvable
order~$(n-1)$ subextension of~$\W$.  This subextension does not
include~$\W_{(n)}$.  We have also made the definition
\begin{equation}
	\Wn^{\mu\nu} \ldef {\W_n}^{\mu\nu}.
	\eqlabel{Wndef}
\end{equation}
Equation~\eqref{cascondsubext} is a Casimir condition: it says that~$\Casi$ is
also a Casimir of~$\Wt$. We now proceed to solve~\eqref{Axeqb} for the case
where~$\Wn$ is nonsingular. In \secref{singWn} we will solve the
singular~$\Wn$ case. We will see that in both cases \eqref{cascondsubext}
follows from \eqref{Axeqb}.

\subsection{Nonsingular $\Wn$}
\seclabel{nsingWn}

\index{Casimir!for nonsingular~$\Wn$|(}
The simplest case occurs when~$\Wn$ has an inverse, which we will
call~$\Wni_{\mu\nu}$.  Then Eq.~\eqref{Axeqb} has solution
\begin{equation}
	\Casi_{\dcom\tau\sigma} =
		\coW^\mu_{\tau\sigma}\, \Casi_{\dcom\mu n}
		+ \Wni_{\tau\sigma}\, \Casi_{\dcom 0 n}\, ,
	\eqlabel{nonsingsol}
\end{equation}
where
\begin{equation}
	\coW^\mu_{\tau\sigma} \ldef \Wni_{\tau\nu}\,{\Wt_\sigma}^{\nu\mu}.
	\eqlabel{coextdef}
\end{equation}
We now verify that~$\coW^\mu_{\tau\sigma} = \coW^\mu_{\sigma\tau}$, as
required by the symmetry of the left-hand side of~\eqref{nonsingsol}.
\[
\begin{split}
	\coW^\mu_{\tau\sigma} &= \Wni_{\tau\nu}\,{\Wt_\kappa}^{\nu\mu}
		\,{\delta_\sigma}^\kappa \\
	&= \Wni_{\tau\nu}\,{\Wt_\kappa}^{\nu\mu}
		\,\Wn^{\rho\kappa}\,\Wni_{\sigma\rho} \\
	&= \Wni_{\tau\nu}\,\Bigl(\,\sum_{\kappa=1}^{n}{\Wt_\kappa}^{\nu\mu}
		\,{\W_n}^{\rho\kappa}\Bigr)\,\Wni_{\sigma\rho},
\end{split}
\]
where we used the fact that~\hbox{${\W_n}^{\rho n}=0$} to extend the sum.
Then we can use the commutativity property~\eqref{Wjacob} to
interchange~$\rho$ and~$\nu$,
\[
\begin{split}
	\coW^\mu_{\tau\sigma} &= \Wni_{\tau\nu}\,
		\Bigl(\,\sum_{\kappa=1}^{n}{\Wt_\kappa}^{\rho\mu}
		\,{\W_n}^{\nu\kappa}\Bigr)\,\Wni_{\sigma\rho}\\
	&= \Wni_{\tau\nu}\,{\W_n}^{\nu\kappa}
		\,\Wni_{\sigma\rho}\,{\Wt_\kappa}^{\rho\mu}\\
	&= {\delta_\tau}^{\kappa}
		\,\coW_{\sigma\kappa}^{\mu}\\
	&= \coW_{\sigma\tau}^{\mu},
\end{split}
\]
which shows that~$\coW$ is symmetric in its lower indices.

In~\eqref{nonsingsol}, it is clear that the~$n$th variable is ``special'';
this suggests that we try the following form for the Casimir:
\begin{equation}
	\Casi(\fv^0,\fv^1,\dots,\fv^n) =
	\sum_{i \ge 0}
	\ag^{(i)}(\fv^0,\fv^1,\dots,\fv^{n-1})\,\af_{i}(\fv^n),
	\eqlabel{Casform}
\end{equation}
where~$\af$ is arbitrary and~$\af_i$ is the $i$th derivative of~$\af$ with
respect to its argument.  One immediate advantage of this form is that
\eqref{cascondsubext} follows from \eqref{Axeqb}.  Indeed, taking a derivative
of \eqref{Axeqb} with respect to~$\fv^\lambda$, inserting~\eqref{Casform}, and
equating derivatives of~$\af$ leads to
\[
	{\Wn}^{\nu\mu}\, \ag^{(i)}_{\dcom\mu\sigma\lambda} =
		{\Wt_\sigma}^{\nu\mu}\, \ag^{(i+1)}_{\dcom\mu\lambda},
\]
where we have used \eqref{C0seq0}.  Since the left-hand side is symmetric
in~$\lambda$ and~$\sigma$ then so is the right-hand side, and
\eqref{cascondsubext} is satisfied.

Now, inserting the form of the Casimir~\eqref{Casform} into the
solution~\eqref{nonsingsol}, we can equate derivatives of~$\af$ to obtain
for~\hbox{$\tau,\sigma=1,\dots,n-1$},
\begin{equation}
\begin{split}
	\ag^{(0)}_{\dcom\tau\sigma} &= 0,\\
	\ag^{(i)}_{\dcom\tau\sigma} &=
		\coW^\mu_{\tau\sigma}\, \ag^{(i-1)}_{\dcom\mu} +
		\Wni_{\tau\sigma}\,\ag^{(i-1)}_{\dcom 0}, \quad i \ge 1.
	\eqlabel{gcondi}
\end{split}
\end{equation}
The first condition, together with~\eqref{C0seq0}, says that~$\ag^{(0)}$
is linear in~$\fv^0,\dots\fv^{n-1}$.  There are no other conditions
on~$\ag^{(0)}$, so we can obtain~$n$ independent solutions by choosing
\begin{equation}
	\ag^{(0)\nu} = \fv^\nu, \ \ \ \nu=0,\dots,n-1.
	\eqlabel{nsingsolzero}
\end{equation}
The equation for~$\ag^{(1)\nu}$ is
\begin{equation}
	\ag^{(1)\nu}_{\dcom\tau\sigma} =
	\begin{cases}
		\Wni_{\tau\sigma}& \nu = 0, \\
		\coW^\nu_{\tau\sigma}& \nu = 1,\dots,n-1.
	\end{cases}
	\eqlabel{nsingsolone}
\end{equation}
Thus~$\ag^{(1)\nu}$ is a quadratic polynomial (the arbitrary linear part does
not yield an independent Casimir, so we set it to zero).  Note
that~$\ag^{(1)\nu}$ does not depend on~$\fv^0$
since~$\tau,\sigma=1,\dots,n-1$.  Hence, for~$i>1$ we can drop
the~$\ag^{(i-1)}_{\dcom 0}$ term in~\eqref{gcondi}.  Taking derivatives
of~\eqref{gcondi}, we obtain
\begin{equation}
	\ag^{(i)\nu}_{\dcom\tau_1\tau_2\dots\tau_{(i+1)}} =
		\coW^{\mu_1}_{\tau_1\tau_2}\,
		\coW^{\mu_2}_{\mu_1\tau_3}\cdots
		\coW^{\mu_{(i-1)}}_{\mu_{(i-2)}\tau_{i}}
		\,\ag^{(1)\nu}_{\dcom\mu_{(i-1)}\tau_{(i+1)}}.
	\eqlabel{nonsingsolni}
\end{equation}
We know the series will terminate because the~$\Wt^{(\mu)}$, and hence
the~$\coW_{(\mu)}$, are nilpotent.  The solution to \eqref{nonsingsolni} is
\begin{equation}
	\ag^{(i)\nu} = \frac{1}{(i+1)!}\,\,
		\agc^{(i)\nu}_{\tau_1\tau_2\dots\tau_{(i+1)}}\,
		\fv^{\tau_1}\fv^{\tau_2}\cdots\fv^{\tau_{(i+1)}}\,,
		\ \ \ \ i > 1,
	\eqlabel{Cascoeff}
\end{equation}
where the constants~$\agc$ are defined by
\begin{equation}
	\agc^{(i)\nu}_{\tau_1\tau_2\dots\tau_{(i+1)}} \ldef
		\coW^{\mu_1}_{\tau_1\tau_2}\,
		\coW^{\mu_2}_{\mu_1\tau_3}\cdots
		\coW^{\mu_{(i-1)}}_{\mu_{(i-2)}\tau_{i}}
		\,\ag^{(1)\nu}_{\dcom\mu_{(i-1)}\tau_{(i+1)}}.
	\eqlabel{agcdef}
\end{equation}
In summary, the~$\ag^{(i)}$'s of~\eqref{Casform} are
given by~\eqref{nsingsolzero},~\eqref{nsingsolone}, and~\eqref{Cascoeff}.

Because the left-hand side of~\eqref{nonsingsolni} is symmetric in all its
indices, we require
\begin{equation}
	\coW^{\mu}_{\tau\sigma}\,\coW^{\nu}_{\mu\lambda} =
	\coW^{\mu}_{\tau\lambda}\,\coW^{\nu}_{\mu\sigma}, \qquad i>1.
	\eqlabel{coextcond}
\end{equation}
This is straightforward to show, using~\eqref{Wjacob} and the symmetry
of~$\coW$:
\[
\begin{split}
	\coW^{\mu}_{\tau\sigma}\,\coW^{\nu}_{\mu\lambda} &=
		\coW^{\mu}_{\sigma\tau}\,\coW^{\nu}_{\lambda\mu} \\
	&= (\Wni_{\sigma\kappa}\,{\Wt_\tau}^{\kappa\mu})\,
		(\Wni_{\lambda\rho}\,{\Wt_\mu}^{\rho\nu})\\
	&= \Wni_{\sigma\kappa}\,\Wni_{\lambda\rho}\,
		{\Wt_\tau}^{\kappa\mu}\,{\Wt_\mu}^{\rho\nu}\\
	&= \Wni_{\sigma\kappa}\,\Wni_{\lambda\rho}\,
		{\Wt_\tau}^{\rho\mu}\,{\Wt_\mu}^{\kappa\nu}\\
	&= \coW^{\mu}_{\tau\lambda}\,\coW^{\nu}_{\mu\sigma}
\end{split}
\]
If we compare this to~\eqref{Wjacob}, we see that~$\coW$ satisfies all the
properties of an extension, except with the dual indices.  Thus we will
call~$\coW$ the \emph{coextension}\index{coextension} of~$\Wt$ with respect
to~$\Wn$.  Essentially,~$\Wn$ serves the role of a metric that allows us to
raise and lower indices.  The formulation presented here is, however, not
covariant.  We have not been able to find a covariant formulation of the
coextension, which is especially problematic for the singular~$\Wn$ case
(\secref{singWn}).  Since the coextension depends strongly on the
lower-triangular form of the~$\W^{(\mu)}$'s, it may well be that a covariant
formulation does not exist.

For a solvable extension we simply restrict~$\nu > 0$ and the above treatment
still holds.  We conclude that the Casimirs of the solvable part of a
semidirect extension are Casimirs of the full extension.\index{Casimir!for
solvable extensions}  We have also shown,
for the case of nonsingular~$\Wn$, that the number of independent Casimirs is
equal to the order of the extension.
\index{Casimir!for nonsingular~$\Wn$|)}

\subsection{Singular $\Wn$}
\seclabel{singWn}

\index{Casimir!for singular~$\Wn$|(}
In general,~$\Wn$ is singular and thus has no inverse.  However, it always has
a (symmetric and unique) pseudoinverse~$\Wni_{\mu\nu}$\index{pseudoinverse}
such that
\begin{align}
	\Wni_{\mu\sigma}\,\Wn^{\sigma\tau}\,\Wni_{\tau\nu}
		&= \Wni_{\mu\nu},\eqlabel{pseudoinv1}\\
	\Wn^{\mu\sigma}\,\Wni_{\sigma\tau}\,\Wn^{\tau\nu}
		&= \Wn^{\mu\nu}.
	\eqlabel{pseudoinv2}
\end{align}
The pseudoinverse is also known as the strong generalized inverse or the
Moore--Penrose inverse~\cite{Osta}. It follows from~\eqref{pseudoinv1}
and~\eqref{pseudoinv2} that the matrix operator
\[
	{\Proj^\nu}_\tau \ldef \Wn^{\nu\kappa}\,\Wni_{\kappa\tau}
\]
projects onto the range of~$\Wn$.  The system~\eqref{Axeqb} only has a
solution if the following solvability condition is satisfied:
\begin{equation}
	{\Proj^\nu}_\tau\,
		\l({\Wt_\sigma}^{\tau\mu} \Casi_{\dcom\mu n}
			+ {\delta^\tau}_\sigma\, \Casi_{\dcom 0 n}\r)
	= {\Wt_\sigma}^{\nu\mu} \Casi_{\dcom\mu n}
		+ {\delta^\nu}_\sigma\, \Casi_{\dcom 0 n};
	\eqlabel{solvcondz}
\end{equation}
that is, the right-hand side of~\eqref{Axeqb} must live in the range of~$\Wn$.

If~\hbox{$\Casi_{\dcom 0 n}\ne 0$}, the quantity~${\Wt_\sigma}^{\nu\mu}\,
\Casi_{\dcom\mu n} + {\delta^\nu}_\sigma\, \Casi_{\dcom 0 n}$ has rank equal
to~$n$, because the quantity~${\Wt_\sigma}^{\nu\mu}\,\Casi_{\dcom\mu n}$ is
lower-triangular (it is a linear combination of lower-triangular matrices).
Thus, the projection operator must also have rank~$n$.  But then this implies
that~$\Wn$ has rank~$n$ and so is nonsingular, which contradicts the
hypothesis of this section.  Hence,~\hbox{$\Casi_{\dcom 0 n} = 0$} for the
singular~$\Wn$ case, which together with \eqref{C0seq0} means that a Casimir
that depends on~$\fv^0$ can only be of the form~$\Casi = \af(\fv^0)$.
However, since~$\fv^0$ is not an eigenvector of the~$\W^{(\mu)}$'s, the only
possibility is~$\Casi = \fv^0$, the trivial linear case mentioned in
\secref{cascond}.

The solvability condition \eqref{solvcondz} can thus be rewritten as
\begin{equation}
	\l({\Proj^\nu}_\tau\,{\Wt_\sigma}^{\tau\mu}
		- {\Wt_\sigma}^{\nu\mu}\r) \Casi_{\dcom\mu n} = 0.
	\eqlabel{solvcond}
\end{equation}
An obvious choice would be to
require~\hbox{${\Proj^\nu}_\tau\,{\Wt_\sigma}^{\tau\mu} =
{\Wt_\sigma}^{\nu\mu}$}, but this is too strong.  We will derive a weaker
requirement shortly.

By an argument similar to that of~\secref{nsingWn}, we now assume~$\Casi$ is
of the form
\begin{equation}
	\Casi(\fv^1,\dots,\fv^n) =
	\sum_{i \ge 0} \ag^{(i)}(\fv^1,\dots,\fv^{n-1})\,\af_{i}(\fv^n),
	\eqlabel{singCas}
\end{equation}
where again~$\af_i$ is the $i$th derivative of~$f$ with respect to its
argument.  As in \secref{nsingWn}, we only need to show \eqref{Axeqb}, and
\eqref{cascondsubext} will follow.  The number of independent solutions of
\eqref{Axeqb} is equal of the rank of~$\Wn$.  The choice
\begin{equation}
	\ag^{(0)\nu} = {\Proj^\nu}_{\rho}\,\fv^\rho, \ \ \ \nu=1,\dots,n-1,
	\eqlabel{singsolzero}
\end{equation}
provides the right number of solutions because the rank of~$\Proj$ is equal to
the rank of~$\Wn$.  It also properly specializes to
\eqref{nsingsolzero} when~$\Wn$ is nonsingular, for then~${\Proj^\nu}_\rho =
{\delta^{\,\nu}}_\rho$.

The solvability condition~\eqref{solvcond} with this form for the Casimir
becomes
\begin{equation}
	\l({\Proj^\nu}_\tau\,{\Wt_\sigma}^{\tau\mu}
		- {\Wt_\sigma}^{\nu\mu}\r) \ag^{(i)\nu}_{\dcom\mu} = 0,\ \ \
	i \ge 0.
	\eqlabel{solvcondg}
\end{equation}
For~$i=0$ the condition can be shown to simplify to
\[
	{\Proj^\nu}_\tau\,{\Wt_\sigma}^{\tau\mu}
	= {\Wt_\sigma}^{\nu\tau}\,{\Proj^\mu}_\tau,
\]
or to the equivalent matrix form
\begin{equation}
	\Proj\,\Wt_{(\sigma)} = \Wt_{(\sigma)}\,\Proj,
	\eqlabel{solvcond0}
\end{equation}
since~$\Proj$ is symmetric~\cite{Osta}.

Equation \eqref{Axeqb} becomes
\begin{align*}
	{\Wn}^{\kappa\mu} \ag^{(0)\nu}_{\dcom\mu\sigma} &= 0,\\
	{\Wn}^{\kappa\mu} \ag^{(i)\nu}_{\dcom\mu\sigma} &=
		{\Wt_\sigma}^{\kappa\mu} \ag^{(i-1)\nu}_{\dcom\mu},\qquad
		i > 0.
\end{align*}
If \eqref{solvcond} is satisfied, we know this has a solution given by
\[
	\ag^{(i)\nu}_{\dcom\lambda\sigma} =
		\Wni_{\lambda\rho}\,{\Wt_\sigma}^{\rho\mu}\,
		\ag^{(i-1)\nu}_{\dcom\mu} + \l({\delta_\lambda}^\mu
		- \Wni_{\lambda\rho}\,\Wn^{\rho\mu}\r)
		\ae^{(i-1)\nu}_{\mu\sigma} ,\ \ \ i > 0,
\]
where~$\ae$ is arbitrary, and~\hbox{$({\delta_\lambda}^\mu -
\Wni_{\lambda\rho}\,\Wn^{\rho\mu})$} projects onto the null space of~$\Wn$.
The left-hand side is symmetric in~$\lambda$ and~$\sigma$, but not the
right-hand side.  We can symmetrize the right-hand side by an appropriate
choice of the null eigenvector,
\[
	\ae^{(i)\nu}_{\lambda\sigma} \ldef
		\Wni_{\sigma\rho}\,{\Wt_\lambda}^{\rho\mu}\,
		\ag^{(i)\nu}_{\dcom\mu},\ \ \ i \ge 0,
\]
in which case
\[
	\ag^{(i)\nu}_{\dcom\lambda\sigma} =
		\coW^\mu_{\lambda\sigma}\,
		\ag^{(i-1)\nu}_{\dcom\mu},\ \ \ i > 0,
\]
where
\begin{equation}
	\coW^\nu_{\lambda\sigma} \ldef
		\Wni_{\sigma\rho}\,{\Wt_\lambda}^{\rho\nu}
		+ \Wni_{\lambda\rho}\,{\Wt_\sigma}^{\rho\nu}
		- \Wni_{\lambda\rho}\,\Wni_{\sigma\kappa}\,
		\Wn^{\rho\mu}\,{\Wt_\mu}^{\kappa\nu}\,,
	\eqlabel{coextension}
\end{equation}
which is symmetric in~$\lambda$ and~$\sigma$.  Equation~\eqref{coextension}
also reduces to \eqref{coextdef} when~$\Wn$ is nonsingular, for then the null
eigenvector vanishes.  The full solution is thus given in the same manner as
\eqref{nonsingsolni} by
\begin{equation}
	\ag^{(i)\nu} = \frac{1}{(i+1)!}\,\,
		\agc^{(i)\nu}_{\tau_1\tau_2\dots\tau_{(i+1)}}\,
		\fv^{\tau_1}\fv^{\tau_2}\cdots\fv^{\tau_{(i+1)}}\,,
		\ \ \ \ i > 0,
	\eqlabel{Cascoeffsing}
\end{equation}
where the constants~$\agc$ are defined by
\begin{equation}
	\agc^{(i)\nu}_{\tau_1\tau_2\dots\tau_{(i+1)}} \ldef
		\coW^{\mu_1}_{\tau_1\tau_2}\,
		\coW^{\mu_2}_{\mu_1\tau_3}\cdots
		\coW^{\mu_{(i-1)}}_{\mu_{(i-2)}\tau_{i}}
		\,\coW^{\mu_{i}}_{\mu_{(i-1)}\tau_{(i+1)}}
		\,{\Proj^{\,\nu}}_{\mu_{i}}\,,
	\eqlabel{agcdefsing}
\end{equation}
and~$\ag^{(0)}$ is given by \eqref{singsolzero}.

The~$\coW$'s must still satisfy the coextension condition \eqref{coextcond}.
Unlike the nonsingular case this condition does not follow directly and is an
extra requirement in addition to the solvability condition \eqref{solvcondg}.
Note that only the~$i=0$ case, Eq. \eqref{solvcond0}, needs to be satisfied,
for then \eqref{solvcondg} follows.  Both these conditions are
coordinate-dependent, and this is a drawback.  Nevertheless, we have found in
obtaining the Casimir invariants for the low-order brackets that if these
conditions are not satisfied, then the extension is a direct sum and the
Casimirs can be found by the method of \secref{Casdirprod}.  However, this has
not been proved rigorously.
\index{Casimir!for singular~$\Wn$|)}

\section{Examples}
\seclabel{Casex}

We now illustrate the methods developed for finding Casimirs with a few
examples.  First we treat our prototypical case of CRMHD, and give a physical
interpretation of invariants. Then, we derive the Casimir invariants for
Leibniz extensions of arbitrary order. Finally, we give an example involving a
singular~$\Wn$.

\subsection{Compressible Reduced MHD}
\seclabel{CRMHDCas}

\index{Casimir!for compressible reduced MHD|(}
The~$\W$ tensors representing the bracket for CRMHD (see \secref{CRMHD}) were
given in \secref{matCRMHD}. We have~$n=3$, so from \eqref{Wndef} we get
\begin{equation}
	\Wn = \begin{pmatrix} 0 & -\crmhdbeta \\ -\crmhdbeta & 0
		\end{pmatrix},\ \ \
	\Wni = \Wn^{-1} = \begin{pmatrix} 0 & -\crmhdbeta^{-1} \\
		-\crmhdbeta^{-1} & 0 \end{pmatrix}.
	\eqlabel{CRMHDWni}
\end{equation}
In this case, the coextension is trivial: all three matrices~$\coW^{(\nu)}$
defined by \eqref{coextdef} vanish. Using \eqref{Casform} and
\eqref{nsingsolzero}, with~$\nu=1$ and~$2$, the Casimirs for the solvable part are
\[
	\Casi^1 = \fv^1\,\afii(\fv^3) = \pvel\,\afii(\magf),\ \ \
	\Casi^2 = \fv^2\,\afiii(\fv^3) = \pres\,\afiii(\magf),
\]
and the Casimir associated with the eigenvector~$\fv^3$ is
\[
	\Casi^3 = \afiv(\fv^3) = \afiv(\magf).
\]
Since~$\Wn$ is nonsingular we also get another Casimir from the semidirect sum
part,
\[
	\Casi^0 = \fv^0\,\afi(\fv^3)
		- \frac{1}{\crmhdbeta}\,\fv^1\,\fv^2\,\afi'(\fv^3)
	= \vort\,\afi(\magf)
		- \frac{1}{\crmhdbeta}\,\pres\,\pvel\,\afi'(\magf).
\]

\index{Casimir!physical interpretation}
The physical interpretation of the invariant~$\Casi^3$ is given in
Morrison~\cite{Morrison1987} and Thiffeault and
Morrison~\cite{Thiffeault1998}.  This invariant implies the preservation of
contours of~$\magf$, so that the value~$\magf_0$ on a contour labels that
contour for all times. This is a consequence of the lack of dissipation and
the divergence-free nature of the velocity. Substituting~\hbox{$\Casi^3(\magf)
= \magf^k$} we also see that all the moments of the magnetic flux are
conserved.  By choosing~\hbox{$\Casi^3(\magf) = \heavyside(\magf(\xv) -
\magf_0)$}, a heavyside function, and inserting into \eqref{formcas}, it
follows that the area inside of any $\magf$-contour is conserved.

To understand the Casimirs~$\Casi^1$ and~$\Casi^2$, we also let
$\afii(\magf)=\heavyside(\magf-\magf_0)$ in~$\Casi^1$. In this case we have
\[
	\Cas^1[\pvel\,;\magf] = \int_{\fdomain}\pvel\,\afii(\magf)\d^2x
	= \int_{\magfcont_0}\,\pvel(\xv)\d^2x,
\]
where~$\magfcont_0$ represents the (not necessarily connected) region of
$\fdomain$ enclosed by the contour $\magf=\magf_0$ and~$\pd\magfcont_0$ is
its boundary. By the interpretation we gave of~$\Casi^3$, the contour
$\pd\magfcont_0$ moves with the fluid. So the total value of~$\pvel$ inside of
a~$\magf$-contour is conserved by the flow. The same is true of the
pressure~$\pres$. (See Thiffeault and Morrison~\cite{Thiffeault1998} for an
interpretation of these invariants in terms of relabeling symmetries, and a
comparison with the rigid body.)

The total pressure and parallel velocity inside of any $\magf$-contour are
preserved.  To understand $\Casi^4$, we use the fact
that~$\vort=\lapl\elecp$ and integrate by parts to obtain
\[
	\Cas^4[\vort,\pvel,\pres,\magf] = -\int_\fdomain
		\l(\grad\elecp\cdot\grad\magf
		+ \frac{\pvel\,p}{\crmhdbeta}\r)\afi'(\magf)\d^2x.
\]
The quantity in parentheses is thus invariant inside of any $\magf$-contour.
It can be shown that this is a remnant of the conservation by the full MHD
model of the cross helicity,
\[
	V = \int_\fdomain {\mathbf{v}}\cdot{\mathbf{B}}\d^2x\,,
\]
at second order in the inverse aspect ratio, while the conservation
of~$\Cas^1[\pvel\,;\magf]$ is a consequence of preservation of this quantity
at first order. Here ${\mathbf{B}}$ is the magnetic field.  The
quantities~$\Cas^3[\magf]$ and~$\Cas^2[\pres\,;\magf]$ they are, respectively,
the first and second order remnants of the preservation of helicity,
\[
	W = \int_\fdomain {\mathbf{A}}\cdot{\mathbf{B}}\d^2x,
\]
where ${\mathbf{A}}$ is the magnetic vector potential.
\index{Casimir!for compressible reduced MHD|)}

\subsection{Leibniz Extension}
\seclabel{Casleib}

\index{Casimir!for Leibniz extension|(}
We first treat the nilpotent case.  The Leibniz extension of \secref{Leibniz}
can be characterized by
\begin{equation}
	{\W_\lambda}^{\mu\nu} = {\delta_\lambda}^{\mu+\nu}\,,
	\ \ \ \mu,\nu, \lambda = 1,\dots,n,
	\tag{\ref{eq:sLeib}}
\end{equation}
where the tensor~$\delta$ is an ordinary Kronecker delta.  Upon restricting
the indices to run from~$1$ to~$n-1$ (the tilde notation of \secref{cassoln}),
we have
\[
	\Wn^{\mu\nu} = {\Wt_n}^{\mu\nu} = {\delta_n}^{\mu+\nu}\,,
	\ \ \ \mu,\nu = 1,\dots,n-1.
\]
The matrix~$\Wn$ is nonsingular with inverse equal to
itself:~\hbox{$\Wni_{\mu\nu} =
\delta_{\mu+\nu}^{\,\,n}$}.  The coextension of~$\Wt$\index{coextension!for
Leibniz extension} is thus
\[
	\coW^\mu_{\tau\sigma} = \sum_{\nu=1}^{n-1}\Wni_{\tau\nu}\,
		{\Wt_\sigma}^{\nu\mu}
	= \sum_{\nu=1}^{n-1}\delta^n_{\tau+\nu}\,
		{\delta_\sigma}^{\nu+\mu}
	= \delta^{\mu+n}_{\tau+\sigma}\,.
\]
Equation \eqref{agcdef} becomes
\[
\begin{split}
	\agc^{(i)\nu}_{\tau_1\tau_2\dots\tau_{(i+1)}} &=
		\coW^{\mu_1}_{\tau_1\tau_2}\,
		\coW^{\mu_2}_{\mu_1\tau_3}\cdots
		\coW^{\mu_{(i-1)}}_{\mu_{(i-2)}\tau_{i}}\,
		\coW^{\nu}_{\mu_{(i-1)}\tau_{(i+1)}}\\
	&= \delta^{\mu_1+n}_{\tau_1+\tau_2}\,
		\delta^{\mu_2+n}_{\mu_1+\tau_3}\cdots
		\delta^{\mu_{(i-1)+n}}_{\mu_{(i-2)}+\tau_{i}}\,
		\delta^{\nu+n}_{\mu_{(i-1)}+\tau_{(i+1)}}\\
	&= \delta^{\nu+in}_{\tau_1+\tau_2+\cdots+\tau_{(i+1)}}\,,
		\qquad \nu = 1,\dots,n-1,
\end{split}
\]
which, as required, this is symmetric under interchange of the~$\tau_i$. Using
\eqref{Casform}, \eqref{nsingsolzero}, \eqref{nsingsolone}, and
\eqref{Cascoeff} we obtain the~$n-1$ Casimir invariants
\begin{equation}
	\Casi^\nu(\fv^1,\dots,\fv^n) =
	\sum_{i \ge 0}
	\frac{1}{(i+1)!}\,\,
	{\delta^{\nu+in}_{\tau_1+\tau_2+\cdots+\tau_{(i+1)}}}\,
	\fv^{\tau_1}\cdots\fv^{\tau_{(i+1)}}\,
	\af^\nu_{i}(\fv^n),
	\eqlabel{CasiLeib}
\end{equation}
for~$\nu=1,\dots,n-1$.  The superscript~$\nu$ on~$\af$ indicates that the
arbitrary function is different for each Casimir, and recall the subscript~$i$
denotes the~$i$th derivative with respect to~$\fv^n$.  The~$n$th invariant is
simply~$\Casi^\nu(\fv^n) = \af^n(\fv^n)$, corresponding to the null
eigenvector in the system.  Thus there are~$n$ independent Casimirs, as stated
in \secref{nsingWn}.

For the Leibniz semidirect sum case, since~$\Wn$ is nonsingular, there will be
an extra Casimir given by \eqref{CasiLeib} with~$\nu=0$, and the~$\tau_i$ sums
run from~$0$ to~$n-1$.  This is the same form as the~$\nu=1$ Casimir of the
order~$(n+1)$ nilpotent extension.

For the $i$th term in \eqref{CasiLeib}, the maximal value of any~$\tau_j$ is
achieved when all but one (say, $\tau_1$) of the~$\tau_j$ are equal to~$n-1$,
their maximum value.  In this case we have
\[
	\tau_1+\tau_2+\cdots+\tau_{i+1} = \tau_1 + i(n-1) = \nu + i n,
\]
so that~$\tau_1 = i+\nu$.  Hence, the~$i$th term depends only
on~\hbox{$\l(\fv^{\nu+i},\dots,\fv^n\r)$}, and the~$\nu$th Casimir depends
on~\hbox{$\l(\fv^\nu,\dots,\fv^n\r)$}.  Also,
\[
	\max{\l(\tau_1+\cdots+\tau_{i+1}\r)} = (i+1)(n-1) = \nu + i n,
\]
which leads to~\hbox{$\max i = n - \nu - 1$}.  Thus the sum \eqref{CasiLeib}
terminates, as claimed in \secref{nsingWn}.  We rewrite \eqref{CasiLeib} in
the more complete form
\[
	\Casi^\nu(\fv^\nu,\dots,\fv^n) =
	\sum_{k = 1}^{n-\nu}
	\frac{1}{k!}\,\,
	{\delta^{\nu+(k-1)n}_{\tau_1+\tau_2+\cdots+\tau_{k}}}\,
	\fv^{\tau_1}\cdots\fv^{\tau_{k}}\,
	\af^\nu_{k-1}(\fv^n),
\]
for~$\nu=0,\dots,n$. \tabref{CasimirLeibn5} gives the~$\nu=1$ Casimirs up to
order~$n=5$.

\begin{table}
\begin{center}
\begin{tabular}{ll} \hline
\rule[0in]{0em}{1.1em}$n$ & Invariant \\[3pt] \hline

%
%
\rule[0in]{0em}{1.4em}1 & $\af(\fv^1)$
\\[9pt]

%
%
2 & $\fv^1\af(\fv^2)$
\\[9pt]

%
%
3 & $\fv^1\af(\fv^3) + \frac{1}{2}{(\fv^2)^2}\af'(\fv^3)$
\\[9pt]

%
%
4 & $\fv^1\af(\fv^4) + \fv^2\fv^3\af'(\fv^4)
	+ \frac{1}{3!}(\fv^3)^3\af''(\fv^4)$
\\[9pt]

%
%
5 & $\fv^1\af(\fv^5) + \l(\fv^2\fv^4
		+ \frac{1}{2}(\fv^3)^2\r)\af'(\fv^5)
		+ \frac{1}{2}\fv^3(\fv^4)^2\af''(\fv^5)
		+ \frac{1}{4!}(\fv^4)^4\af'''(\fv^5)$
\\[9pt] \hline

\end{tabular}
\end{center}

\caption{Casimir invariants for Leibniz extensions up to order~$n=5$
($\nu=1$).  The primes denote derivatives.}
\tablabel{CasimirLeibn5}

\end{table}
\index{Casimir!for Leibniz extension|)}

\subsection{Singular~$\Wn$}
\seclabel{singWnex}

Now consider the~$n=4$ extension from \secref{neqfour}, \caseref{n4-10}c. We
have
\[
	\Wt_{(2)} = \begin{pmatrix} 1 & 0 & 0 \\ 0 & 0 & 0 \\ 0 & 0 & 0
		    \end{pmatrix},\ \ \ \
	\Wn = \begin{pmatrix} 0 & 0 & 1 \\ 0 & 0 & 0 \\ 1 & 0 & 0
		\end{pmatrix},
\]
with~$\Wt_{(1)} = \Wt_{(3)} = 0$. The pseudoinverse of~$\Wn$ is $\Wni=\Wn$ and
the projection operator is
\[
	{\Proj^\nu}_\tau \ldef \Wn^{\nu\kappa}\,\Wni_{\kappa\tau} = 
		\begin{pmatrix} 1 & 0 & 0 \\ 0 & 0 & 0 \\ 0 & 0 & 1
		\end{pmatrix}.
\]
The solvability condition \eqref{solvcond0} is obviously satisfied. We build
the coextension given by \eqref{coextension}, which in matrix form is
\[
	\coW^{(\nu)} = \Wt^{(\nu)}\,\Wni + (\Wt^{(\nu)}\,\Wni)^T
		- \Wni\,\Wn\,\Wt^{(\nu)}\,\Wni,
\]
to obtain
\[
	\coW^{(1)} = \begin{pmatrix}
		0 & 0 & 0 \\ 0 & 0 & 1 \\ 0 & 1 & 0\end{pmatrix},\ \ \
	\coW^{(2)} = \coW^{(3)} = 0.
\]
These are symmetric and obviously satisfy \eqref{coextcond}, so we have a good
coextension. Using \eqref{singCas}, \eqref{singsolzero},
\eqref{Cascoeffsing}, and \eqref{agcdefsing} we can write, for~$\nu=1$ and~$3$,
\begin{align*}
	\Casi^1 &= \fv^1\afi(\fv^4) + \fv^2\,\fv^3\afi'(\fv^4),\\
	\Casi^3 &= \fv^3\afii(\fv^4).
\end{align*}
This extension has two null eigenvectors, so from \secref{Casdirprod} we also
have the Casimir~$\afiii(\fv^2,\fv^4)$. The functions~$\afi$, $\afii$, and
$\afiii$ are arbitrary, and the prime denotes differentiation with respect to
argument.

\section{Casimir Invariants for Low-order Extensions}
\seclabel{caslowdim}

Using the techniques developed so far, we now find the Casimir invariants for
the low-order extensions classified in \secref{lowdimext}.  We first find the
Casimir invariants for the solvable extensions, since these are also
invariants for the semidirect sum case.  Then, we obtain the extra Casimir
invariants for the semidirect case, when they exist.

\subsection{Solvable Extensions}

Now we look for the Casimirs of solvable extensions.
As mentioned in \secref{localcas}, the Casimirs associated with null
eigenvectors (the only kind of eigenvector for solvable extensions) are
actually conserved locally.  We shall still write them in the
form~$\Casi=f(\fv^n)$, where~$\Casi$ is as in
\eqref{formcas}, so they have the correct form as invariants for the
semidirect case of \secref{lowsemicasi} (for which they are no longer locally
conserved).

\subsubsection{n=1}

Since the bracket is Abelian, any function~$\Casi=\Casi({\fv}^{1})$ is a
Casimir.

\subsubsection{n=2}

For the Abelian case we have~$\Casi=\Casi(\fv^1,\fv^2)$.  The only other case
is the Leibniz extension,
\[
	\Casi(\fv^1,\fv^2) = \fv^1\afi(\fv^2) + \afii(\fv^2).
\]

\subsubsection{n=3}
\seclabel{Casn3}

As shown in \secref{neqthree}, there are four cases.  \caseref{n4-00} is the
Abelian case, for which any function~$\Casi=\Casi(\fv^1,\fv^2,\fv^3)$ is a
Casimir.  \caseref{n4-01} is essentially the solvable part of the CRMHD
bracket, which we treated in~\secref{CRMHDCas}.  \caseref{n4-10} is a direct
sum of the Leibniz extension for~$n=2$, which has the bracket
\[
	\lpb(\alpha_1,\alpha_2)\com(\beta_1,\beta_2)\rpb
		= (0,\lpb\alpha_1\com\beta_1\rpb),
\]
with the Abelian algebra~$\lpb\alpha_3\com\beta_3\rpb=0$.  Hence, the Casimir
invariant is the same as for the~$n=2$ Leibniz extension with the
extra~$\fv^3$ dependence of the arbitrary function (see \secref{Casdirprod}).
Finally, \caseref{n4-11} is the Leibniz Casimir.  These results are summarized
in \tabref{Casimirn3}.

Cases~\ref{case:n4-00} and~\ref{case:n4-10} are trivial extensions, that is,
the cocycle appended to the $n=2$ case vanishes.  The procedure of then adding
$\fv^n$ dependence to the arbitrary function works in general.

\begin{table}

\begin{center}
\begin{tabular}{ll} \hline
\rule[0in]{0em}{1.1em}Case & Invariant \\[3pt] \hline

%
%
\rule[0in]{0em}{1.4em}1 & $\Casi(\fv^1,\fv^2,\fv^3)$
\\[9pt]

%
%
2 & $\fv^1\afi(\fv^3) + \fv^2\afii(\fv^3) + \afiii(\fv^3)$
\\[9pt]

%
%
3 & $\fv^1\afi(\fv^2) + \afii(\fv^2,\fv^3)$
\\[9pt]

%
%
4 & $\fv^1\afi(\fv^3) + \frac{1}{2}(\fv^2)^2\afi'(\fv^3)
	+ \fv^2\afii(\fv^3) + \afiii(\fv^3)$
\\[9pt] \hline

\end{tabular}
\end{center}

\caption{Casimir invariants for solvable extensions of order~$n=3$.}
\tablabel{Casimirn3}

\end{table}

\subsubsection{n=4}
\seclabel{Casn4}

As shown in \secref{neqfour}, there are nine cases to consider.  We
shall proceed out of order, to group together similar Casimir invariants.

Cases~\ref{case:n4-00}a,~\ref{case:n4-01},~\ref{case:n4-10}a,
and~\ref{case:n4-11}a are trivial extensions, and as mentioned in
\secref{Casn3} they involve only addition of $\fv^4$ dependence to
their~$n=3$ equivalents.  \caseref{n4-10}b is a direct sum of two~$n=2$
Leibniz extensions, so the Casimirs add.

\caseref{n4-10}c is the semidirect sum of the $n=2$ Leibniz extension with an
Abelian algebra defined by $\lpb(\alpha_3,\alpha_4)\com(\beta_3,\beta_4)\rpb =
(0,0)$, with action given by
\[
	\rho_{(\alpha_1,\alpha_2)}(\beta_3,\beta_4)
		= (0,\lpb\alpha_1\com\beta_3\rpb).
\]
The Casimir invariants for this extension were derived in \secref{singWnex}.

\caseref{n4-10}d has a nonsingular~$\Wn$, so the techniques of
\secref{nsingWn} can be applied directly.

Finally, \caseref{n4-11}b is the~$n=4$ Leibniz extension, the Casimir
invariants of which were derived in \secref{Casleib}. The invariants are all
summarized in \tabref{Casimirn4}.

\begin{table}
\begin{center}

\begin{tabular}{ll} \hline
\rule[0in]{0em}{1.1em}Case & Invariant \\[3pt] \hline

%
%
\rule[0in]{0em}{1.4em}1a & $\Casi(\fv^1,\fv^2,\fv^3,\fv^4)$
\\[9pt]

%
%
1b & $\fv^1\afi(\fv^4) + \fv^2\afii(\fv^4) + \fv^3\afiii(\fv^4)
	+ \afiv(\fv^4)$
\\[9pt]

%
%
2 & $\fv^1\afi(\fv^3) + \fv^2\afii(\fv^3) + \afiii(\fv^3,\fv^4)$
\\[9pt]

%
%
3a & $\fv^1\afi(\fv^2) + \afii(\fv^2,\fv^3,\fv^4)$
\\[9pt]

%
%
3b & $\fv^1\afi(\fv^2) + \fv^3\afii(\fv^4) + \afiii(\fv^2,\fv^4)$
\\[9pt]

%
%
3c & $\fv^1\afi(\fv^4) + \fv^2\fv^3\afi'(\fv^4) + \fv^3\afii(\fv^4)
	+ \afiii(\fv^2,\fv^4)$
\\[9pt]

%
%
3d & $\fv^1\afi(\fv^4) + \frac{1}{2}(\fv^2)^2\afi'(\fv^4) +
	\fv^3\afii(\fv^4) + \fv^2\afiii(\fv^4) + \afiv(\fv^4)$
\\[9pt]

%
%
4a & $\fv^1\afi(\fv^3) + \frac{1}{2}(\fv^2)^2\afi'(\fv^3) +
	\fv^2\afii(\fv^3) + \afiii(\fv^3,\fv^4)$
\\[9pt]

%
%
4b & $\fv^1\afi(\fv^4) + \fv^2\fv^3\afi'(\fv^4)
	+ \frac{1}{3!}(\fv^3)^3\afi''(\fv^4)$\\[4pt]  &
	$\mbox{} + \fv^2\afii(\fv^4) + \frac{1}{2}(\fv^3)^2\afii'(\fv^4)
	+ \fv^3\afiii(\fv^4) + \afiv(\fv^4)$
\\[9pt] \hline

\end{tabular}
\end{center}

\caption{Casimir invariants for solvable extensions of order~$n=4$.}
\tablabel{Casimirn4}

\end{table}

\subsection{Semidirect Extensions}
\seclabel{lowsemicasi}

Now that we have derived the Casimir invariants for solvable extensions, we
look at extensions involving the semidirect sum of an algebra with these
solvable extensions.  We label the new variable (the one which acts on the
solvable part) by~$\fv^0$.  In \secref{nsingWn} we showed that the Casimirs of
the solvable part were also Casimirs of the full extension.  We also concluded
that a necessary condition for obtaining a new Casimir (other than the linear
case~$\Casi(\fv^0) = \fv^0$) from the semidirect sum was that~$\det \W_{(n)}
\ne 0$.  We go through the solvable cases and determine the Casimirs associated
with the semidirect extension, if any exist.

\subsubsection{n=1}

There is only one solvable extension, so upon appending a semidirect part we
have
\[
	\W_{(0)} = \begin{pmatrix} 1 & 0 \\ 0 & 0 \end{pmatrix},\ \ \ \
	\W_{(1)} = \begin{pmatrix} 0 & 1 \\ 1 & 0 \end{pmatrix}.
\]
Since~$\det \W_{(1)}\ne 0$, we expect another Casimir.  In fact this extension
is of the semidirect Leibniz type and has the same Casimir form as the~$n=2$
solvable Leibniz (\secref{Casleib}) extension.  Thus, the new Casimir is
just~$\fv^0\afsd(\fv^1)$.

\subsubsection{n=2}

Of the two possible extensions only the Leibniz one
satisfies~$\det \W_{(2)}\ne 0$.  The Casimir is thus
\[
	\CasiSD = \fv^0\afsd(\fv^2)+\frac{1}{2}(\fv^1)^2\afsd'(\fv^2).
\]

\subsubsection{n=3}

Cases~\ref{case:n4-01} and~\ref{case:n4-11} have a nonsingular~$\W_{(3)}$.
The Casimir for \caseref{n4-01} is
\[
	\CasiSD = \fv^0\afsd(\fv^3) + \fv^1\fv^2\afsd'(\fv^3),
\]
and for \caseref{n4-11} it is of the Leibniz form
\[
	\CasiSD = \fv^0\afsd(\fv^3) + \fv^1\fv^2\afsd'(\fv^3)
		+\frac{1}{3!}(\fv^2)^3\afsd''(\fv^3).
\]

\subsubsection{n=4}

Cases \ref{case:n4-00}b, \ref{case:n4-10}d, and \ref{case:n4-11}b have a
nonsingular~$\W_{(4)}$.  The Casimirs are shown in \tabref{Casimirsemn5}.

\begin{table}
\begin{center}
\begin{tabular}{ll} \hline
\rule[0in]{0em}{1.1em}Case & Invariant \\[3pt] \hline

%
%
\rule[0in]{0em}{1.4em}1b & $\fv^0\afsd(\fv^4) + \l(\fv^1\fv^3
		+ \frac{1}{2}(\fv^2)^2\r)\afsd'(\fv^4)$
\\[9pt]

%
%
3d & $\fv^0\afsd(\fv^4) + \l(\fv^1\fv^2
		+ \frac{1}{2}(\fv^3)^2\r)\afsd'(\fv^4)
		+ \frac{1}{3!}(\fv^2)^3\afsd''(\fv^4)$
\\[9pt]

%
%
4b & $\fv^0\afsd(\fv^4) + \l(\fv^1\fv^3
		+ \frac{1}{2}(\fv^2)^2\r)\afsd'(\fv^4)
		+ \frac{1}{2}\fv^2(\fv^3)^2\afsd''(\fv^4)
		+ \frac{1}{4!}(\fv^3)^4\afsd'''(\fv^4)$
\\[9pt] \hline

\end{tabular}
\end{center}
\caption{Casimir invariants for semidirect extensions of order~$n=5$. These
extensions also possess the corresponding Casimir invariants in
\tabref{Casimirn4}.}
\tablabel{Casimirsemn5}

\end{table}

\chapter{Stability}
\chlabel{stability}

In this chapter we discuss the general problem of stability of steady
solutions of Lie--Poisson systems, for different classes of Hamiltonians.  We
first define, in \secref{genstab}, what we mean by a steady solution being
stable.  We review the different types of stability and discuss how they are
related.  In~\secref{energycasimir} we discuss the energy-Casimir method for
finding sufficient conditions for stability, and demonstrate its use by a few
examples.  The energy-Casimir method for fluids uses an infinite-dimensional
analogue of Lagrange multipliers to find constrained extrema of the
Hamiltonian (extrema of the free energy).

In \secref{dynaccess} we turn to a different method of establishing stability,
that of dynamical accessibility.  The technique involves restricting the
variations of the energy to lie on the symplectic leaves of the system.  It is
more general that the energy-Casimir method since it yields all equilibria of
the equations of motion.  The dynamical accessibility method is closely
related to the energy-Casimir method, which we will see is reflected in the
fact that the concept of coextension of \chref{casinv} is used in the
solution.

For the different types of extensions, we derive as general a result as
possible, and then specialize to particular forms of the bracket and
Hamiltonian, until usable stability conditions are obtained.  We will treat
CRMHD in detail, using both the energy-Casimir and dynamical accessibility
methods.

\section{The Many Faces of Stability}
\seclabel{genstab}

\index{stability|(}
A somewhat universally accepted definition of stability is as follows:
Let~$\fv_{\mathrm{e}}$ be an equilibrium solution of the (not necessarily
Hamiltonian) system
\begin{equation}
	\dotfv = \rhs(\fv),
	\eqlabel{thesystem}
\end{equation}
i.e.,~\hbox{$\rhs(\fv_{\mathrm{e}}) = 0$}.  The system is to said to be
\emph{nonlinearly stable}, or simply \emph{stable}, if for every
neighborhood~$U$ of~$\fv_{\mathrm{e}}$ there is a neighborhood~$V$
of~$\fv_{\mathrm{e}}$ such that trajectories~$\fv(t)$ initially in~$V$ never
leave~$U$ (in finite time).

In terms of a norm~$\l\|\cdot\r\|$, this definition is equivalent to demanding
that for every~\hbox{$\epsilon>0$}, there is a~\hbox{$\delta>0$} such that
if~\hbox{$\l\|\fv(0)-\fveq\r\| < \delta$},
then~\hbox{$\l\|\fv(t)-\fveq\r\| < \epsilon$} for all~\hbox{$t>0$}.

We also consider the linearized system,
\begin{equation}
	\fd\dotfv = {\l.
		\lang\fd\fv\com\frac{\fd\rhs}{\fd\fv}\rang\r|
		}_{\fv=\fveq},
	\eqlabel{thelinsystem}
\end{equation}
where~$\fd\fv$ is an infinitesimal perturbation.  From this we define the
formally self-adjoint linear operator~$\fsaop$ by
\begin{equation}
	\lang\fd\eta\com\fsaop\,\fd\zeta\rang \ldef
		{\l.\lang\fd\eta\com\frac{\fd^2\rhs}{\fd\eta\fd\zeta}
			\,\fd\zeta\rang\r|}_{\fv=\fveq}.
	\eqlabel{fsaop}
\end{equation}

From this definition we distinguish four basic types of stability:

\begin{itemize}
\item \emph{Spectral stability}.  The linearized system~\eqref{thelinsystem} is
spectrally stable if the spectrum of the linear operator~$\fsaop$ defined
by~\eqref{fsaop} has no eigenvalue with a positive real part.  A special case
is \emph{neutral stability}, for which the spectrum is purely imaginary.
Hamiltonian systems are neutrally stable if they are spectrally stable.
\item \emph{Linear stability}.  If the linearized system~\eqref{thelinsystem}
is stable according to the above definition, then the system~\eqref{thesystem}
is said to be linearly stable (or linearized stable).  This implies spectral
stability.
\item \emph{Formal stability} (Dirichlet criterion).  The equilibrium is
formally stable if we can find a conserved quantity whose first variation
vanishes when evaluated at the equilibrium, and whose second variation is
positive (or negative) definite when evaluated at the same equilibrium.  In
finite dimensions, this implies nonlinear stability.  When the system is
Hamiltonian and separable (i.e., it can be written as a sum of kinetic and
potential energy), this criterion becomes Lagrange's theorem.
\item \emph{Nonlinear stability}.  This is just the nonlinear stability of
the full system as defined above.  Note that this only implies that there
exists a sufficiently small neighborhood~$V$ such that trajectories never
leave~$U$.  It does not imply absence of \emph{finite-amplitude instability},
called nonlinear instability by some authors, which says that the system is
unstable for large enough perturbations.
\end{itemize}

\figref{stabrel} summarizes the relationships between the various
types of stability.  See Siegel and Moser~\cite{Siegel}, Holm et
al.~\cite{Holm1985}, or Morrison~\cite{Morrison1998} for examples and
counterexamples of these relationships.\index{stability|)}
\begin{figure}
\[
\SelectTips{cm}{12}\xymatrix@M=8pt{
\framebox[1.75in][c]{\rule[-0.75em]{0em}{2em}Spectral Stability}
	\mathbf{\ar@2{<-}[dd]} & &
\framebox[1.75in][c]{\rule[-0.75em]{0em}{2em}Linearized Stability}
	\ar@2{->}[ll] \ar@2{<-}[dd] \\ \\
\framebox[1.75in][c]{\rule[-0.75em]{0em}{2em}Nonlinear Stability}
	\ar@2{<-}[rr]_{\txt{finite-\\ dim}} & &
\framebox[1.75in][c]{\rule[-0.75em]{0em}{2em}Formal Stability} \\
}
\]
\caption{Relationship between the different types of stability.}
\figlabel{stabrel}
\end{figure}
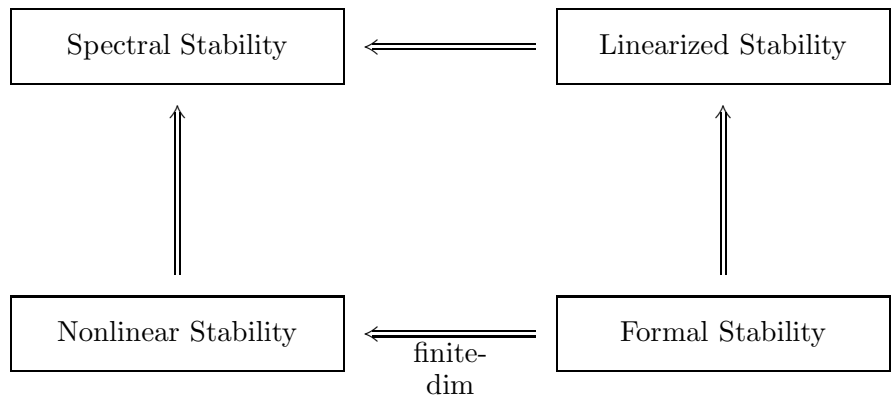

We have stated that formal stability implies nonlinear stability for
finite-dimensional systems.  Before discussing this point, we prove a
stability theorem for finite-dimensional systems that has its origins with
Lagrange.  It was proved in a less general form than presented here by
Dirichlet~\cite{Dirichlet}, and was subsequently generalized by
Liapunov.\index{Dirichlet's theorem}

The theorem is as follows.  If the system~\eqref{thesystem}, in finite
dimensions, has a constant of the motion~$\constmot$ that has a relative
extremum in the strong sense~\cite[p.~13]{Gelfand} at the equilibrium
point~\hbox{$\fv=\fveq$}, then the equilibrium solution is stable.%
\footnote{In finite dimensions a strong minimum is just a minimum with respect
to the usual Euclidean norm,~\hbox{$\l\|\fv\r\|=\l|\fv\r|$}.}

We follow the proof of Siegel and Moser~\cite[p.~208]{Siegel}.  See also
Hirsch and Smale for a thorough treatment~\cite{Hirsch}.  Since~$\constmot$
can be replaced by~$-\constmot$, we can assume it has a minimum without loss
of generality.  By the strong minimum hypothesis, there exists
a~\hbox{$\varrho > 0$} such that
\begin{equation}
	\constmot(\fveq) < \constmot(\fv) \text{\ whenever}
	\l\|\fv - \fveq\r\| < \varrho,
	\eqlabel{strongmin}
\end{equation}
for some norm~$\l\|\cdot\r\|$.  Now, let
\[
	{\mathfrak{M}}_\varepsilon \ldef
		\l\{ \fv \mid \l\|\fv - \fveq\r\| < \varepsilon \r\},
	\quad 0 < \varepsilon < \varrho,
\]
be a ball of radius~$\varepsilon$ around the equilibrium point.
Let~$\mu(\varepsilon)$ be the minimum value of~$\constmot$ on the surface of
the ball~${\mathfrak{M}}_\varepsilon$,
\[
	\mu(\varepsilon) \ldef \min_{\l\|\fv - \fveq\r\| = \varepsilon}
		\constmot(\fv).
\]
Using the strong minimum hypothesis,~\eqref{strongmin}, we have
\[
	\constmot(\fv) < \mu(\varepsilon),\ \
		\text{for}\ \fv \in {\mathfrak{M}}_\varepsilon.
\]
Now consider a trajectory with initial conditions~$\fv(0)$
in~${\mathfrak{M}}_\varepsilon$.  Then
\[
	\constmot(\fv(t)) = \constmot(\fv(0)) < \mu(\varepsilon).
\]
But by continuity this implies~\hbox{$\fv(t) \in {\mathfrak{M}}_\varepsilon$}
since otherwise we would have had \hbox{$\constmot(\fv(t)) \ge
\mu(\varepsilon)$} at some point in the trajectory.  Thus,~$\fv(t)$ lies
in~${\mathfrak{M}}_\varepsilon$ whenever~$\fv(0)$ does.  We then have
stability, because~${\mathfrak{M}}_\varepsilon$ is a neighborhood of~$\fveq$
and we can make~$\varepsilon$ as small as we want.

\index{stability!formal vs nonlinear}
In finite dimensions, positive or negative definiteness of the second
variation of~$\constmot$ is sufficient for the strong minimum
requirement~\eqref{strongmin}.  In infinite dimensions this is not the
case~\cite{Arnold1969a,Ball1984,Finn1987,Gelfand,Holm1985,McIntyre1987,%
Shepherd1992}.  Further convexity arguments must be made, as done for several
physical systems in Holm~\etal~\cite{Holm1985}.  Another crucial requirement,
which is immediate in finite dimensions, is that the invariant~$\constmot$ be
\emph{continuous} in the norm~$\l\|\cdot\r\|$.  In general an
infinite-dimensional minimum will not necessarily satisfy this
condition~\cite{Gelfand,Holm1985}.

Ball and Marsden~\cite{Ball1984} give an example from elasticity theory of a
system that is formally stable but is nonlinearly unstable.  Finn and
Sun~\cite{Finn1987} discuss additional requirements for nonlinear stability of
an ideal fluid in a gravitational field (for an exponential atmosphere), which
is formally stable.  One does not know how stringent these requirements
are---they could be far from the actual instability threshold.  We take the
viewpoint here that establishing definiteness of the second
variation---showing formal stability---is a good indicator of stability.
Indeed, formal stability is often used to mean stability, as is the case
with~\hbox{$\delta W$}\index{stability!delta W@$\delta W$ criterion} stability
criteria in MHD, which are actually second-order variations of the potential
energy.  For the Grad--Shafranov equilibria of reduced MHD (no flow), the
sufficient conditions for formal stability are the same as for nonlinear
stability~\cite[pp.~41--43]{Holm1985}.

It will be the topic of future work to try and make these general stability
conditions more rigorous by making more stringent convexity arguments.
Certainly formal stability implies linearized stability, since the second
variation of the constant of motion provides a norm (conserved by the
linearized dynamics) that can be used to establish stability of the linearized
system.

Finally, note that Dirichlet's theorem does \emph{not} imply that if~$\FrE$
does not have an extremum at~$\fveq$, then the system is unstable. It
gives a sufficient, but not necessary, condition for stability of an
equilibrium.

\section{The Energy-Casimir Method}
\seclabel{energycasimir}

\index{energy-Casimir method}
The energy-Casimir method has a long history which dates back to
Fjortoft~\cite{Fjortoft1950}, Newcomb~\cite{Newcomb1958}, Kruskal and
Oberman~\cite{Kruskal1958}, Fowler~\cite{Fowler1963}, and
Gardner~\cite{Gardner1963}, but is usually called ``Arnold's method'' or
``Arnold's theorem''~\cite{Arnold1965a,Arnold1965b,Arnold1966a,%
Arnold1969b,Arnold1969a}.\index{Arnold's theorem} We illustrate the method for
a Lie--Poisson system.  The equations of motion for the field variables~$\fv$
in terms of a given Hamiltonian~$\Ham$ are
\begin{equation}
	\dotfv = -\lpb \frac{\fd \Ham}{\fd \fv}\com\,\fv\rpb^\dagger.
	\tag{\ref{eq:motion}}
\end{equation}
This can be rewritten
\[
	\dotfv = -\lpb \frac{\fd\Ham}{\fd \fv}
		+ \frac{\fd\Cas}{\fd \fv}
		\com\,\fv\rpb^\dagger,
\]
where~$C$ is any function of the Casimirs.
It follows that if
\[
	\fd(\Ham + \Cas)[\fveq] \rdef \fd\FrE[\fveq] = 0,
\]
then~$\fveq$ is an equilibrium of the system.  We call~$\FrE$ the free
energy\index{free energy}.  The free energy~$\FrE$ is a constant of the motion
whose first variation vanishes at an equilibrium point.  Therefore, if we can
show it also has a strong extremum at that point then we have proved
stability, by the theorem of Dirichlet.  Showing that~$\fd^2\FrE$ is definite
(that is, showing formal stability) is almost sufficient to show stability, in
the sense discussed at the end of \secref{genstab}.

We now apply the energy-Casimir method to compressible reduced MHD.  We will
give more examples in \secref{dynaccess} when we introduce the method of
dynamical accessibility, which is more general and includes the energy-Casimir
result as a special case.



\subsection{CRMHD Stability}

\index{compressible reduced MHD!stability|(}
The free energy functional~$\FrE$ is built from the
Hamiltonian~\eqref{CRMHDHam} and the Casimir invariants found in
\secref{CRMHDCas},
\[
	\FrE[\vort,\pvel,\pres,\magf] \ldef \Ham + \Cas,
\]
where
\[
	\Cas = \lang\afi(\magf) + \pvel\,\afii(\magf)
		+ \pres\,\afiii(\magf) + \l(\vort\,\afiv(\magf)
		- \binv\,\pres\,\pvel\,\afiv'(\magf)\r)\rang
\]
is a combination of the Casimirs of the system.  We use the same angle
brackets as for the pairing, without the comma, to denote an integral over the
fluid domain (we assume that we have identified~$\LieA$ and~$\LieA^*$).

\subsubsection{Equilibrium Solutions}

We seek equilibria of the system that extremize the free energy.  The first
variation of~$\FrE$ yields
\begin{multline*}
	\fd\FrE = \biggl\langle
	\l\lgroup-\elecp + \afiv(\magf)\r\rgroup\fd\vort
	+ \l\lgroup\pvel + \afii(\magf)
		- \binv\,\pres\,\afiv'(\magf)\r\rgroup
		\fd\pvel\\
	+ \l\lgroup\binv(\pres-2\crmhdbeta\,x)
		+ \afiii(\magf)
		- \binv\,\pvel\,\afiv'(\magf)\r\rgroup\fd\pres
		\\
	\phantom{\fd\FrE = \biggl\langle}
		+ \l\lgroup-\ecurrent + \afi'(\magf) + \pvel\,\afii'(\magf)
		+ \pres\,\afiii'(\magf)
		+ \l(\vort\,\afiv'(\magf)
		- \binv\,\pres\,\pvel\,\afiv''(\magf)\r)
	\r\rgroup\fd\magf
	\biggr\rangle.
\end{multline*}
An equilibrium solution~$(\vorteq,\pveleq,\preseq,\magfeq)$ for
which~\hbox{$\fd\FrE=0$} must therefore satisfy
\begin{align}
	\elecpeq &= \phib(\magfeq),\eqlabel{equirel1} \\
	\pveleq &=  \binv\,\preseq\,\phib'(\magfeq) - \afii(\magfeq),
		\eqlabel{equirel2} \\
	\preseq &= \pveleq\,\phib'(\magfeq)
		+ \crmhdbeta(2x - \afiii(\magfeq)),
		\eqlabel{equirel3} \\
	\ecurrenteq &= \afi'(\magfeq) + \pveleq\,\afii'(\magfeq)
		+ \preseq\,\afiii'(\magfeq) + \vorteq\,\phib'(\magfeq)
		- \binv\,\preseq\,\pveleq\,\phib''(\magfeq),
		\eqlabel{equirel4}
\end{align}
where we have defined~$\phib(\magf)\ldef k(\magf)$. 

Since~\hbox{$\elecpeq = \phib(\magfeq)$}, we have~\hbox{$\grad\elecpeq = 
\phib'(\magfeq)\,\grad\magfeq$}. Hence,~$\velperpeq =
\phib'(\magfeq)\,\Bperpeq$, so the perpendicular (poloidal) 
velocity and magnetic field are collinear at an equilibrium.

We can use~\eqref{equirel2} and~\eqref{equirel3} to solve
for~$\pveleq$ and~$\preseq$,
\begin{equation}
	\l(\begin{array}{c}
		\pveleq \\[6pt] \preseq
	\end{array}\r) =
	\l(\frac{|\phib'(\magfeq)|^2}{\crmhdbeta} - 1\r)^{-1}
	\l(\begin{array}{c}
		\afii(\magfeq) + (\afiii(\magfeq) - 2x)\,\phib'(\magfeq)
			\\[6pt]
		\afii(\magfeq)\,\phib'(\magfeq)
			+ \crmhdbeta\,(\afiii(\magfeq) - 2x)
	\end{array}\r),
	\eqlabel{pvesol}
\end{equation}
except where~${|\phib'(\magfeq)|^2} = {\crmhdbeta}$.  This singularity
represents a resonance\index{resonance} in the system, about which we will say
more later.  Equation~\eqref{pvesol} implies
\begin{align*}
	\l(\grad\pveleq - 2\phib'(\magfeq)
		\l(1 - \binv{|\phib'(\magfeq)|^2}\r)^{-1} \unvx\r)
		\times \grad\magfeq &= 0,\\
	\l(\grad\preseq - 2\crmhdbeta
		\l(1 - \binv\,{|\phib'(\magfeq)|^2}\r)^{-1} \unvx\r)
		\times \grad\magfeq &= 0.
\end{align*}

An important class of equilibria are given by
\[
	\phib(\magfeq) = \alfvenc^{-1}\,\magfeq(x,y),
\]
where~$\alfvenc$ is a constant.  We call those \emph{\Alfvenic}\
solutions. (The true \Alfven\ \index{alfven solutions@\Alfven\
solutions}solutions are the particular case with~$c=\pm1$.)  We then
have~\hbox{$\vorteq\,\phib'(\magfeq) =
\ecurrenteq/\alfvenc^2$}, and so from~\eqref{equirel4}
\begin{equation}
	\l(1-\frac{1}{\alfvenc^2}\r)\ecurrenteq = \afi'(\magfeq)
		+ \pveleq\,\afii'(\magfeq)
		+ \preseq\,\afiii'(\magfeq).
	\eqlabel{alfvenicJ}
\end{equation}
Note that, because of~\eqref{pvesol}, the right-hand side of~\eqref{alfvenicJ}
depends explicitly on~$x$, unless we have
\begin{equation}
	\afii'(\magfeq) = -\crmhdbeta\,\alfvenc\,\afiii'(\magfeq),
	\eqlabel{xindepcond}
\end{equation}
in which case~\eqref{alfvenicJ} simplifies to
\begin{equation}
	\l(1-\frac{1}{\alfvenc^2}\r)\ecurrenteq(\magfeq) = \afi'(\magfeq)
		- \afii(\magfeq)\,\afii'(\magfeq).
	\eqlabel{alfvenicJnox}
\end{equation}
Such an equation, with no explicit independence on~$x$, has an analogue in
low-beta reduced MHD, but cannot occur for a system like high-beta reduced
MHD~\cite[p.~59]{Hazeltine1985c} without a vanishing pressure gradient. Here,
with CRMHD, we can eliminate the~$x$ dependence because we can set up an
equilibrium gradient in the parallel velocity which cancels the pressure
gradient.

If in~\eqref{alfvenicJ} we let
\[
	\afi'(\magfeq) - \afii(\magfeq)\,\afii'(\magfeq)
	= \l(1-\frac{1}{\alfvenc^2}\r)\exp(-2\magfeq),
\]
then we have the particular solution
\begin{equation}
	\magfeq(x,y) = \ln(a\cosh y + \sqrt{a^2 - 1}\, \cos x).
	\eqlabel{catseye}
\end{equation}
This solution, the Kelvin--Stuart cat's eye\index{Kelvin--Stuart cat's eye}
formula~\cite{Bondeson1983,Finn1977,Pritchett1979}, is plotted in
\figref{catseye}.
\begin{figure}
\centerline{\psfig{file=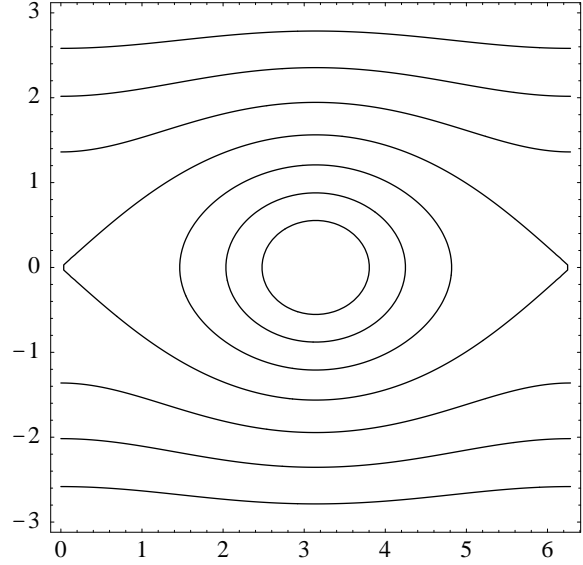,width=3in}}
\caption{Contour plot of the magnetic flux~$\magfeq(x,y)$ for the cat's eye
solution~\eqref{catseye}, with~\hbox{$a=1.5$}.}
\figlabel{catseye}
\end{figure}

\subsubsection{Formal Stability}

The second variation of~$\FrE$ is given by
\begin{multline*}
	\fd^2\FrE = \biggl\langle
		-\fd\vort\,(\lapl)^{-1}\fd\vort + |\fd\pvel|^2
		+ \frac{1}{\crmhdbeta}
		|\fd\pres|^2 - \fd\magf\,(\lapl)^{-1}\fd\magf
		+ 2\afiv'(\magf)\,\fd\vort\,\fd\magf\\
	\phantom{\fd^2\FrE = \biggl\langle}
		+ \l\lgroup\afi''(\magf) + \pvel\,\afii''(\magf)
		+ \pres\,\afiii''(\magf)
		+ \vort\,\afiv''(\magf) - \binv\,\pres\,\pvel\,\afiv'''(\magf)
		\r\rgroup|\fd\magf|^2\\
	\phantom{\fd^2\FrE = \biggl\langle}
		+ 2\l\lgroup\afii'(\magf) 
		- \binv\,\pres\,\afiv''(\magf)\r\rgroup
		\fd\magf\,\fd\pvel
		+ 2\l\lgroup\afiii'(\magf) 
		- \binv\,\pvel\,\afiv''(\magf)\r\rgroup
		\fd\magf\,\fd\pres\\
	- 2\binv\,\afiv'(\magf)\,\fd\pvel\,\fd\pres
	\biggr\rangle.
\end{multline*}
We want to determine when this is
non-negative. Using~\hbox{$\fd\vort=\lapl\fd\elecp$}, we can write
\begin{eqnarray*}
	&&\lang|\grad\fd\elecp|^2 + |\grad\fd\magf|^2
		+ 2\afiv'(\magf)\,(\lapl\fd\elecp)\,\fd\magf\rang
		\\
	&&\hspace{1cm}\mbox{} = \lang|\grad\fd\elecp|^2 + |\grad\fd\magf|^2
		- 2\grad(\afiv'(\magf)\,\fd\magf)\cdot\grad\fd\elecp\rang
		\\
	&&\hspace{1cm}\mbox{} = \lang|\grad\fd\elecp
			- \grad(\afiv'(\magf)\,\fd\magf)|^2
		- |\grad(\afiv'(\magf)\,\fd\magf)|^2
		+ |\grad(\fd\magf)|^2\rang,
\end{eqnarray*}
which, after expanding the~$|\grad(\afiv'(\magf)\,\fd\magf)|^2$ term, becomes
\begin{multline}
	\lang|\grad\fd\elecp|^2 + |\grad\fd\magf|^2
		+ 2\afiv'(\magf)\,(\lapl\fd\elecp)\,\fd\magf\rang
		= \\ \bigl\langle|\grad\fd\elecp
			- \grad(\afiv'(\magf)\,\fd\magf)|^2
	+ (1 - |\afiv'(\magf)|^2)|\grad\fd\magf|^2
		\\ + \afiv'(\magf)\,\lapl\afiv'(\magf)\,|\fd\magf|^2
		\bigr\rangle,
	\eqlabel{coolident}
\end{multline}
so that the second variation, evaluated at the
equilibrium solution~\eqref{equirel1}--\eqref{equirel4}, is now
\begin{multline}
	\fd^2\FrE_\equil = \biggl\langle
		|\grad\fd\elecp - \grad(\phib'(\magfeq)\,\fd\magf)|^2
		+ (1 - |\phib'(\magfeq)|^2)|\grad\fd\magf|^2
		+ |\fd\pvel|^2 + \frac{1}{\crmhdbeta} |\fd\pres|^2
		\\
	+ 2\l\lgroup\afii'(\magfeq) 
		- \binv\,\preseq\,\phib''(\magfeq)\r\rgroup
		\fd\magf\,\fd\pvel
	+ 2\l\lgroup\afiii'(\magfeq) 
		- \binv\,\pveleq\,\phib''(\magfeq)\r\rgroup
		\fd\magf\,\fd\pres\\
	+ \magfvarii(x,y)\,|\fd\magf|^2
		 - 2\binv\,\phib'(\magfeq)\,\fd\pvel\,\fd\pres
	\biggr\rangle,
	\eqlabel{CRMHDd2F}
\end{multline}
where
\begin{multline*}
	\magfvarii(x,y) \ldef \afi''(\magfeq) + \pveleq\,\afii''(\magfeq)
		+ \preseq\,\afiii''(\magfeq)\\
	+ \vorteq\,\phib''(\magfeq)
		- \binv\,\preseq\,\pveleq\,\phib'''(\magfeq)
		+ \phib'(\magfeq)\,\lapl\phib'(\magfeq).
\end{multline*}
For positive-definiteness of~\eqref{CRMHDd2F}, we require
\begin{equation}
	|\phib'(\magfeq)| \le 1.
	\eqlabel{phireqi}
\end{equation}
If we have equality in~\eqref{phireqi}, then we obtain a \emph{family} of
marginally stable equilibria, the \Alfven\ solutions.

Assuming~\eqref{phireqi} is satisfied, a sufficient condition for stability is
to show that the~$(\fd\pvel,\fd\pres,\fd\magf)$ part of the second variation
is non-negative. We thus demand the quadratic form represented by the
symmetric matrix
\[
	\l(\begin{array}{ccc}
	1 & -\binv\,\phib'(\magfeq) &
		\afii'(\magfeq) - \binv\,\preseq\,\phib''(\magfeq) \\[8pt]
	-\binv\,\phib'(\magfeq) & \binv &
		\afiii'(\magfeq) - \binv\,\pveleq\,\phib''(\magfeq) \\[8pt]
	\afii'(\magfeq) - \binv\,\preseq\,\phib''(\magfeq) &
		\afiii'(\magfeq) - \binv\,\pveleq\,\phib''(\magfeq) &
		\magfvarii(x,y)
	\end{array}\r)
\]
be non-negative. A necessary and sufficient condition for this is that the
\emph{principal minors} of the matrix be non-negative.  The principal minors
are simply the determinants of the submatrices of increasing size along the
diagonal.  Thus, the first two principal minors are
\begin{align*}
	\pminor_1 &= |1| > 0,\\
	\pminor_2 &= \l|\begin{array}{cc}
		1 & -\binv\,\phib'(\magfeq) \\
		-\binv\,\phib'(\magfeq) & \binv \\
	\end{array}\r| = \binv\l(1
		- \frac{|\phib'(\magfeq)|^2}{\crmhdbeta}\r) \ge 0,
\end{align*}
and the third is just the determinant of the matrix,
\begin{multline*}
	\pminor_3 = \pminor_2\l(\magfvarii(x,y)
		- \l[\afii'(\magfeq) - \binv\,\preseq\,\phib''(\magfeq)
			\r]^2\r)\\
	- \l[\afiii'(\magfeq)
		+ \binv\,\afii'(\magfeq)\,\phib'(\magfeq)
		- \binv (\pveleq
		+ \binv\,\preseq\,\phib'(\magfeq))\r]^2 \ge 0.
\end{multline*}
Combining~\eqref{phireqi} with the requirement~\hbox{$\pminor_2\ge 0$}, we
have
\begin{equation}
	|\phib'(\magfeq)|^2 \le \min(1,\crmhdbeta).
	\eqlabel{phireq}
\end{equation}
According to this condition, for~\hbox{$\crmhdbeta < 1$} CRMHD is \emph{less}
stable than the RMHD case. This is a direct manifestation of the nontrivial
cocycle in the bracket: there is a new resonance,\index{resonance} associated
with the \emph{acoustic} resonance, so-named because at that point the flow
velocity equals the ion-acoustic speed (proportional
to~$2\,\Telec$).\index{acoustic resonance} We will see in \secref{VC} that new
resonances are a generic feature of Lie--Poisson systems with cocycles.

The condition that~$\pminor_3$ be non-negative is of a
more complicated form. For the \Alfvenic\ case, with~\hbox{$\phib(\magfeq) =
\alfvenc^{-1}\,\magfeq(x,y)$}, and assuming condition~\eqref{xindepcond}, so
that~$\ecurrenteq=\ecurrenteq(\magfeq)$, the condition~\hbox{$\pminor_3 \ge
0$} simplifies to
\[
	\pminor_2\l(1 - \frac{1}{\alfvenc^2}\r)\,\ecurrenteq'(\magfeq) \ge 0.
\]
Since~\hbox{$\pminor_2 \ge 0$} and, by~\eqref{phireq}, \hbox{$1/\alfvenc^2 \le
\min(1,\crmhdbeta)$}, we can simply write
\begin{equation}
	\ecurrenteq'(\magfeq) \ge 0.
	\eqlabel{alfvenicstab}
\end{equation}
Hence, for~\hbox{$\crmhdbeta \ge 1$}, \Alfvenic\ solutions have the same
stability characteristics as for RMHD.
\index{compressible reduced MHD!stability|)}

\section{Dynamical Accessibility}
\seclabel{dynaccess}

\index{dynamical accessibility|(}
We turn now to a different method of finding equilibria and ascertaining their
stability. Finding the solutions for which the first variation of the free
energy vanishes yields some, but not all of the equilibria of the equations of
motion. For example, this method fails to detect the static equilibrium of the
heavy top~\cite{Morrison1998}. For the 2-D Euler system, the equilibria it
yields are those for which the streamfunction is a monotonic function of the
vorticity, but there are equilibria which do not have this form.  This is tied
to the rank-changing of the cosymplectic form: there are equilibria that
arise because the bracket itself vanishes~\cite{Morrison1998}.  The method of
dynamical accessibility was used by Morrison and Pfirsch to examine the
stability of the Vlasov--Maxwell
system~\cite{Morrison1989,Morrison1990}. Isichenko~\cite{Isichenko1998} made
use of a similar method to study hydrodynamic stability, based on ideas of
Arnold~\cite{Arnold1966b}.

We first explain the method of dynamically accessible variations, and then
apply it to extensions.  We derive general results for pure semidirect
extensions and extensions with a nonsingular~$\Wn$.  For both cases, we
examine several different types of Hamiltonians.

\subsection{The Method}
\seclabel{dynacmethod}

Consider a perturbation defined as
\begin{equation}
	\fd\fv_\dynac \ldef \lPB\dynacgen\com\fv\rPB,
	\eqlabel{dynacpert}
\end{equation}
with the perturbation given in terms of the generating function~$\dynacg$ by
\[
	\dynacgen \ldef \lang\fv\com\dynacg\rang.
\]
The~$\dynacg$ are arbitrary ``constant'' functions (i.e., they do not depend
on~$\fv$, but do depend on~$\xv$). We call~\eqref{dynacpert} a
\emph{dynamically accessible} perturbation. The first-order variation of the
Casimir invariant of the bracket is given by
\begin{equation}
	\fd\Cas_\dynac = \lang\fd\fv_\dynac\com\frac{\fd \Cas}{\fd\fv}\rang
		= \lang\lPB\dynacgen\com\fv\rPB
		\com\frac{\fd\Cas}{\fd\fv}\rang.
	\eqlabel{fdCda}
\end{equation}
If we now assume that the bracket~$\lPB\com\rPB$ is of the Lie--Poisson type
(Eq.~\eqref{LPB}), we have
\[
	\fd\Cas_\dynac
		= \lang{\lpb\dynacg
		\com\fv\rpb}^\dagger\com\frac{\fd \Cas}{\fd\fv}\rang
		= \lang\fv\com\lpb\dynacg
		\com\frac{\fd \Cas}{\fd\fv}\rpb\rang
		= \lPB\dynacgen\com\Cas\rPB = 0.
\]
Hence, to first order, \emph{Casimirs are unchanged by a dynamically
accessible perturbation}. The first-order variation of the Hamiltonian is
\[
	\fd\Ham_\dynac = \fd\FrE_\dynac = \lang{\lpb\dynacg
		\com\fv\rpb}^\dagger\com\frac{\fd \Ham}{\fd\fv}\rang
	= -\lang{\lpb\frac{\fd \Ham}{\fd\fv}
		\com\fv\rpb}^\dagger\com\dynacg\rang.
\]
The variation of the Hamiltonian and of the free energy are the same because
they differ only by Casimirs. If we look for equilibrium solutions by
requiring that~$\fd\Ham_\dynac=0$ for all~$\dynacg$, we obtain
\[
	{\lpb\frac{\fd \Ham}{\fd\fv}(\fveq) \com\fveq\rpb}^\dagger = 0,
	\eqlabel{eomeq}
\]
which is equivalent to looking for steady solutions of the equation of
motion~\eqref{motion}.

Because we want to establish formal stability, we have to take second-order
dynamically accessible variations that preserve the Casimirs. If we denote the
second-order part of the dynamically accessible variation
by~$\fd^2\fv_\dynac$, and the first and second order generating functions
by~$\dynacg^{(1)}$ and~$\dynacg^{(2)}$, we have
\[
\begin{split}
	\fd^2\Cas_\dynac &= \half\lang\fd\fv_\dynac\com
		\frac{\fd^2\Cas}{\fd\fv\,\fd\fv}\,\fd\fv_\dynac\rang
		+ \lang\fd^2\fv_\dynac\com\frac{\fd\Cas}{\fd\fv}\rang
		\\
	&= \half\lang\lPB\dynacgen^{(1)}\com\fv\rPB\com
		\frac{\fd^2\Cas}{\fd\fv\,\fd\fv}\,
		\lPB\dynacgen^{(1)}\com\fv\rPB\rang
		+ \lang\fd^2\fv_\dynac\com\frac{\fd\Cas}{\fd\fv}\rang
		\\
	&= \half\lang\lPB\dynacgen^{(1)}\com\fv\rPB\com
		\frac{\fd}{\fd\fv}
		\lang\lPB\dynacgen^{(1)}\com\fv\rPB\!\com
			\frac{\fd\Cas}{\fd\fv}\rang
		- \lpb\dynacg^{(1)}\!\com\frac{\fd\Cas}{\fd\fv}\rpb\rang
		+ \lang\fd^2\fv_\dynac\com\frac{\fd\Cas}{\fd\fv}\rang
		\\
	&= -\half\lang{\lpb\dynacg^{(1)}\com
			\lPB\dynacgen^{(1)}\com\fv\rPB\rpb}^\dagger\com
		\frac{\fd\Cas}{\fd\fv}\rang
		+ \lang\fd^2\fv_\dynac\com\frac{\fd\Cas}{\fd\fv}\rang
		\\
	&= \lang\fd^2\fv_\dynac - \half{\lPB\dynacgen^{(1)}\com
			{\lPB\dynacgen^{(1)}\com\fv\rPB}
			\rPB}\com
		\frac{\fd\Cas}{\fd\fv}\rang.
\end{split}
\]
We made use of the fact that~\eqref{fdCda} vanishes identically.  In order
for~\hbox{$\fd^2\Cas_\dynac$} to be zero, we can set
\begin{equation}
\begin{split}
	\fd^2\fv_\dynac &= \lPB\dynacgen^{(2)}\com\fv\rPB
		+ \half{\lPB\dynacgen^{(1)}\com
		{\lPB\dynacgen^{(1)}\com\fv\rPB}\rPB}\\
	&= {\lpb\dynacg^{(2)}\com\fv\rpb}^\dagger
		+ \half{\lpb\dynacg^{(1)}\com
		{\lpb\dynacg^{(1)}\com\fv\rpb}^\dagger
		\rpb}^\dagger.
	\eqlabel{d2fvda}
\end{split}
\end{equation}
The second-order dynamically accessible variation of~$\Ham$ is
\begin{align*}
	\fd^2\Ham_\dynac &= \half\lang\fd\fv_\dynac\com
		\frac{\fd^2\Ham}{\fd\fv\,\fd\fv}\,\fd\fv_\dynac\rang
		+ \lang\fd^2\fv_\dynac\com\frac{\fd\Ham}{\fd\fv}\rang
		\\
	&= \half\lang\lPB\dynacgen^{(1)}\com\fv\rPB\com
		\frac{\fd^2\Ham}{\fd\fv\,\fd\fv}
		\lPB\dynacgen^{(1)}\com\fv\rPB\rang\\
	&\phantom{=} + \lang\lPB\dynacgen^{(2)}\com\fv\rPB
		+ \half{\lPB\dynacgen^{(1)}\com
		{\lPB\dynacgen^{(1)}\com\fv\rPB}\rPB}
		\com\frac{\fd\Ham}{\fd\fv}\rang,
\end{align*}
which upon using~\eqref{d2fvda} becomes
\begin{align*}
	\fd^2\Ham_\dynac &= \half\lang{\lpb\dynacg^{(1)}\com\fv\rpb}^\dagger
		\com\frac{\fd^2\Ham}{\fd\fv\,\fd\fv}\,
		{\lpb\dynacg^{(1)}\com\fv\rpb}^\dagger\rang\\
	&\phantom{=} + \lang{\lpb\dynacg^{(2)}\com\fv\rpb}^\dagger
		+ \half{{\lpb\dynacg^{(1)}\com
		{{\lpb\dynacg^{(1)}\com\fv\rpb}^\dagger}\rpb}^\dagger}
		\com\frac{\fd\Ham}{\fd\fv}\rang
\end{align*}
The piece involving~$\dynacg^{(2)}$ can be written as
\[
	\lang{\lpb\dynacg^{(2)}\com\fv\rpb}^\dagger
		\com\frac{\fd\Ham}{\fd\fv}\rang
	= -\lang{\lpb\frac{\fd\Ham}{\fd\fv}\com\fv\rpb}^\dagger
		\com\dynacg^{(2)}\rang,
\]
which vanishes when evaluated at an equilibrium of the equations of
motion~\eqref{eomeq}. Hence, for purposes of testing stability we may neglect
the second-order generating function entirely. We therefore drop the
superscripts on~$\dynacgen$ and~$\dynacg$, and write
\begin{equation}
\begin{split}
	\fd^2\Ham_\dynac &=
		\half\lang{\lpb\dynacg\com\fv\rpb}^\dagger\com
		\frac{\fd^2\Ham}{\fd\fv\,\fd\fv}\,
		{\lpb\dynacg\com\fv\rpb}^\dagger\rang
		+ \half\lang{{\lpb\dynacg\com
		{{\lpb\dynacg\com\fv\rpb}^\dagger}\rpb}^\dagger}
		\com\frac{\fd\Ham}{\fd\fv}\rang\\
	&= \half\lang{\lpb\dynacg\com\fv\rpb}^\dagger\com
		\frac{\fd^2\Ham}{\fd\fv\,\fd\fv}\,
		{\lpb\dynacg\com\fv\rpb}^\dagger
		+ {\lpb\dynacg\com\frac{\fd\Ham}{\fd\fv}\rpb}\rang
	\eqlabel{dynactwoHam}
\end{split}
\end{equation}
To more easily determine sufficient stability conditions, we want to
write~\eqref{dynactwoHam} as a function of~$\fd\fv_\dynac$.
(Then~\eqref{dynactwoHam} will be a quadratic form in~$\fd\fv_\dynac$.)  We
now show that this is always possible.  This is a generalization of a proof by
Arnold~\cite{Arnold1966a} for 2-D Euler.

Assume that we have a dynamically accessible variation given in terms of a
second generating function~$\dynacg'$,
\[
	{\fd'}\fv_\dynac = {\lpb{\dynacg}'\com\fv\rpb}^\dagger,\qquad
	{\fd'}^2\fv_\dynac = \half{\lpb{\dynacg}'\com
		{\lpb{\dynacg}'\com\fv\rpb}^\dagger\rpb}^\dagger,
\]
such that~\hbox{$\fd\fv_\dynac = {\fd'}\fv_\dynac$}.  Then the difference in
the second order variation of the energy is
\begin{align}
	2\fd^2\Ham_\dynac - 2{\fd'}^2\Ham_\dynac &=
		\lang{\lpb\dynacg\com\fv\rpb}^\dagger\com
		\frac{\fd^2\Ham}{\fd\fv\,\fd\fv}\,
		{\lpb\dynacg\com\fv\rpb}^\dagger
		+ {\lpb\dynacg\com\frac{\fd\Ham}{\fd\fv}\rpb}\rang\nonumber\\
	& \phantom{=} - \lang{\lpb\dynacg'\com\fv\rpb}^\dagger\com
		\frac{\fd^2\Ham}{\fd\fv\,\fd\fv}\,
		{\lpb\dynacg'\com\fv\rpb}^\dagger
		+ {\lpb\dynacg'\com\frac{\fd\Ham}{\fd\fv}\rpb}\rang\nonumber\\
	&= \lang{\lpb\dynacg\com\fv\rpb}^\dagger\com
		{\lpb\dynacg\com\frac{\fd\Ham}{\fd\fv}\rpb}\rang
		- \lang{\lpb\dynacg\com\fv\rpb}^\dagger\com
		{\lpb\dynacg'\com\frac{\fd\Ham}{\fd\fv}\rpb}\rang.
	\eqlabel{d2Hdiff}
\end{align}
Using~\eqref{cobracket} and the Jacobi identity in~$\LieA$, we have that for
any~\hbox{$\alpha,\beta,\gamma \in \LieA$} and~\hbox{$\fv \in \LieA^*$},
\[
\begin{split}
	\lang{\lpb\alpha\com\fv\rpb}^\dagger\com\lpb\beta\com\gamma\rpb\rang
	&= \lang\fv\com\lpb\alpha\com\lpb\beta\com\gamma\rpb\rpb\rang\\
	&= -\lang\fv\com\l(
		\lpb\beta\com\lpb\gamma\com\alpha\rpb\rpb
		+ \lpb\gamma\com\lpb\alpha\com\beta\rpb\rpb\r)\rang\\
	&= \lang{\lpb\beta\com\fv\rpb}^\dagger\com
		\lpb\alpha\com\gamma\rpb\rang
		- \lang{\lpb\gamma\com\fv\rpb}^\dagger\com
		\lpb\alpha\com\beta\rpb\rang.
\end{split}
\]
Making use of this identity in the last term of~\eqref{d2Hdiff}, we get
\begin{multline*}
	2\fd^2\Ham_\dynac - 2{\fd'}^2\Ham_\dynac =
		\lang{\lpb\dynacg\com\fv\rpb}^\dagger\com
		{\lpb\dynacg\com\frac{\fd\Ham}{\fd\fv}\rpb}\rang
		- \lang{\lpb\dynacg'\com\fv\rpb}^\dagger\com
		{\lpb\dynacg\com\frac{\fd\Ham}{\fd\fv}\rpb}\rang\\
		+ \lang{\lpb\frac{\fd\Ham}{\fd\fv}\com\fv\rpb}^\dagger\com
		{\lpb\dynacg\com\dynacg'\rpb}\rang.
\end{multline*}
The first two terms cancel, and from~\eqref{motion} we are left with
\[
	2\fd^2\Ham_\dynac - 2{\fd'}^2\Ham_\dynac =
		-\lang\dotfv\com{\lpb\dynacg\com\dynacg'\rpb}\rang,
\]
which vanishes at an equilibrium of the equations of motion, for
any~$\dynacg$,~$\dynacg'$.  We conclude that~$\fd^2\Ham_\dynac$ depends
on~$\dynacg$ only through~$\fd\fv_\dynac$.  Thus, it is always possible to
rewrite~$\fd^2\Ham_\dynac$ in terms of only the dynamically accessible
perturbations~$\dynacg$.
\index{dynamical accessibility|)}

\subsection{2-D Euler}

\index{dynamical accessibility!and 2-D Euler}
An equilibrium of the equation of motion for 2-D Euler (see
\secref{twodfluid}) satisfies~\hbox{$\lpb\streamfeq\com\vorteq\rpb = 0$}.
The most general equilibrium solution can thus be written
\[
	\streamfeq = \Streamf(\afu(\xv));\ \ \ \ \vorteq = \Vort(\afu(\xv)),
\]
where~$\afu(\xv)$ is an arbitrary function. Contrary to the energy-Casimir
result, neither the function~$\Streamf$ or~$\Vort$ need be invertible (i.e.,
monotonic in their argument).

We can then examine stability by taking the dynamically accessible second
variation of the energy. This is given by~\eqref{dynactwoHam}
with~\hbox{${\lpb\dynacg\com\fveq\rpb}^\dagger =
-{\lpb\dynacg\com\vorteq\rpb}$},
\[
\begin{split}
	\fd^2\Ham_\dynac[\vorteq] &=
		\half\lang{\lpb\dynacg\com\vorteq\rpb}^\dagger\com
		(-\invlapl)\,{\lpb\dynacg\com\vorteq\rpb}^\dagger
		- {\lpb\dynacg\com\streamfeq\rpb}\rang\\
	&= \half\lang{\lpb\dynacg\com\vorteq\rpb}\com
		(-\invlapl)\,{\lpb\dynacg\com\vorteq\rpb}
		+ {\lpb\dynacg\com\streamfeq\rpb}\rang\\
	&= \half\lang{|\grad\fd\streamf_\dynac|}^2
		+ {\lpb\dynacg\com\vorteq\rpb}{\lpb\dynacg\com
			\streamfeq\rpb}\rang
		\\
	&= \half\lang{|\grad\fd\streamf_\dynac|}^2
		+ \Streamf'(u)\,\Vort'(u){\lpb\dynacg\com\afu\rpb}^2\rang,
\end{split}
\]
where~\hbox{$\lapl\fd\streamf_\dynac \ldef \fd\vort_\dynac$}. A sufficient
condition for~$\fd^2\Ham_\dynac[\vorteq]$ to be non-negative is
\begin{equation}
	\Streamf'(u)\,\Vort'(u) \ge 0,
	\eqlabel{Eulerstab}
\end{equation}
that is, the derivatives of~\hbox{$\Streamf$} and~\hbox{$\Vort$} must have
opposite signs. The energy-Casimir result is recovered by
letting~\hbox{$\Streamf(u)=u$}, for then we have~\hbox{$\vorteq =
\Vort(\streamfeq)$} and the stability condition is the usual Rayleigh
criterion,~\hbox{$\Vort'(\streamfeq) \ge 0$}. The stability
result~\eqref{Eulerstab} obtained using the dynamical accessibility method is
more general.

\subsection{Reduced MHD}
\seclabel{dynacRMHD}

\index{reduced MHD!stability|(}
\index{dynamical accessibility!and reduced MHD}
The equations of motion and bracket for RMHD are described in
\secref{lowbetaRMHD}. The dynamical variables
are~\hbox{$(\fv^0,\fv^1)=(\vort,\magf)$}.

\subsubsection{Equilibrium Solutions}

We must first determine equilibrium solutions~$(\vorteq,\magfeq)$ of the
equations of motion~\eqref{RMHDeom}, which must satisfy
\[
\begin{split}
	\l[\vorteq,\elecpeq\r] + \l[\magfeq,\ecurrenteq\r] &= 0,\\
	\l[\magfeq,\elecpeq\r] &= 0.
\end{split}
\]
To satisfy the second of these conditions we must have~\hbox{$\elecpeq =
\Elecp(\afu)$}, \hbox{$\magfeq = \Magf(\afu)$}, with~$\afu=\afu(\xv)$. Using
the fact that, for any~$\afii(\xv)$ and~$\afi(\afu(\xv))$,
\begin{equation}
	\lpb\afii(\xv)\com\afi(\afu)\rpb 
	= \afi'(\afu)\lpb\afii(\xv)\com\afu\rpb
	= \lpb\afi'(\afu)\,\afii(\xv)\com\afu\rpb,
	\eqlabel{cuteid}
\end{equation}
the first equilibrium condition can be written as
\[
	\l[\Elecp'(\afu)\,\vorteq - \Magf'(\afu)\,\ecurrenteq\com\afu\r] = 0.
\]
This is solved by
\begin{equation}
	\ecurrenteq = \frac{\Upsilon'(\afu)
		+ \Elecp'(\afu)\,\vorteq}{\Magf'(\afu)},
	\eqlabel{Jdynaceq}
\end{equation}
where~$\Upsilon(\afu)$ is an arbitrary function.  Note that this does
\emph{not} necessarily imply that~$\vorteq$ or~$\ecurrenteq$ are functions
of~$\afu$ only.

\subsubsection{Formal Stability}

Using the coadjoint bracket for extensions~\eqref{cobrakext}, the dynamically
accessible perturbations are given by
\[
\begin{split}
	\fd\vort_\dynac &= {\lpb\dynacg_0\com\vort\rpb}^\dagger
		+ {\lpb\dynacg_1\com\magf\rpb}^\dagger
		= -{\lpb\dynacg_0\com\vort\rpb}
		- {\lpb\dynacg_1\com\magf\rpb},
		\\
	\fd\magf_\dynac &= {\lpb\dynacg_0\com\magf\rpb}^\dagger
		 = -{\lpb\dynacg_0\com\magf\rpb}.
\end{split}
\]
The second-order variation of the Hamiltonian,~\eqref{dynactwoHam}, is
\begin{align}
	\fd^2\Ham_\dynac[\vorteq\,;\magfeq] &= \half\lang\fd\vort_\dynac\com
		(-\invlapl)\,\fd\vort_\dynac
		- {\lpb\dynacg_0\com\elecpeq\rpb}\rang\nonumber\\
	&\phantom{=}\qquad\qquad + \half\lang\fd\magf_\dynac\com
		(-\lapl)\,\fd\magf_\dynac
		- {\lpb\dynacg_0\com\ecurrenteq\rpb}
		- {\lpb\dynacg_1\com\elecpeq\rpb}\rang\nonumber\\
	&= \half\lang|\grad\fd\elecp_\dynac|^2
		+ |\grad\fd\magf_\dynac|^2\rang\nonumber\\
	&\phantom{=}\qquad\qquad - \half\lang{\lpb\dynacg_0\com\elecpeq\rpb}\,
			\fd\vort_\dynac
		+ {\lpb\dynacg_0\com\ecurrenteq\rpb}\,\fd\magf_\dynac
		+ {\lpb\dynacg_1\com\elecpeq\rpb}\,\fd\magf_\dynac\rang
		\nonumber\\
	&= \half\lang|\grad\fd\elecp_\dynac|^2
		+ |\grad\fd\magf_\dynac|^2
		+ \frac{\Elecp'}{\Magf'}
		\,\fd\magf_\dynac\,\fd\vort_\dynac
		\rang\nonumber\\
	&\phantom{=}\qquad\qquad
		- \half\lang{\lpb\dynacg_0\com\ecurrenteq\rpb}\,\fd\magf_\dynac
		+ \Elecp'\,
		{\lpb\dynacg_1\com\afu\rpb}\,\fd\magf_\dynac\rang,
\end{align}
where we have defined~\hbox{$\lapl\fd\elecp_\dynac \ldef
\fd\vort_\dynac$}. Now we use
\[
\begin{split}
	{\lpb\dynacg_1\com\afu\rpb} &=
		\frac{1}{\Magf'}
		\l(\Magf'\,{\lpb\dynacg_1\com\afu\rpb}
		+ {\lpb\dynacg_0\com\vorteq\rpb}\r)
		- \frac{1}{\Magf'}
		\,{\lpb\dynacg_0\com\vorteq\rpb}\\
	&= -\frac{1}{\Magf'}\l(\fd\vort_\dynac
		+ {\lpb\dynacg_0\com\vorteq\rpb}\r),
\end{split}
\]
to get
\begin{align*}
	\fd^2\Ham_\dynac[\vorteq\,;\magfeq] &=
		\half\lang|\grad\fd\elecp_\dynac|^2
		+ |\grad\fd\magf_\dynac|^2
		+ 2\,\frac{\Elecp'}{\Magf'}
		\,\fd\magf_\dynac\,\fd\vort_\dynac
		\rang\\
	&\phantom{=}\qquad\qquad + \half\lang
		\l(\frac{\Elecp'}{\Magf'}\,
		{\lpb\dynacg_0\com\vorteq\rpb}
		- {\lpb\dynacg_0\com\ecurrenteq\rpb}\r)
		\fd\magf_\dynac\rang\\
	&= \half\lang|\grad\fd\elecp_\dynac|^2
		+ |\grad\fd\magf_\dynac|^2
		+ 2\,\frac{\Elecp'}{\Magf'}
		\,\fd\magf_\dynac\,\fd\vort_\dynac
		\rang\\
	&\phantom{=}\qquad\qquad - \half\lang
		\l(
		\vorteq{\lpb\dynacg_0\com\frac{\Elecp'}{\Magf'}\rpb}
		+ {\lpb\dynacg_0\com\frac{\Upsilon'}{\Magf'}\rpb}\r)
		\fd\magf_\dynac\rang,
\end{align*}
where we substituted~\eqref{Jdynaceq} to eliminate~$\ecurrenteq$.  To simplify
the notation, we define the differential operator~$\Dafu$ by
\begin{equation}
	\Dafu\afi(\afu) \ldef
		\frac{1}{\Magf'(\afu)}\,\frac{d}{d\afu}\,\afi(\afu)\,,
	\eqlabel{Dafudef}
\end{equation}
so that
\begin{align}
	\fd^2\Ham_\dynac[\vorteq\,;\magfeq] &=
		\half\lang|\grad\fd\elecp_\dynac|^2
		+ |\grad\fd\magf_\dynac|^2
		+ 2\,\Dafu\Streamf
		\,\fd\magf_\dynac\,\lapl\fd\elecp_\dynac
		\rang\nonumber\\
	&\phantom{=}\qquad\qquad - \half\lang
		\l(
		\vorteq\,{\lpb\dynacg_0\com\Dafu\Streamf\rpb}
		+ {\lpb\dynacg_0\com\Dafu\Upsilon\rpb}\r)
		\fd\magf_\dynac\rang.
	\eqlabel{d2Hdai}
\end{align}
Note that the first angle bracket in~\eqref{d2Hdai} is the same
as~\eqref{coolident}, with~$\afiv'$ replaced by~$\Dafu\Streamf$. Hence, we can
use identity~\eqref{coolident} to obtain
\begin{multline}
	\fd^2\Ham_\dynac = \half\lang|\grad\fd\elecp_\dynac
		- \grad(\Dafu\Streamf\,\fd\magf_\dynac)|^2
		+ \l(1 - |\Dafu\Streamf|^2\r)\,|\grad\fd\magf_\dynac|^2
		\rang\\
	+ \half\lang\l(\Dafu\Streamf\,\lapl(\Dafu\Streamf)
		+ \vorteq\,\Dafu^2{\Streamf}
		+ \Dafu^2{\Upsilon}\r)|\fd\magf_\dynac|^2\rang.
	\eqlabel{RMHDd2H}
\end{multline}
Sufficient conditions for the perturbation energy~\eqref{RMHDd2H} to be
non-negative are~\cite{Hazeltine1984}
\begin{gather}
	|\Dafu\Streamf| \le 1, \eqlabel{RMHDstabcond0}\\
	\Dafu\Streamf\,\lapl(\Dafu\Streamf) + \lapl\Streamf\,\Dafu^2{\Streamf}
			+ \Dafu^2{\Upsilon} \ge 0.
	\eqlabel{RMHDstabcond}
\end{gather}
In the second expression we have substituted~\hbox{$\vorteq = \lapl\Streamf$}.
The first condition says that~\hbox{$|\Elecp'(\afu)| \le |\Magf'(\afu)|$},
that is, the gradient of the magnetic flux is greater or equal to the gradient
of the electric potential.  This is a similar condition to~\eqref{phireqi},
and says that the flow needs to be \emph{sub-\Alfvenic}\ to be formally
stable~\cite{Kent1968}.  This is due to the well-known fact that the magnetic
field provides a restoring force for perturbations of the flow, so that a
large enough magnetic field can potentially stabilize the system (but not
necessarily so, because the magnetic field can also have a destabilizing
effect~\cite{Chen1991}).  Indeed, condition~\eqref{RMHDstabcond0} is actually
\emph{necessary} for positive-definiteness of~$\fd^2\Ham_\dynac$.  If we
choose~\hbox{$\fd\elecp_\dynac = \Dafu\Streamf\,\fd\magf_\dynac$}
in~\eqref{RMHDd2H}, then the first term vanishes.  We can then pick a
variation of~$\fd\magf_\dynac$ with as steep a gradient as we want, while
maintaining the value of~$\fd\magf_\dynac$ bounded~\cite[p.~103]{Gelfand}.
This means that the~\hbox{$|\grad\fd\magf_\dynac|^2$} term can always be made
to dominate, so that we \emph{require}~\hbox{$|\Dafu\Streamf| \le 1$} for
positive-definiteness of~$\fd^2\Ham_\dynac$.
\index{reduced MHD!stability|)}

This places a limitation on the method of dynamical accessibility: if we want
to satisfy~\hbox{$|\Dafu\Streamf| = \Streamf'/\Magf' \le 1$} everywhere, then
on their domain of definition \emph{the zeros of~$\Magf$ must also be zeros
of~$\Streamf$} with equal or higher multiplicity.  (However, the
function~$\Streamf$ could potentially have \emph{more} zeros than~$\Magf$.)

The simplest case is when~$\Magf$ has no zeros, but then~$\Magf(\afu)$ is
invertible, and we can recover the energy-Casimir result by solving
for~\hbox{$\afu=\afu(\Magf)$}.  In practice, this inversion may be difficult,
and using the dynamical accessibility method is often easier.

As an example, we will derive equilibria for \emph{magnetic islands with
flow}\index{magnetic islands|(}.  Consider the RMHD equilibrium
relation~\eqref{Jdynaceq}, multiplied by~$\Magf'(\afu)$,
\begin{equation}
	\Magf'(\afu)\ecurrenteq - \Elecp'(\afu)\,\vorteq = \Upsilon'(\afu).
	\eqlabel{Jdynaceq0}
\end{equation}
where~\hbox{$\ecurrenteq = \lapl\Magf(\afu)$}.  Using the fact that
\begin{equation}
	\vorteq = \lapl\Streamf(\afu) = \Streamf'(\afu)\,\lapl\afu
		+ \Streamf''(\afu)\,|\grad\afu|^2,
	\eqlabel{vortitu}
\end{equation}
and the analogous relation for~$\ecurrenteq$, we can rewrite~\eqref{Jdynaceq0}
as
\[
	\l((\Magf')^2-(\Streamf')^2\r)\,\lapl\afu
		+ \l(\Magf'\,\Magf'' - \Streamf'\,\Streamf''\r)
		\,|\grad\afu|^2 = \Upsilon'(\afu),
\]
or equivalently
\begin{equation}
	\l((\Magf')^2-(\Streamf')^2\r)\,\lapl\afu
		+ \half\l((\Magf')^2-(\Streamf')^2\r)'
		\,|\grad\afu|^2 = \Upsilon'(\afu).
	\eqlabel{Jdynaceq1}
\end{equation}
We can get rid of the~$|\grad\afu|^2$ term, and make the equation easier to
solve, by choosing
\[
	(\Magf')^2 - (\Streamf')^2 = \knograd^2.
\]
(Choosing a different sign for the right-hand side would lead to solutions
with~\hbox{$\Dafu\Streamf>1$}.)  An obvious solution is
\begin{align}
	\Magf'(\afu) &= \knograd \cosh(\nunograd\afu),
	\eqlabel{Magfpcosh}\\
	\Streamf'(\afu) &= \knograd \sinh(\nunograd\afu).
	\eqlabel{Streamfpsinh}
\end{align}
These satisfy~\hbox{$|\Dafu\Streamf| = |\tanh(\nunograd\afu)| < 1$},
condition~\eqref{RMHDstabcond0}.

Equation~\eqref{Jdynaceq1} becomes
\begin{equation}
	\lapl\afu = \knograd^{-2}\,\Upsilon'(\afu),
	\eqlabel{Jdynaceq2}
\end{equation}
to be solved for~$\afu(\xv)$.  This equation has the same form
as~\eqref{alfvenicJnox}, which was an equation for~$\magfeq(\xv)$, so it has
the same Kelvin--Stuart cat's eye\index{Kelvin--Stuart cat's eye} solution,
\[
	\afu(x,y) = \ln(a\cosh y + \sqrt{a^2 - 1}\, \cos x),
\]
with~\hbox{$\Upsilon'(\afu) = \knograd^2\,\exp(-2\afu)$}.  The difference is
that now the physical variables are given in terms of~$\afu$
by~\eqref{Magfpcosh} and~\eqref{Streamfpsinh}, so that the electric potential
(and so the flow velocity) does not necessarily vanish, as opposed to the
usual magnetic island solutions, which are recovered in the
limit~\hbox{$\nunograd = 0$}.  The stability of the islands with flow could be
very different, since now~\hbox{$\Streamf' \ne 0$} in~\eqref{d2Hdai}.
However, as for the usual magnetic islands, the sufficient
condition~\eqref{RMHDstabcond} is not satisfied, so that stability must be
determined by test perturbations, or by direct numerical
simulation~\cite{Bondeson1983,Finn1977,Holm1985,Pritchett1979}.
\index{magnetic islands|)}

\subsection{Pure Semidirect Sum}
\seclabel{SDPstab}

\index{dynamical accessibility!and semidirect sums}
We now treat the general stability of the pure semidirect sum structure, with
no cocycles (see \secref{semisimple}). This structure is given simply by
the~\hbox{$n+1 \times n+1$} matrices~\hbox{$\W^{(0)} = I$}, and
\hbox{$\Wt^{(\mu)}=0$}, \hbox{$\mu=1,\dots,n$}. We denote the 0th field
variable by~\hbox{$\fv^{0} = \fvz$}, and the remaining~$n$ variables
by~\hbox{$\fv^1,\dots,\fv^n$}.

\subsubsection{Equilibrium Solutions}

An equilibrium~\hbox{$(\fvzeq,\l\{\fveq^\mu\r\})$} of the equations of motion
for a pure semidirect extension satisfies
\begin{align}
	\dotfvz_\equil &= -{\lpb\Ham_{\dcom 0}\com\fvzeq\rpb}^\dagger
	- \sum_{\mu=1}^n{\lpb\Ham_{\dcom\mu}
			\com\fveq^\mu\rpb}^\dagger = 0,
			\eqlabel{SDPeq1} \\
	\dotfv_\equil^\mu &= -{\lpb\Ham_{\dcom 0}\com\fveq^\mu\rpb}^\dagger
		= 0,
	\ \ \ \mu=1,\dots,n.
	\eqlabel{SDPeq2}
\end{align}
To unclutter the notation, we assume that the first and second derivatives of
the Hamiltonian~$\Ham$ are evaluated at the
equilibrium~\hbox{$(\fvzeq,\l\{\fveq^\mu\r\})$}, unless otherwise noted.

We now specialize the bracket to the 2-D canonical one, Eq.~\eqref{canibrak},
so that~${\lpb\com\rpb}^\dagger = -{\lpb\com\rpb}$.  To satisfy
condition~\eqref{SDPeq2}, we require
\begin{equation}
	\Ham_{\dcom 0} = -\Streamf(\afu),\qquad \fveq^\mu = \Fv^\mu(\afu),
		\quad \mu=1,\dots,n,
	\eqlabel{SDPeqlm}
\end{equation}
for arbitrary functions~$\Streamf$,~$\Fv^\mu$, and~$\afu=\afu(\xv)$.  (The
choice of the minus sign for the definition of~$\Streamf$ is purely a
convention to agree with the sign of the streamfunction in 2-D Euler, for
which~\hbox{$\Ham_{\dcom 0} = \fd\Ham/\fd\vort = -\streamf$}.)
Condition~\eqref{SDPeq1} is then
\[
	-\lpb\Streamf(\afu)\com\fvzeq\rpb
		+ \sum_{\mu=1}^n\lpb\Ham_{\dcom\mu}\com\Fv^\mu(\afu)\rpb = 0,
\]
or, using~\eqref{cuteid},
\[
	\Bigl[\afu\com\Streamf'(\afu)\,\fvzeq
		+ \sum_{\mu=1}^n\Ham_{\dcom\mu}\,{\Fv^\mu}'(\afu)\Bigr] = 0,
\]
which has solution
\begin{equation}
	\Streamf'(\afu)\,\fvzeq
		+ \sum_{\mu=1}^n\Ham_{\dcom\mu}\,{\Fv^\mu}'(\afu)
		= \Upsilon'(\afu).
	\eqlabel{SDPeqlmfvz}
\end{equation}
Equation~\eqref{SDPeqlmfvz} should be compared with~\eqref{Jdynaceq}, the
equivalent solution for reduced MHD, for which~$n=1$ and~\hbox{$\Ham_{\dcom 1}
= \fd\Ham/\fd\magf = -\ecurrent$}.

Now that we have the equilibria, using~\eqref{cobrakext} we write down the
dynamically accessible perturbations
\begin{align}
	\fd\fvz_\dynac &= {\lpb\dynacg_0\com\fvz\rpb}^\dagger
		+ \sum_{\nu=1}^n {\lpb\dynacg_\nu\com\fv^\nu\rpb}^\dagger,
	\eqlabel{SDPfvzfuncvar} \\
	\fd\fv^\mu_\dynac &= {\lpb\dynacg_0\com\fv^\mu\rpb}^\dagger,
		\qquad \mu = 1,\dots,n,
	\eqlabel{SDPfvmufuncvar}
\end{align}
and from~\eqref{dynactwoHam} we get the second-order dynamically accessible
variation of the Hamiltonian,
\begin{multline*}
	\fd^2\Ham_\dynac = \half\Bigl\langle\fd\fvz_\dynac\com
		\Ham_{\dcom 00}\,\fd\fvz_\dynac
		+ \sum_{\mu=1}^n \Ham_{\dcom 0\mu}\,\fd\fv^\mu_\dynac
		+ \lpb\dynacg_0\com\Ham_{\dcom 0}\rpb\Bigr\rangle\\
	+ \sum_{\mu=1}^n\half\Bigl\langle\fd\fv^\mu_\dynac\com
		\sum_{\nu=1}^n \Ham_{\dcom \mu\nu}\,\fd\fv^\nu_\dynac
		+ \Ham_{\dcom \mu 0}\,\fd\fvz_\dynac
		+ \lpb\dynacg_0\com\Ham_{\dcom\mu}\rpb
		+ \lpb\dynacg_\mu\com\Ham_{\dcom 0}\rpb\Bigr\rangle.
\end{multline*}
Because the second-order functional derivative is formally a self-adjoint
operator, we have the identity
\[
	\lang\fd\fvz_\dynac\com\Ham_{\dcom 0\mu}\,\fd\fv^\mu_\dynac\rang
	= \lang\fd\fv^\mu_\dynac\com\Ham_{\dcom \mu 0}\,\fd\fvz_\dynac\rang,
\]
which we use in~$\fd^2\Ham_\dynac$ to combine two terms and obtain
\begin{multline}
	\fd^2\Ham_\dynac = \half\Bigl\langle\fd\fvz_\dynac\com
		\Ham_{\dcom 00}\,\fd\fvz_\dynac
		+ 2\sum_{\mu=1}^n \Ham_{\dcom 0\mu}\,\fd\fv^\mu_\dynac
		+ \lpb\dynacg_0\com\Ham_{\dcom 0}\rpb\Bigr\rangle\\
	+ \sum_{\mu=1}^n\half\Bigl\langle\fd\fv^\mu_\dynac\com
		\sum_{\nu=1}^n \Ham_{\dcom \mu\nu}\,\fd\fv^\nu_\dynac
		+ \lpb\dynacg_0\com\Ham_{\dcom\mu}\rpb
		+ \lpb\dynacg_\mu\com\Ham_{\dcom 0}\rpb\Bigr\rangle.
	\eqlabel{SDPHsecorder}
\end{multline}
Using the equilibrium solution~\eqref{SDPeqlm}, the dynamically accessible
variations given by~\eqref{SDPfvmufuncvar} can be rewritten
\[
	\fd\fv^\mu_\dynac = -{\Fv^\mu}'(\afu)\,{\lpb\dynacg_0\com\afu\rpb}
\]
Observe that the perturbations of the~$\fv^\mu$ are not independent: they all
depend on a single generating function,~$\dynacg_0$.  We choose to write all
the variations in terms of~$\fd\Fv^n_\dynac$.  We define
\[
	\magf(\xv) \ldef \fv^n(\xv), \qquad\qquad
		\Magf(\afu) \ldef \Fv^n(\afu),
\]
to explicitly show the special role of~$\fv^n$.  Then we have
\begin{equation}
	\fd\fv^\mu_\dynac = \frac{{\Fv^\mu}'(\afu)}{{\Magf}'(\afu)}
		\,\fd\magf_\dynac
		= \Dafu\Fv^\mu\,\fd\magf_\dynac,
	\eqlabel{magfdynac}
\end{equation}
where we have used the previous definition of the operator~$\Dafu$,
\begin{equation}
	\Dafu\afi(\afu) \ldef
		\frac{1}{\Magf'(\afu)}\,\frac{d}{d\afu}\,\afi(\afu)\,.
	\tag{\ref{eq:Dafudef}}
\end{equation}
Note that~\hbox{$\Dafu\Fv^n=\Dafu\Magf=1$}.  We could have chosen any field
instead of~$\fv^n$, but in \secref{WnStab} this particular choice will prove
advantageous due to the lower-triangular structure of our extensions.

The dynamically accessible variations must obey the constraints of the system,
that is they must lie on the coadjoint orbits\index{coadjoint orbit}.  We
have already discussed briefly this property of the semidirect sum
in~\secref{semidext}.

\index{rigid body}
\index{semidirect sum!and dynamical accessibility}
To illustrate the situation we consider the equations of motion for a
finite-dimensional semidirect sum, specifically a semidirect sum of the
rotation group~$\SOthree$ (associated with our old friend the rigid body)
with~$\reals^3$ (see \secref{lpexample}).  We take~$\fvz$ to be~$\ell$, the
angular momentum vector, with Hamiltonian~$\Ham$ given by the usual kinetic
energy, Eq.~\eqref{rbHam}.  The variables~$\fv^\mu$ are three-vectors, and
their equations of motion are given in terms of the bracket~\eqref{rbcobrak}
by
\begin{equation}
\begin{split}
	\dotfv^\mu &= -{\lpb \Ham_{\dcom 0}\com \fv^\mu\rpb}^\dagger\\
		 &= \l(I^{-1}\ell\r)\times\fv^\mu\ .
	\eqlabel{rbadveq}
\end{split}
\end{equation}
Note the angular momentum~$\ell$ is analogous to the
vorticity~$\vort$,\index{vorticity!analogy with angular momentum}
and~$I^{-1}\ell$ is analogous to the
streamfunction~\hbox{$\streamf=\invlapl\vort$}.\index{streamfunction}
\index{angular momentum!analogy with vorticity} Equation~\eqref{rbadveq}
says that the vector~$\fv^\mu$ is rotating with the rigid body, keeping its
length constant (the length of~$\fv^\mu$ is a Casimir).  Thus, each~$\fv^\mu$
can be used to describe a point in the rigid body, such as the center of
gravity.  Adding a coupling term to the Hamiltonian can provide us with, for
instance, a description of the heavy top\index{rigid body!heavy top} in a
gravitational field, but this would not change the form of~\eqref{rbadveq}.
The point is that the~$\fv^\mu$ are constrained to rotate rigidly, and the
dynamically accessible perturbations must obey the same constraint---they must
depend on the perturbation applied to~$\ell$, but by themselves there are no
dynamically accessible perturbations that allow the~$\fv^\mu$ to change length
or rotate independently.  Physically, this makes sense, because we are not
allowing the rigid body to have other degrees of freedom than the rotational
ones.  If we did, we would have to rethink our description, which would lead
to different dynamically accessible perturbations; but within the confines of
rigidity those perturbations make sense.\index{dynamical
accessibility!underlying physics}

The situation in infinite dimensions is analogous to the rigid body.  Here the
typical case is an ideal fluid with passive scalars:\index{advection!as
semidirect sum} \index{semidirect sum!and advection} we
take~$\fvz=\vort(\xv)$, the vorticity, and a Hamiltonian of the
form~\hbox{$\Ham[\vort]=-\half\lang\streamf\com\vort\rang$}.  The equations of
motion for the~$\fv^\mu(\xv)$ are given by
\begin{equation}
	\dotfv^\mu = -{\lpb \Ham_{\dcom 0}\com \fv^\mu\rpb}^\dagger
		=  -{\lpb \streamf\com \fv^\mu\rpb}\,.
	\eqlabel{scadveq}
\end{equation}
Thus, the~$\fv^\mu(\xv)$ are advected along by the fluid.  The~$\fv^\mu(\xv)$
can be used to describe passive scalars, since they do not enter the
Hamiltonian (they do not affect the flow itself).  An interaction term
in~$\Ham$ could describe, for example, the effect of temperature on the flow
in the Boussinesq approximation, but this would not modify~\eqref{scadveq}:
only the equation for~$\dotvort$ would change.  Much like for the rigid body,
the quantities~$\fv^\mu$ are constrained to move with the fluid, regardless of
the form of the Hamiltonian.  This is also true for the dynamically accessible
perturbations of the~$\fv^\mu$, which must then be induced by the perturbation
on~$\vort$.

\subsubsection{Formal Stability}

We now try to rewrite the second-order variation of the
Hamiltonian~\eqref{SDPHsecorder} only in terms of dynamically accessible
variations. We have, from~\eqref{SDPeqlm},
\[
\begin{split}
	\lpb\dynacg_0\com\Ham_{\dcom 0}\rpb &=
		-\Streamf'(\afu)\,\lpb\dynacg_0\com\afu\rpb\\
	&= \Dafu\Streamf(\afu)\,\fd\magf_\dynac.
\end{split}
\]
From the second line of~\eqref{SDPHsecorder}, we can write
\begin{equation}
\begin{split}
	\sum_{\mu=1}^n\Bigl\langle\fd\fv^\mu_\dynac\com
		\lpb\dynacg_\mu\com\Ham_{\dcom 0}\rpb\Bigr\rangle
	&= -\sum_{\mu=1}^n\Bigl\langle\fd\magf_\dynac\,\Dafu{\Fv^\mu}
		\,\Streamf'
		\lpb\dynacg_\mu\com\afu\rpb\Bigr\rangle\\
	&= -\sum_{\mu=1}^n\Bigl\langle
		\fd\magf_\dynac\,\Dafu\Streamf
		\lpb\dynacg_\mu\com{\Fv^\mu}\rpb\Bigr\rangle\\
	&= \Bigl\langle\Dafu\Streamf\,
		(\fd\fvz_\dynac + \lpb\dynacg_0\com\fvzeq\rpb)
		\,\fd\magf_\dynac\Bigr\rangle,
\end{split}
	\eqlabel{SDPd2Hterm1}
\end{equation}
where we have made use of~\eqref{SDPfvzfuncvar} and~\eqref{magfdynac}. Finally,
we have
\[
\begin{split}
	\sum_{\mu=1}^n\Bigl\langle\fd\fv^\mu_\dynac\com
		\lpb\dynacg_0\com\Ham_{\dcom\mu}\rpb\Bigr\rangle &=
		\sum_{\mu=1}^n\Bigl\langle\fd\magf_\dynac\,\Dafu{\Fv^\mu}
		\lpb\dynacg_0\com\Ham_{\dcom\mu}\rpb\Bigr\rangle\\
	&= \sum_{\mu=1}^n\Bigl\langle\fd\magf_\dynac
		\lpb\dynacg_0\com\Ham_{\dcom\mu}\,\Dafu{\Fv^\mu}\rpb
		- \fd\magf_\dynac\,\Ham_{\dcom\mu}
			\lpb\dynacg_0\com\Dafu{\Fv^\mu}\rpb
		\Bigr\rangle,
\end{split}
\]
in which we make use of~\eqref{SDPeqlmfvz} to obtain
\begin{equation}
\begin{split}
	\sum_{\mu=1}^n\Bigl\langle\fd\fv^\mu_\dynac\com
		\lpb\dynacg_0\com\Ham_{\dcom\mu}\rpb\Bigr\rangle
		&= \Bigl\langle\fd\magf_\dynac\Bigl(
		\lpb\dynacg_0\com\Dafu\Upsilon - \Dafu\Streamf\,\fvzeq\rpb
		- \sum_{\mu=1}^n\Ham_{\dcom\mu}({\Dafu\Fv^\mu})'
		\lpb\dynacg_0\com\afu\rpb\Bigr)\Bigr\rangle\\
	&= \Bigl\langle
		\bigl(\Dafu^2\Streamf\,\fvzeq
		+ \sum_{\mu=1}^n\Ham_{\dcom\mu}\Dafu^2{\Fv^\mu}
		- \Dafu^2\Upsilon\bigr)
		|\fd\magf_\dynac|^2\Bigr\rangle\\
	& \phantom{= \Bigl\langle}\qquad\qquad\qquad\qquad
		- \Bigl\langle\Dafu\Streamf\,
		\lpb\dynacg_0\com\fvzeq\rpb\,\fd\magf_\dynac\Bigr\rangle,
\end{split}
	\eqlabel{SDPd2Hterm2}
\end{equation}
The last term in~\eqref{SDPd2Hterm2} cancels part of~\eqref{SDPd2Hterm1}, and
we get
\begin{multline}
	\fd^2\Ham_\dynac = \half\Bigl\langle
		\fd\fvz_\dynac\,\Ham_{\dcom 00}\,\fd\fvz_\dynac
		+ 2\sum_{\mu=1}^n \fd\fvz_\dynac\,
			\Ham_{\dcom 0\mu}\,\fd\fv^\mu_\dynac
		+ 2\Dafu\Streamf\,\fd\fvz_\dynac\,\fd\magf_\dynac\\
	+ \sum_{\mu,\nu=1}^n\fd\fv^\mu_\dynac\,
		\Ham_{\dcom \mu\nu}\,\fd\fv^\mu_\dynac
		+ \Bigl(\Dafu^2\Streamf\,\fvzeq
		+ \sum_{\mu=1}^n\Ham_{\dcom\mu}{\Dafu^2\Fv^\mu}
		- \Dafu^2\Upsilon\Bigr)|\fd\magf_\dynac|^2\Bigr\rangle.
	\eqlabel{SDPd2H}
\end{multline}
Further progress cannot be made without assuming some particular form for the
second-order functional derivative operator of~$\Ham$.

\subsubsection{Hamiltonian without operators}

The simplest case we can study is when~$\Ham$ contains no differential or
integral operators.  Then~$\Ham_{\dcom\mu\nu}$ is just a symmetric matrix.
Using~\eqref{magfdynac}, we can simplify~\eqref{SDPd2H} to
\begin{multline*}
	\fd^2\Ham_\dynac = \half\Bigl\langle
		\Ham_{\dcom 00}\,|\fd\fvz_\dynac|^2
		+ 2\Bigl(\sum_{\mu=1}^n 
			\Ham_{\dcom 0\mu}\,{\Dafu\Fv^\mu}
		+ \Dafu\Streamf
		\Bigr)\fd\fvz_\dynac\,\fd\magf_\dynac\\
	+ \Bigl(\sum_{\mu,\nu=1}^n{\Dafu\Fv^\mu}\,
		\Ham_{\dcom \mu\nu}\,{\Dafu\Fv^\nu}
		+ \Dafu^2\Streamf\,\fvzeq
		+ \sum_{\mu=1}^n\Ham_{\dcom\mu}{\Dafu^2\Fv^\mu}
		- \Dafu^2\Upsilon\Bigr)|\fd\magf_\dynac|^2\Bigr\rangle.
\end{multline*}
This can be rewritten as a quadratic form,
\[
	\fd^2\Ham_\dynac = \half
	\begin{pmatrix}\fd\fvz_\dynac & \fd\magf_\dynac\end{pmatrix}
	\quadform
	\begin{pmatrix}\fd\fvz_\dynac \\ \fd\magf_\dynac\end{pmatrix},
\]
where~$\quadform$ is the~$2\times 2$ matrix
\[
	\quadform \ldef \l(\begin{array}{c|c}
	\Ham_{\dcom 00}
	& \Ham_{\dcom 0\mu}\,{\Dafu\Fv^\mu} + \Dafu\Streamf\\[4pt]
	\Ham_{\dcom 0\mu}\,{\Dafu\Fv^\mu} + \Dafu\Streamf
	& {\Dafu\Fv^\mu}\,
		\Ham_{\dcom \mu\nu}\,{\Dafu\Fv^\nu}
		+ \Dafu^2\Streamf\,\fvzeq
		+ \Ham_{\dcom\mu}{\Dafu^2\Fv^\mu}
		- \Dafu^2\Upsilon
	\end{array}\r).
\]
We assume repeated indices are summed from~$1$ to~$n$.  The
matrix~$\quadform$ is non-negative if and only if its principal minors are
non-negative, i.e.,
\begin{align}
	\Ham_{\dcom 00} &\ge 0,
		\eqlabel{H00ge0}\\
	\det \quadform &\ge 0.
		\eqlabel{Qge0}
\end{align}
Hence, to have formal stability it is imperative to have that the energy
associated with the perturbation of~$\fvz$ be non-negative.  Also note that
the contribution of~\hbox{$(\Ham_{\dcom 0\mu}\,{\Dafu\Fv^\mu} +
\Dafu\Streamf)$} is always destabilizing.  For an equilibrium without flow
(\hbox{$\Dafu\Streamf\equiv 0$}) and with~\hbox{$\Ham_{\dcom 0\nu}=0$},
condition~\eqref{Qge0} reduces to
\[
	{\Dafu\Fv^\mu}\,
		\Ham_{\dcom \mu\nu}\,{\Dafu\Fv^\nu}
		+ \Ham_{\dcom\mu}{\Dafu^2\Fv^\mu}
		- \Dafu^2\Upsilon\ge 0.
	\eqlabel{Hamnoopnoflow}
\]

\subsubsection{Advected Scalars}

\index{dynamical accessibility!and advection}
\index{advection!stability}
We now treat the problem of advection of scalars.  We shall not restrict
ourselves to passive advection, and the form we choose for~$\Ham$ is general
enough to encompass systems with \emph{generalized vorticities},\footnote{Also
called the \emph{potential vorticity}.} such as the quasigeostrophic
equations~\cite{Holm1998,Weinstein1983}.

Let~$\gvort$ denote the generalized vorticity, related to the stream
function~$\streamf$ by
\begin{equation}
	\gvort = \lapl\streamf - \Fgvort\,\streamf + \fgvort,
	\eqlabel{genvort}
\end{equation}
for some given functions~$\Fgvort(\xv)$ and~$\fgvort(\xv)$.
Taking~\hbox{$\fv^0 = \fvz = q$}, we consider a Hamiltonian
\[
\begin{split}
	\Ham &= \lang\half\l(|\grad\streamf|^2 + \Fgvort\,\streamf^2\r)
		+ \Vpot(\xv,\gvort,\fv^1,\dots,\fv^n)\rang\\
	&= \lang\half(\gvort-\fgvort)
		(\Fgvort-\lapl)^{-1}(\gvort-\fgvort)
		+ \Vpot(\xv,\gvort,\fv^1,\dots,\fv^n)\rang,
\end{split}
\]
where~$\Vpot$ does not contain any operators. We have the first derivatives
\[
	\Ham_{\dcom 0} = -\streamf + \Vpot_{\dcom 0}\,,\qquad\qquad
	\Ham_{\dcom\mu} = \Vpot_{\dcom\mu}\,,
\]
and the second derivative operators
\begin{align*}
	\Ham_{\dcom 00} &= (\Fgvort-\lapl)^{-1} + \Vpot_{\dcom 00}\,,\\
	\Ham_{\dcom\mu\nu} &= \Vpot_{\dcom\mu\nu}\,,\\
	\Ham_{\dcom 0\nu} &= \Vpot_{\dcom 0\nu}\,.
\end{align*}
Using identity~\eqref{coolident}, we can rewrite the first line of the second
dynamically accessible variation of the energy~\eqref{SDPd2H} as
\begin{align}
	\half\Bigl\langle
		\fd\gvort_\dynac&
		\,((\Fgvort-\lapl)^{-1} + \Vpot_{\dcom 00})
		\,\fd\gvort_\dynac + 2 \l(\Vpot_{\dcom 0\mu}\,
			{\Dafu\Fv^\mu}
		+ \Dafu\Streamf\r)\fd\gvort_\dynac\,\fd\magf_\dynac
		\Bigr\rangle
		\nonumber\\
	&= \half\Bigl\langle
		\fd\streamf_\dynac\,(\Fgvort-\lapl)\,\fd\streamf_\dynac
		+ \Vpot_{\dcom 00}\,|\fd\gvort_\dynac|^2
		 - 2 \Kfunc(\afu)\,\fd\magf_\dynac\,(\Fgvort-\lapl)
		\fd\streamf_\dynac\Bigr\rangle\nonumber\\
	&= \half\Bigl\langle
		|\grad\fd\streamf_\dynac
		- \grad(\Kfunc\,\fd\magf_\dynac)|^2
		- \Kfunc^2\,|\grad\fd\magf_\dynac|^2
		+ \Vpot_{\dcom 00}\,|\fd\gvort_\dynac|^2\nonumber\\
	&\phantom{= \half\Bigl\langle}
		+ \Fgvort\,|\fd\streamf_\dynac - \Kfunc
			\,\fd\magf_\dynac|^2
		+ \Kfunc\l(\lapl\Kfunc
			- \Fgvort\,\Kfunc\r)|\fd\magf_\dynac|^2
		\Bigr\rangle
	\eqlabel{SDPd2Hfirstline}
\end{align}
where
\begin{equation}
	\Kfunc(\afu) \ldef \Vpot_{\dcom 0\mu}\,{\Dafu\Fv^\mu}(\afu)
		+ \Dafu\Streamf(\afu).
	\eqlabel{Kdef}
\end{equation}
The term proportional to~$|\grad\fd\magf_\dynac|^2$ in~\eqref{SDPd2Hfirstline}
is negative definite unless we require an equilibrium
with~\hbox{$\Kfunc(\afu) \equiv 0$}, that is
\[
	\Vpot_{\dcom 0\mu}\,{\Dafu\Fv^\mu}(\afu) + \Dafu\Streamf(\afu) = 0.
\]
Using~\hbox{$\Kfunc \equiv 0$} in~\eqref{SDPd2Hfirstline} and writing
out the rest of~\eqref{SDPd2H}, we obtain
\begin{multline*}
	\fd^2\Ham_\dynac = \half\Bigl\langle
		|\grad\fd\streamf_\dynac|^2
		+ \Fgvort\,|\fd\streamf_\dynac|^2
		+ \Vpot_{\dcom 00}\,|\fd\gvort_\dynac|^2\\
	+ \Bigl(\Dafu^2\Streamf\,\gvorteq
		+ {\Dafu\Fv^\mu}\,\Vpot_{\dcom \mu\nu}\,
			{\Dafu\Fv^\nu}
		+ \Vpot_{\dcom\mu}\,{\Dafu^2\Fv^\mu}
		- \Dafu^2\Upsilon\Bigr)|\fd\magf_\dynac|^2\Bigr\rangle.
\end{multline*}
For Hamiltonians with~$\Vpot_{\dcom 0\mu} = 0$, the only equilibria for which
we can demonstrate formal stability are ones without flow.  If we assume this
is the case, then from~\eqref{SDPeqlmfvz} equilibria
satisfy~\hbox{$\Vpot_{\dcom\mu}\,{\Dafu\Fv^\mu}(\afu) = \Dafu\Upsilon(\afu)$}.
Note that~$\fgvort$ in~\eqref{genvort} enters the stability expression
through~\hbox{$\gvorteq = \lapl\Streamf - \Fgvort\,\Streamf + \fgvort$}.

Combining~\eqref{SDPd2Hfirstline} with the rest of~\eqref{SDPd2H} we have the
the sufficient conditions for stability
\begin{gather*}
	\Fgvort \ge 0,\\
	\Vpot_{\dcom 00} \ge 0,\\
	{\Dafu\Fv^\mu}\,\Vpot_{\dcom \mu\nu}\,{\Dafu\Fv^\nu}
		+ \Vpot_{\dcom\mu}{\Dafu^2\Fv^\mu}
		- \Dafu^2\Upsilon \ge 0,
\end{gather*}
where we have assumed~$\Dafu\Streamf\equiv 0$ so
that~\hbox{$\Kfunc(\afu) =
\Vpot_{\dcom 0\mu}\,{\Dafu\Fv^\mu}(\afu)$}.  This is the same stability
condition as for a Hamiltonian without operators,~\eqref{Hamnoopnoflow},
because we have chose a form of the Hamiltonian which decouples the operator
part (kinetic energy) and the potential, so we get a Lagrange-theorem-like
condition on the potential.

\subsubsection{RMHD-like System}

Another case of interest, a generalization of the RMHD system of
\secreftwo{lowbetaRMHD}{dynacRMHD}, involves a Hamiltonian of the form
\begin{equation}
	\Ham = \half\lang\l(|\grad\streamf|^2 + \Fgvort\,\streamf^2\r)
		+ 2\Vpot(\xv,\gvort,\fv^1,\dots,\fv^{n-1},\magf)
		+ |\grad\magf|^2\rang.
	\eqlabel{RMHDlikeHam}
\end{equation}
Here~$\gvort$,~$\Fgvort$, and~$\streamf$ are as in the previous section
in~\eqref{genvort}.  As before, we have labeled~$\fv^n$ by~$\magf$ as a
reminder of its distinguished role: it enters the Hamiltonian as a
gradient. (In this section greek indices run from~$1$ to~$n-1$.)  The first
derivatives of~$\Ham$ are given by
\begin{gather}
	\Ham_{\dcom 0} = -\streamf + \Vpot_{\dcom 0}\,,\qquad
	\Ham_{\dcom\mu} = \Vpot_{\dcom\mu}\,,\qquad
	\Ham_{\dcom n} =  -\ecurrent + \Vpot_{\dcom n}\,,\qquad
	\eqlabel{RMHDlikeHam1stdiff}
\end{gather}
and the second derivative operators are
\begin{alignat}{2}
	\Ham_{\dcom 00} &= (\Fgvort-\lapl)^{-1} + \Vpot_{\dcom 00} &
	\qquad \Ham_{\dcom 0n} &=  \Vpot_{\dcom 0n} \nonumber\\
	\Ham_{\dcom\mu\nu} &= \Vpot_{\dcom\mu\nu} & \qquad
		\Ham_{\dcom \mu n} &= \Vpot_{\dcom \mu n}
	\eqlabel{RMHDlikeHam2nddiff} \\
	\Ham_{\dcom 0\nu} &= \Vpot_{\dcom 0\nu} & \qquad
		\Ham_{\dcom nn} &= -\lapl + \Vpot_{\dcom nn}\ \nonumber.
\end{alignat}
The quantity~\hbox{$\ecurrent \ldef \lapl\magf$} is analogous to the electric
current in RMHD.  As before, we use~$\Fv^\mu(\afu)$ to denote the equilibrium
solution of~$\fv^\mu$ for~$\mu=1,\dots,n$, and the equilibrium solution
of~$\fv^n$ is written~\hbox{$\fv_\equil^n = \Magf(\afu)$}.  Also as done
previously, we use the relation
\begin{equation}
	\fd\fv_\dynac^\mu = \Dafu\Fv^\mu\,\fd\magf_\dynac\,,
	\tag{\ref{eq:magfdynac}}
\end{equation}
where~$\Dafu$ is defined by~\eqref{Dafudef}.  Adding
the~$|\grad\fd\magf_\dynac|^2$ contribution to~\eqref{SDPd2Hfirstline}, we
obtain
\begin{align}
	\half\Bigl\langle&
		\fd\gvort_\dynac
		\,((\Fgvort-\lapl)^{-1} + \Vpot_{\dcom 00})
		\,\fd\gvort_\dynac
		+ 2 \Kfunc\,\fd\gvort_\dynac\,\fd\magf_\dynac
		+ |\grad\fd\magf_\dynac|^2
		\Bigr\rangle
		\nonumber\\
	&= \half\Bigl\langle
		\fd\streamf_\dynac\,(\Fgvort-\lapl)\,\fd\streamf_\dynac
		+ \Vpot_{\dcom 00}\,|\fd\gvort_\dynac|^2
		- 2 \Kfunc\,\fd\magf_\dynac\,(\Fgvort-\lapl)
			\fd\streamf_\dynac + |\grad\fd\magf_\dynac|^2
		\Bigr\rangle\nonumber\\
	&= \half\Bigl\langle
		|\grad\fd\streamf_\dynac
		- \grad(\Kfunc\,\fd\magf_\dynac)|^2
		+ \l(1 - \Kfunc^2\r)|\grad\fd\magf_\dynac|^2
		+ \Vpot_{\dcom 00}\,|\fd\gvort_\dynac|^2\nonumber\\
	&\phantom{= \half\Bigl\langle}\qquad\qquad
		+ \Fgvort\,|\fd\streamf_\dynac - \Kfunc
			\,\fd\magf_\dynac|^2
		+ \Kfunc\l(\lapl\Kfunc
			- \Fgvort\,\Kfunc\r)|\fd\magf_\dynac|^2
		\Bigr\rangle
	\eqlabel{RMHDliked2H}
\end{align}
where~$\Kfunc$ is defined by~\eqref{Kdef}.  The energy provided by the
new~$|\grad\fd\magf_\dynac|^2$ term in the Hamiltonian (magnetic line-bending
energy in MHD) allows us to have formally stable equilibria
provided~\hbox{$\Kfunc^2 \le 1$}.  Thus, in contrast to the system in the
previous section, there exist formally stable equilibria with flow even for a
potential with~{$\Vpot_{\dcom 0\mu}=0$}.

\subsection{Nonsingular~$\Wn$}
\seclabel{WnStab}

\index{dynamical accessibility!and general extensions}
Now that we have demonstrated the procedure for obtaining equilibria and
determining their stability for brackets with no cocycles (\secref{SDPstab}),
we are in a position to deal with the more complicated case of an arbitrary
semidirect-type extensions with a nonsingular~$\W_{(n)}=\Wn$. We shall make
heavy use of the concept of coextension introduced in~\secref{nsingWn}.

\subsubsection{Equilibrium Solutions}

First we must look for equilibria of the equations of motion, which
from~\eqref{motion} and~\eqref{cobrakext} are
\begin{align}
	\dotfvz_\equil &= 0 = -{\lpb\Ham_{\dcom 0}\com\fvzeq\rpb}^\dagger
		- {\lpb\Ham_{\dcom\mu}\com\fveq^\mu\rpb}^\dagger
		- {\lpb\Ham_{\dcom n}\com\magfeq\rpb}^\dagger,
	\eqlabel{WnMotionfvz}\\
	\dotfv_\equil^\mu &= 0 = -{\lpb\Ham_{\dcom 0}\com\fveq^\mu\rpb}^\dagger
		- {\Wt_{\lambda}}^{\mu\nu}\,
		{\lpb\Ham_{\dcom\nu}\com\fveq^\lambda\rpb}^\dagger
		- {\Wn}^{\mu\nu}\,
		{\lpb\Ham_{\dcom\nu}\com\magfeq\rpb}^\dagger,
	\eqlabel{WnMotionfvmu}\\
	\dotmagf_\equil &= 0 = -{\lpb\Ham_{\dcom 0}\com\magfeq\rpb}^\dagger.
	\eqlabel{WnMotionfvn}
\end{align}
Unless otherwise noted, in this section all greek indices take values from~$1$
to~$n-1$, and repeated indices are summed. The tensors~$\Wt$ were defined
in~\secref{cassoln}: they are the subtensors of~$\W$ with indices restricted
from~$1$ to~$n-1$. They form a solvable extension. We have also made use of
the definition~\hbox{$\Wn^{\mu\nu} \ldef {\W_{(n)}}^{\mu\nu}$}.  As in
\secref{SDPstab}, we have set the variable~$\fv^n$ apart and labeled it
by~$\magf$, but now it does actually play a distinguished role in the solution
of the problem, as it did in~\secref{cassoln}.  Also note that the derivatives
of the Hamiltonian are implicitly evaluated at the
equilibrium~$(\fvzeq,\{\fveq^\mu\},\magfeq)$.

We now specialize the bracket to the 2-D canonical one, given
by~\eqref{canicobrak}.  Equation~\eqref{WnMotionfvn} is satisfied if
\begin{eqnarray}
	\Ham_{\dcom 0} = -\Streamf(\afu),\ \ \magfeq = \Magf(\afu),
	\eqlabel{WnfvnEqlm}
\end{eqnarray}
for functions~$\Streamf$ and~$\Magf$, and
some~$\afu=\afu(\xv)$. Equation~\eqref{WnMotionfvmu} is quite a bit dicier to
solve. The trick is to use the lower-triangular form of the~$\Wt^{(\mu)}$ to
solve for the~$\Ham_{\dcom\nu}$.  We multiply~\eqref{WnMotionfvmu}
by~\hbox{$\Wni \ldef \Wn^{-1}$}, and use~\eqref{WnfvnEqlm}, to obtain
\[
	-{\lpb\Streamf(\afu)\com{\Wni_{\tau\mu}}\,\fveq^\mu\rpb}
		+ {\Wni_{\tau\mu}}\,{\Wt_{\lambda}}^{\mu\nu}\,
		{\lpb\Ham_{\dcom\nu}\com\fveq^\lambda\rpb}
		+ {\Wni_{\tau\mu}}\,{\Wn}^{\mu\nu}\,
		{\lpb\Ham_{\dcom\nu}\com\Magf(\afu)\rpb} = 0,
\]
or, using the definition~\eqref{coextdef} of the
coextension,~\hbox{$\coW^\nu_{\tau\lambda}
\ldef {\Wt_\tau}^{\nu\mu}\,\Wni_{\mu\lambda}$},
\begin{equation}
	{\lpb\Ham_{\dcom\tau}\,\Magf'(\afu)
		+ \Streamf'(\afu)\,{\Wni_{\tau\mu}}\,\fveq^\mu\com\afu\rpb}
		+ \coW^\nu_{\tau\lambda}\,
		{\lpb\Ham_{\dcom\nu}\com\fveq^\lambda\rpb} = 0.
	\eqlabel{Wneqlmcond2}
\end{equation}
Since the~$\Wt^{(\mu)}$'s are lower-triangular, the~$\coW^{(\nu)}$'s have the
form
\[
	\coW^{(\nu)} = \l(\begin{array}{cc}
		0 & 0 \\
		0 & \fbox{\rule[-2ex]{0ex}{4ex}\phantom{xxxx}}
	\end{array}\r),\ \ \nu=1,\dots,n-1,
\]
where the box represents a square~\hbox{$(n-\nu-1)$}-dimensional symmetric
matrix of possibly nonzero elements.  There are never any nonvanishing
elements in the first row of~$\coW^{(\nu)}$, so setting~\hbox{$\tau=1$}
in~\eqref{Wneqlmcond2} gives
\begin{equation}
	{\lpb\Ham_{\dcom 1}\,\Magf'(\afu)
		+ \Streamf'(\afu)\,{\Wni_{1\mu}}\,\fveq^\mu\com\afu\rpb}
		= 0.
	\eqlabel{H1sol}
\end{equation}
We write the solution as
\[
	\Ham_{\dcom 1} = \Hafu_1(\afu)
		- \Dafu\Streamf(\afu)\,{\Wni_{1\mu}}\,\fveq^\mu,
\]
where~$\Hafu_1(\afu)$ is an arbitrary function and the operator~$\Dafu$ is
defined by~\eqref{Dafudef}.  Equation~\eqref{Wneqlmcond2} with~\hbox{$\tau=2$}
is
\[
	{\lpb\Ham_{\dcom 2}\,\Magf'(\afu)
		+ \Streamf'(\afu)\,{\Wni_{2\mu}}\,\fveq^\mu\com\afu\rpb}
		+ \coW^1_{2\lambda}\,
		{\lpb\Ham_{\dcom 1}\com\fveq^\lambda\rpb} = 0.
\]
If we then substitute in the solution for~$\Ham_{\dcom 1}$, Eq.~\eqref{H1sol},
we have
\begin{equation}
	{\lpb\Ham_{\dcom 2}\,\Magf'(\afu)
		+ \Streamf'(\afu)\,{\Wni_{2\mu}}\,\fveq^\mu\com\afu\rpb}
		+ \coW^1_{2\lambda}\,
		{\lpb\Hafu_1(\afu)
		- {\Dafu\Streamf(\afu)}\,{\Wni_{1\mu}}\,
			\fveq^\mu\com\fv^\lambda\rpb} = 0.
	\eqlabel{Wneqlmcond2tau2}
\end{equation}
Note that~\hbox{$\coW^1_{2\lambda}\,\Wni_{1\mu} =
\coW^\nu_{2\lambda}\,\Wni_{\nu\mu}
= \Wni_{2\kappa}\,{\W_\lambda}^{\kappa\nu}\,\Wni_{\nu\mu}
= \Wni_{2\kappa}\,\coW_{\lambda\mu}^{\kappa}$} is symmetric in~$\lambda$
and~$\mu$.  Hence,
\[
\begin{split}
	\coW^1_{2\lambda}\,{\lpb
		{\Dafu\Streamf(\afu)}\,{\Wni_{1\mu}}\,
			\fveq^\mu\com\fveq^\lambda\rpb}
	&= \Wni_{2\kappa}\,\coW_{\lambda\mu}^{\kappa}\,{\lpb
		{\Dafu\Streamf(\afu)}\,\fveq^\mu\com\fveq^\lambda\rpb}\\
	&= {\Dafu\Streamf(\afu)}\,
		\Wni_{2\kappa}\,\coW_{\lambda\mu}^{\kappa}\,{\lpb
		\fveq^\mu\com\fveq^\lambda\rpb}
	+ \Wni_{2\kappa}\,\coW_{\lambda\mu}^{\kappa}\,\fveq^\mu\,{\lpb
		{\Dafu\Streamf(\afu)}\com\fveq^\lambda\rpb}\\
	&= \half\,{\Dafu\Streamf'(\afu)}\,
		\Wni_{2\kappa}\,\coW_{\lambda\mu}^{\kappa}\,
		\l(\fveq^\mu\,{\lpb\afu\com\fveq^\lambda\rpb}
		+ \fveq^\lambda\,{\lpb\afu\com\fveq^\mu\rpb}\r)\\
	&= \half\,{\Dafu\Streamf'(\afu)}\,
		\Wni_{2\kappa}\,\coW_{\lambda\mu}^{\kappa}\,
		{\lpb\afu\com\fveq^\lambda\,\fveq^\mu\rpb}
\end{split}
\]
We can now solve~\eqref{Wneqlmcond2tau2} for~$\Ham_{\dcom 2}$, resulting in
\[
	\Ham_{\dcom 2} = \Hafu_2(\afu)
		- \Dafu\Streamf(\afu)\,{\Wni_{2\mu}}\,\fveq^\mu
		+ \coW_{2\lambda}^{\kappa}\,
			\Dafu{\Hafu_\kappa(\afu)}\,\fveq^\lambda
		- \half\,{\Dafu^2\Streamf(\afu)}\,\Wni_{2\kappa}\,
			\coW_{\lambda\mu}^{\kappa}\,\fveq^\lambda\,\fveq^\mu,
\]
where~$\Hafu_2(\afu)$ is another arbitrary function. The procedure carries on
in the same manner for~\hbox{$\tau>2$}, and in general we have
\begin{equation}
	\Ham_{\dcom \tau} = \Hafu_\tau(\afu)
		+ \sum_{m\ge 1} \frac{1}{m!}\,
		\Qt^{(m)}_{\tau\lambda_1\cdots\lambda_m}(\afu)\,
		\fveq^{\lambda_1}\cdots\fveq^{\lambda_m},
	\eqlabel{WnEqlmHam}
\end{equation}
where
\begin{equation}
	\Qt^{(1)}_{\tau\lambda}(\afu) \ldef \Dafu\l(
		\coW^\rho_{\tau\lambda}\,{\Hafu_\rho(\afu)}
		- \Wni_{\tau\lambda}\,\Streamf(\afu)\r),
	\eqlabel{Qt1def}
\end{equation}
and
\begin{multline}
	\Qt^{(m)}_{\tau\lambda_1\cdots\lambda_m}(\afu) \ldef
		\coW^{\tau_1}_{\tau\lambda_1}\,
		\coW^{\tau_2}_{\tau_1\lambda_2}\cdots
		\coW^{\tau_{m-2}}_{\tau_{m-3}\lambda_{m-2}}\,
		\coW^{\tau_{m-1}}_{\tau_{m-2}\lambda_{m-1}}\,\\[5pt]
		\times\Dafu^m
		\l(\coW^\rho_{\tau_{m-1}\lambda_m}\,\Hafu_\rho(\afu)
		- \Wni_{\tau_{m-1}\lambda_m}\,\Streamf(\afu)\r),
	\eqlabel{Qtdef}
\end{multline}
for~$m\ge 2$.  If we define~\hbox{$\Hafu_0(\afu) \ldef -\Streamf(\afu)$}, we
can also write the~$\Qt^{(m)}$ in terms of the~$\agc$ tensors, defined
by~\eqref{agcdef}, as
\[
	\Qt^{(m)}_{\lambda_1\cdots\lambda_m\lambda_{m+1}}(\afu) \ldef
		\sum_{\rho=0}^{n-1}
		\agc^{(m)\rho}_{\lambda_1\cdots\lambda_m\lambda_{m+1}}\,
		\Dafu^m\Hafu_\rho(\afu).
\]
The sum in~$m$ in~\eqref{WnEqlmHam} terminates since the~$\coW_{(\mu)}$ are
nilpotent.  The~$\Qt^{(m)}(\afu)$ are symmetric in all their lower indices.

Note that~\eqref{WnEqlmHam} is not a closed-form solution for the equilibria:
depending on the specific form of the Hamiltonian, the equation may be
straightforward or difficult to solve, or possibly not have any solutions at
all.  The situation is the same as for Eqs.~\eqref{equirel2}
and~\eqref{equirel3} (the energy-Casimir limit for CRMHD), which were solved
for~$\preseq$ and~$\pveleq$ in~\eqref{pvesol}.

The fact that the coextension, which we used to find Casimir invariants in
\chref{casinv}, appears in this calculation is not surprising, since the
energy-Casimir method result is recovered by letting~$\Magf(\afu) = \afu$,
which simply says that~$\Dafu$ is replaced by~$d/d\afu$.

We still have to satisfy~\eqref{WnMotionfvz} ($\dotfvz_\equil=0$) to get an
equilibrium.  Substituting in the results of~\eqref{WnfvnEqlm}
and~\eqref{WnEqlmHam}, we get the condition
\[
	{\lpb\Magf'(\afu)\,\Ham_{\dcom n}
		+ \Streamf'(\afu)\,\fvzeq\com\afu\rpb}
		+ {\lpb\Hafu_\mu(\afu)
		+ \sum_{m\ge 1} \frac{1}{m!}\,
		\Qt^{(m)}_{\mu\lambda_1\cdots\lambda_m}(\afu)\,
		\fveq^{\lambda_1}\cdots\fveq^{\lambda_m}\com\fveq^\mu\rpb}
		= 0.
\]
This can be solved, using the same techniques as for~\hbox{$\Ham_{\dcom
0},\dots,\Ham_{\dcom n-1}$} above, to give
\begin{equation}
	\Ham_{\dcom n} = \Hafu_n - \Dafu\Streamf\,\fvzeq
		+ \Dafu{\Hafu_\mu}\,\fveq^\mu
		+ \sum_{m\ge 1}\Dafu
		\Qt^{(m)}_{\lambda_1\cdots\lambda_m\lambda_{m+1}}\,
		\frac{\fveq^{\lambda_1}\cdots\fveq^{\lambda_{m+1}}}{(m+1)!}.
	\eqlabel{WnEqlmHamn}
\end{equation}
We now have expressions for the equilibria of arbitrary nonsingular
extensions, given by~\eqref{WnfvnEqlm},~\eqref{WnEqlmHam},
and~\eqref{WnEqlmHamn}.  We can proceed to determine their stability.

\subsubsection{Formal Stability}

The dynamically accessible variations are obtained from~\eqref{cobrakext}, and
are just equations \eqref{SDPfvzfuncvar} and~\eqref{SDPfvmufuncvar} modified
appropriately,
\begin{align}
	\fd\fvz_\dynac &= {\lpb\dynacg_0\com\fvz\rpb}^\dagger
		+ {\lpb\dynacg_\mu\com\fv^\mu\rpb}^\dagger
		+ {\lpb\dynacg_n\com\magf\rpb}^\dagger,
		\eqlabel{Wnfvzfuncvar} \\
	\fd\fv^\mu_\dynac &= {\lpb\dynacg_0\com\fv^\mu\rpb}^\dagger
		+ {\Wt_\lambda}^{\mu\nu}\,
			{\lpb\dynacg_\nu\com\fv^\lambda\rpb}^\dagger
		+ {\Wn}^{\mu\nu}\,
			{\lpb\dynacg_\nu\com\magf\rpb}^\dagger,
		\eqlabel{Wnfvmufuncvar}\\
	\fd\magf_\dynac &= {\lpb\dynacg_0\com\magf\rpb}^\dagger.
\end{align}
Notice that unlike the pure semidirect sum case given
by~\eqref{SDPfvzfuncvar} and~\eqref{SDPfvmufuncvar}, the dynamically
accessible variations for~\hbox{$\fv^1,\dots,\fv^n$} are now potentially
\emph{independent}.

We can use expression~\eqref{SDPHsecorder} for~$\fd^2\Ham_\dynac$ of the pure
semidirect sum, modified to admit a cocycle,
\begin{multline}
	\fd^2\Ham_\dynac = \half\Bigl\langle\fd\fvz_\dynac\com
		\Ham_{\dcom 00}\,\fd\fvz_\dynac
		+ 2\Ham_{\dcom 0\mu}\,\fd\fv^\mu_\dynac
		+ 2\Ham_{\dcom 0n}\,\fd\magf_\dynac
		+ \lpb\dynacg_0\com\Ham_{\dcom 0}\rpb\Bigr\rangle\\
	+ \half\Bigl\langle\fd\fv^\mu_\dynac\com
		\Ham_{\dcom \mu\nu}\,\fd\fv^\nu_\dynac
		+ 2\Ham_{\dcom \mu n}\,\fd\magf_\dynac
		+ \lpb\dynacg_0\com\Ham_{\dcom\mu}\rpb
		+ \lpb\dynacg_\mu\com\Ham_{\dcom 0}\rpb
		+ {\W_{\mu}}^{\sigma\tau}\,\lpb\dynacg_\sigma
		\com\Ham_{\dcom\tau}\rpb\Bigr\rangle\\
	+ \half\Bigl\langle\fd\magf_\dynac\com
		\Ham_{\dcom n n}\,\fd\magf_\dynac
		+ \lpb\dynacg_0\com\Ham_{\dcom n}\rpb
		+ \lpb\dynacg_n\com\Ham_{\dcom 0}\rpb
		+ {\Wn}^{\sigma\tau}\,\lpb\dynacg_\sigma
		\com\Ham_{\dcom\tau}\rpb\Bigr\rangle.
	\eqlabel{WnHsecorder}
\end{multline}
As we did for the semidirect sum case, we want to express all the brackets in
terms of dynamically accessible variations.  We know we must be able do this
by the theorem proved at the end of \secref{dynacmethod}.

The starting point is the~$\lpb\dynacg_n\com\Ham_{\dcom 0}\rpb$ term, since it
contains~$\dynacg_n$ and thus can only be expressed in terms
of~$\fd\fvz_\dynac$, given by Eq.~\eqref{Wnfvzfuncvar}.  We do not present the
calculation in detail here because it involves a great deal of algebra, none
of which is very illuminating.  We have to make liberal use of the identity
\[
	\coW^\sigma_{\mu\tau}\,
	\Dafu\Qt^{(m)}_{\sigma\lambda_1\cdots\lambda_m}
	= \Qt^{(m+1)}_{\mu\tau\lambda_1\cdots\lambda_m}\,,
	\quad \text{for}\ m\ge 1,
\]
easily verified from the definition of~$\Qt^{(m)}$, Eq.~\eqref{Qtdef}.

The final form of the second variation of the Hamiltonian is
\begin{multline}
	\fd^2\Ham_\dynac = \half\Bigl\langle\fd\fvz_\dynac\,
		\Ham_{\dcom 00}\,\fd\fvz_\dynac
		+ 2\fd\fvz_\dynac\,\Ham_{\dcom 0\mu}\,\fd\fv^\mu_\dynac
		+ 2\fd\fvz_\dynac\Bigl(\Ham_{\dcom 0n} + \Dafu\Streamf
			\Bigr)\fd\magf_\dynac
		\Bigr\rangle\\
	+ \half\Bigl\langle\fd\fv^\mu_\dynac
		\Bigl(\Ham_{\dcom \mu\nu} - \Qt^{(1)}_{\mu\nu}
		- \sum_{m\ge 1}\frac{1}{m!}\,
		\Qt^{(m+1)}_{\mu\nu\lambda_1\cdots\lambda_m}\,
		\fveq^{\lambda_1}\cdots\fveq^{\lambda_m}\Bigr)
		\fd\fv^\nu_\dynac\Bigr\rangle\\
	+ \Bigl\langle \fd\fv^\mu_\dynac\Bigl(
		\Ham_{\dcom \mu n} - \Dafu\Hafu_\mu
		- \sum_{m\ge 1}\frac{1}{m!}\,\Dafu
		\Qt^{(m)}_{\mu\lambda_1\cdots\lambda_m}\,
		\fveq^{\lambda_1}\cdots\fveq^{\lambda_m}\Bigr)
		\fd\magf_\dynac
		\Bigr\rangle\\
	+ \half\Bigl\langle\fd\magf_\dynac\Bigl(
		\Ham_{\dcom n n} - \Dafu\Hafu_n + \Dafu^2\Streamf\,\fvzeq
		- \Dafu^2\Hafu_\mu\,\fveq^\mu
		- \sum_{m\ge 2}\frac{1}{m!}\,
			\Dafu^2\Qt^{(m-1)}_{\lambda_1\cdots\lambda_m}\,
			\fveq^{\lambda_1}\cdots\fveq^{\lambda_m}
		\Bigr)\fd\magf_\dynac
		\Bigr\rangle.
	\!\!\!\!\eqlabel{WnStabd2H}
\end{multline}
This very general expression allows us to see exactly where the cocycles
modify the energy expression.  Obtaining a useful result out of it is
difficult, so we will do what we usually do: we simplify the problem!  The
case we will treat in more detail is the vanishing coextension case.

\subsection{Vanishing Coextension}
\seclabel{VC}

We consider the case where the coextension~\hbox{$\coW \equiv 0$} but~$\Wn$ is
nonsingular, as is the case for CRMHD (see \secref{CRMHDCas}).  A schematic
representation of this type of extension is shown in \figref{vanishcoextpic}.
Then
\begin{figure}
\centerline{\psfig{file=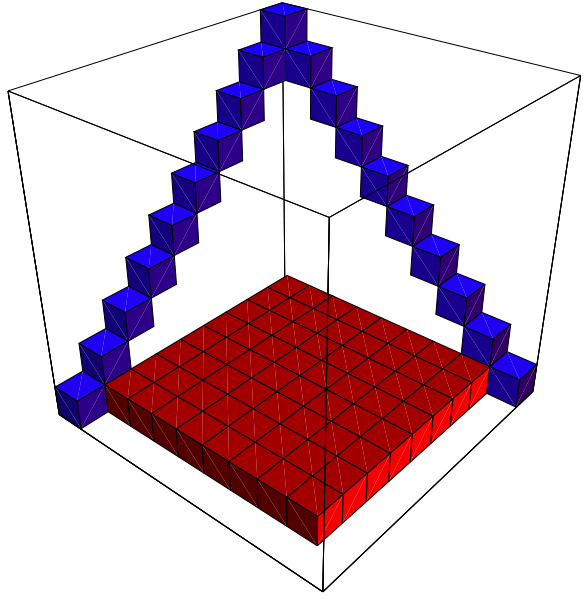,width=3in}}
\caption{Schematic representation of the 3-tensor~$\W$ for a semidirect
extension with \emph{vanishing coextension}~(\hbox{$\coW\equiv 0$}).  The axes
are as in \figref{solvextpic}.  The red cubes represent the~\hbox{$n-1\times
n-1$} matrix~$\Wn^{\mu\nu}$, assumed here nonsingular.  Note that compressible
reduced MHD, in \figref{crmhdextpic}, has this structure.}
\figlabel{vanishcoextpic}
\end{figure}
from~\eqref{Qt1def} we have
\[
	\Qt^{(1)}_{\tau\lambda}(\afu)
		= -\Dafu\Streamf(\afu)\,\Wni_{\tau\lambda},
\]
and from~\eqref{Qtdef} we have~\hbox{$\Qt^{(m)}(\afu)\equiv 0$}
for~\hbox{$m\ge 2$}.  We still have~\hbox{$\magfeq=\Magf(\afu)$}, and the
equilibrium relations~\eqref{WnEqlmHam} and~\eqref{WnEqlmHamn} simplify to
\begin{align}
	\Ham_{\dcom \tau} &= \Hafu_\tau
		- \Dafu\Streamf\,\Wni_{\tau\lambda}\,\fveq^{\lambda},
	\eqlabel{VCeqk}\\
	\Ham_{\dcom n} &= \Hafu_n - \Dafu\Streamf\,\fvzeq
		+ \Dafu{\Hafu_\mu}\,\fveq^\mu
		- \half\,\Dafu^2\Streamf\,\Wni_{\mu\lambda}\,
			\fveq^{\mu}\,\fveq^{\lambda},
	\eqlabel{VCeqn}
\end{align}
where as in \secref{WnStab} the greek indices run from~$1$ to~$n-1$.  The
second order variation of the Hamiltonian, Eq.~\eqref{WnStabd2H}, ``reduces''
to
\begin{multline}
	\fd^2\Ham_\dynac = \half\Bigl\langle\fd\fvz_\dynac\,
		\Ham_{\dcom 00}\,\fd\fvz_\dynac
		+ 2\fd\fvz_\dynac\,\Ham_{\dcom 0\mu}\,\fd\fv^\mu_\dynac
		+ 2\fd\fvz_\dynac\bigl(\Ham_{\dcom 0n} + \Dafu\Streamf
			\bigr)\fd\magf_\dynac\Bigr\rangle\\
	+ \half\Bigl\langle\fd\fv^\mu_\dynac
		\bigl(\Ham_{\dcom \mu\nu} + \Dafu\Streamf\,\Wni_{\mu\nu}\bigr)
		\fd\fv^\nu_\dynac\Bigr\rangle
	+ \Bigl\langle \fd\fv^\mu_\dynac\bigl(
		\Ham_{\dcom \mu n} - \Dafu\Hafu_\mu
		+ \Dafu^2\Streamf\,\Wni_{\mu\lambda}\,
		\fveq^{\lambda}\bigr)\fd\magf_\dynac
		\Bigr\rangle\\
	+ \half\Bigl\langle\fd\magf_\dynac\bigl(
		\Ham_{\dcom n n} - \Dafu\Hafu_n + \Dafu^2\Streamf\,\fvzeq
		- \Dafu^2\Hafu_\mu\,\fveq^\mu
		+ \half\,\Dafu^3\Streamf\,\Wni_{\mu\nu}\,
			\fveq^{\mu}\fveq^{\nu}\bigr)\fd\magf_\dynac
		\Bigr\rangle.
	\eqlabel{VCd2H}
\end{multline}
Again, to make progress we must further specialize the form of the Hamiltonian.

\subsubsection{RMHD-like System}

Let us take the RMHD-like Hamiltonian~\eqref{RMHDlikeHam}.  We first need to
find the equilibria, which we accomplish by substituting~\eqref{RMHDlikeHam}
into the equilibrium conditions~\eqref{WnfvnEqlm},~\eqref{VCeqk}
and~\eqref{VCeqn},
\begin{align}
	-\streamfeq + \Vpot_{\dcom 0} &= -\Streamf(\afu),
	\eqlabel{VCeq0q}\\
	\Vpot_{\dcom\tau} &= \Hafu_\tau(\afu)
		- \Dafu\Streamf(\afu)\,\Wni_{\tau\lambda}\,\fveq^{\lambda},
	\eqlabel{VCeqkq}\\
	-\ecurrenteq + \Vpot_{\dcom n} &= \Hafu_n(\afu)
		- \Dafu\Streamf(\afu)\,\gvorteq
		+ \Dafu{\Hafu_\mu}(\afu)\,\fveq^\mu
		- \half\,\Dafu^2\Streamf(\afu)\,\Wni_{\mu\lambda}\,
			\fveq^{\mu}\,\fveq^{\lambda},
	\eqlabel{VCeqnq}
\end{align}
Since we have not specified the exact dependence of~$\Vpot$ on the~$\fv^\mu$,
we cannot solve these for the~$\fveq^\mu$.  For the pure semidirect sum case,
we had~\hbox{$\fveq^\mu = \Fv(\afu)$}, regardless of the form of the
Hamiltonian.  The presence of the nondegenerate cocycle leads to potentially
much richer equilibria.

For the perturbation energy, we can use the result~\eqref{RMHDliked2H}
in~\eqref{VCd2H} to obtain
\begin{multline}
	\fd^2\Ham_\dynac = \half\Bigl\langle
		|\grad\fd\streamf_\dynac
		- \grad(\Kfunc\,\fd\magf_\dynac)|^2
		+ \l(1 - \Kfunc^2\r)|\grad\fd\magf_\dynac|^2
		+ \Vpot_{\dcom 00}\,|\fd\gvort_\dynac|^2\\
	+ \Fgvort\,|\fd\streamf_\dynac
		- \Kfunc\,\fd\magf_\dynac|^2
		+ 2\Vpot_{\dcom 0\mu}\,\fd\gvort_\dynac\,\fd\fv^\mu_\dynac\\
	+ \bigl(\Vpot_{\dcom \mu\nu} + \Dafu\Streamf\,\Wni_{\mu\nu}\bigr)
		\fd\fv^\mu_\dynac\,\fd\fv^\nu_\dynac
		+ 2\bigl(\Vpot_{\dcom \mu n} - \Dafu\Hafu_\mu
		+ \Dafu^2\Streamf\,\Wni_{\mu\lambda}\,
		\fveq^{\lambda}\bigr)\fd\fv^\mu_\dynac\,\fd\magf_\dynac\\
	+ \l\lgroup\Vpot_{\dcom nn} - \Dafu\Hafu_n + \Dafu^2\Streamf\,\gvorteq
		- \Dafu^2\Hafu_\mu\,\fveq^\mu
		+ \half\,\Dafu^3\Streamf\,\Wni_{\mu\nu}\,
			\fveq^{\mu}\fveq^{\nu} + \Kfunc\l(\lapl\Kfunc
			- \Fgvort\,\Kfunc\r)\r\rgroup\\
	\times|\fd\magf_\dynac|^2
	\Bigr\rangle,
	\eqlabel{afuckd2H}
\end{multline}
where
\[
	\Kfunc(\afu) \ldef \Vpot_{\dcom 0n} + \Dafu\Streamf(\afu).
\]
Immediately we see that the stability conditions~\hbox{$|\Kfunc| \le
1$},~\hbox{$\Fgvort \ge 0$}, and~\hbox{$\Vpot_{\dcom 00} \ge 0$} still hold.
However, until we have a closed form for the equilibria we cannot make
definite stability predictions.  We now proceed to use a more restricted class
of Hamiltonians for which the equilibria can be found explicitly.

\subsubsection{Quadratic Potential}

An important case we can do explicitly is when~$\Vpot$ is quadratic,
\[
	\Vpot = \half\,\fv^\mu\,\Vpotq_{\mu\nu}(\xv)\,\fv^\nu
		+ \Vpotl_\sigma(\xv)\,\fv^\sigma,
\]
where~$\Vpotq$ is a symmetric matrix, in which case we have
\[
	\Vpot_{\dcom \tau} = \Vpotq_{\tau\nu}\,\fv^\nu
		+ \Vpotl_\tau,
\]
and~\hbox{$\Vpot_{\dcom 0} = \Vpot_{\dcom n} = 0$}.  Inserting this
into~\eqref{VCeqkq}, we obtain
\[
	\l(\Vpotq_{\tau\lambda}
		+ \Dafu\Streamf\,\Wni_{\tau\lambda}\r)\fveq^{\lambda}
		 = \Hafu_\tau - \Vpotl_\tau.
	\eqlabel{Vpotsoln}
\]
Assuming~$\Vpotq$ is nondegenerate, the matrix
\begin{equation}
	\modVpot_{\tau\lambda} \ldef \Vpotq_{\tau\lambda}
		+ \Dafu\Streamf\,\Wni_{\tau\lambda}
	\eqlabel{studef}
\end{equation}
will be invertible except possibly at some points.  We denote its inverse
by~$\modVpoti^{\tau\lambda}$, and~\eqref{Vpotsoln} has solution
\begin{equation}
	\fveq^\lambda(\xv) = \modVpoti^{\lambda\tau}
		(\Hafu_\tau(\afu) - \Vpotl_\tau(\xv)).
	\eqlabel{VCfveqsoln}
\end{equation}
We emphasize how different this expression is to the pure semidirect sum
result,~\hbox{$\fveq^\lambda(\xv) = \Fv(\afu)$}.  In~\eqref{VCfveqsoln} the
equilibrium solution~$\fv^\lambda$ can explicitly depend on~$\xv$ through the
Hamiltonian.  This can never occur for equilibria of the pure semidirect sum,
regardless of the form of the Hamiltonian.

The most interesting feature of the new equilibria~\eqref{VCfveqsoln} is the
fact that there are new \emph{resonances}\index{resonance} in the
system---solutions for which~$\modVpoti^{\lambda\tau}$ will blow up.  This is
what occurred for CRMHD in \secref{energycasimir}, where we had a singularity
in the solution~\eqref{pvesol} of~$\pveleq$ and~$\preseq$, associated with the
acoustic resonance.\index{acoustic resonance} As the equilibrium solution
approaches this resonance, we can expect the system to become less stable.

We can use the solution~\eqref{VCfveqsoln} in~\eqref{VCeqnq} to obtain a
closed-form result for~$\ecurrenteq$,
\begin{multline}
	{\ecurrenteq} = -\Hafu_n
		+ \Dafu\Streamf\,(\lapl\Streamf - \Fgvort\,\Streamf + \fgvort)
		- \Dafu{\Hafu_\mu}\,\modVpoti^{\mu\tau}
			(\Hafu_\tau - \Vpotl_\tau)\\
		+ \half\,\Dafu^2\Streamf\,
			(\Hafu_\tau - \Vpotl_\tau)\,
			\modVpoti^{\tau\mu}\,
			\Wni_{\mu\lambda}\,\modVpoti^{\lambda\sigma}
			\,(\Hafu_\sigma - \Vpotl_\sigma),
	\eqlabel{ecurrentsoln0}
\end{multline}
where~\hbox{$\ecurrenteq = \lapl\Magf(\afu)$}.  Using Eq.~\eqref{vortitu}
for~$\vorteq$ and the analogous relation for~$\ecurrenteq$, we have
that~\eqref{ecurrentsoln0} can be rewritten
\begin{multline}
	\frac{\l((\Magf')^2-(\Streamf')^2\r)}{\Magf'}\,\lapl\afu
		+ \frac{\l(\Magf'\,\Magf'' - \Streamf'\,\Streamf''\r)}{\Magf'}
		\,|\grad\afu|^2 =\\ {-\Hafu_n}
		+ \Dafu\Streamf\,(-\Fgvort\,\Streamf + \fgvort)
		- \Dafu{\Hafu_\mu}\,\modVpoti^{\mu\tau}
			(\Hafu_\tau - \Vpotl_\tau)\\
		+ \half\,\Dafu^2\Streamf\,
			(\Hafu_\tau - \Vpotl_\tau)\,
			\modVpoti^{\tau\mu}\,
			\Wni_{\mu\lambda}\,\modVpoti^{\lambda\sigma}
			\,(\Hafu_\sigma - \Vpotl_\sigma).
	\eqlabel{ecurrentsoln}
\end{multline}
This is a nonlinear PDE to be solved for~$\afu(\xv)$ with arbitrary
functions~$\Streamf(\afu)$, $\Magf(\afu)$, and~$\Hafu_\mu(\afu)$, and given
functions~$\modVpoti^{\tau\mu}(\xv)$, $\Vpotl_\sigma(\xv)$, $\Fgvort(\xv)$,
and~$\fgvort(\xv)$.  Needless to say, solving~\eqref{ecurrentsoln} in general
is extremely difficult.  There are, however, classes of solution that can be
obtained analytically.  We now examine one of these special cases.

A particularly simple case are the aforementioned \Alfven\ solutions, for
which
\begin{equation}
	\Magf'(\afu) = \alfvenc\,\Streamf'(\afu),
	\eqlabel{magfeqcstreamf}
\end{equation}
where~$\alfvenc$ is a constant.  We also obtain
\[
	\Dafu\Streamf(\afu) = \frac{1}{\Magf'(\afu)}\,
		\frac{d\Streamf}{d\afu}(\afu)
		= \frac{1}{\alfvenc\,\Streamf'(\afu)}\,\Streamf'(\afu)
		= \alfvenc^{-1},
\]
so that~\hbox{$\Dafu^m\Streamf=0$} for~\hbox{$m>1$}.  Thus, assuming
that~$\Magf'$ and~$\Streamf'$ are proportional for the dynamical accessibility
method is analogous to assuming that~$\Streamf(\magfeq)$ is linear for the
energy-Casimir method.  From~\eqref{magfeqcstreamf}, we might be tempted to
simply write~$\Streamf=\Streamf(\Magf)$, and indeed this is true.  However,
this is not useful because in general we still cannot rewrite~$\afu$ in terms
of~$\Magf$, since~$\Magf=\Magf(\afu)$ may not be invertible.  If~$\Magf$ is
invertible, then we recover the energy-Casimir result completely.

Using~\eqref{magfeqcstreamf} in the equilibrium
condition~\eqref{ecurrentsoln} gives
\begin{equation}
	(1 - \alfvenc^{-2})\l(\Magf'\,\lapl\afu
		+ \Magf''\,|\grad\afu|^2\r) = -\Hafu_n
		- \alfvenc^{-2}\,(\Fgvort\afu - \alfvenc\fgvort)
		- {\Hafu_\mu'}\,\modVpoti^{\mu\tau}
			(\Hafu_\tau - \Vpotl_\tau),
	\eqlabel{VCpropeqcond}
\end{equation}
so that the quadratic term (proportional to~$\Dafu^2\Streamf$) disappears.

Several systems have~$\modVpoti^{\mu\tau}$ independent of~$\xv$.  It may then
also happen that we can choose the~$\Hafu_\mu(\afu)$ such that
\begin{equation}
	{\Hafu_\mu'}\,\modVpoti^{\mu\tau}\Vpotl_\tau
		= \alfvenc^{-2}\,(\Fgvort\afu - \alfvenc\fgvort),
	\eqlabel{VCxindepcond}
\end{equation}
After this is effected, Eq.~\eqref{VCpropeqcond} no longer depends explicitly
on~$\xv$, and has solutions such as the Kelvin--Stuart cat's eye.  This
procedure can be carried out for CRMHD, for which~\hbox{$\Fgvort=\fgvort=0$}.
In that case,~\eqref{VCxindepcond} becomes~\eqref{xindepcond}.

\subsubsection{Stability for Quadratic Potential}

Assuming that we still have the quadratic potential of the previous section,
we now show that the (acoustic) resonance which occurred for CRMHD is a generic
feature of Lie--Poisson systems with cocycles.

We take the energy expression~\eqref{afuckd2H}, 
use the fact the~\hbox{$\Vpot_{\dcom 0} = \Vpot_{\dcom n} = 0$}, and obtain
\begin{multline*}
	\fd^2\Ham_\dynac = \half\Bigl\langle
		|\grad\fd\streamf_\dynac
		- \grad(\Kfunc\,\fd\magf_\dynac)|^2
		+ \l(1 - \Kfunc^2\r)|\grad\fd\magf_\dynac|^2\\
	+ \Fgvort\,|\fd\streamf_\dynac
		- \Kfunc\,\fd\magf_\dynac|^2
	+ \modVpot_{\mu\nu}\,\fd\fv^\mu_\dynac\,\fd\fv^\nu_\dynac
		+ 2\bigl({\Dafu^2\Streamf\,\Wni_{\mu\lambda}\,\fveq^{\lambda}
		- \Dafu\Hafu_\mu}\bigr)\fd\fv^\mu_\dynac\,\fd\magf_\dynac\\
	+ \l\lgroup{\Dafu^2\Streamf\,\gvorteq - \Dafu\Hafu_n}
		- \Dafu^2\Hafu_\mu\,\fveq^\mu
		+ \half\,\Dafu^3\Streamf\,\Wni_{\mu\nu}\,
			\fveq^{\mu}\fveq^{\nu} + \Kfunc\l(\lapl\Kfunc
			- \Fgvort\,\Kfunc\r)\r\rgroup
	|\fd\magf_\dynac|^2
	\Bigr\rangle,
\end{multline*}
where we used the definition of~$\modVpot$, Eq.~\eqref{studef}, and we have
not made any assumptions about the form of~$\Streamf$ and~$\Magf$.  The
equilibrium solutions~$\fveq^{\lambda}$ satisfy~\eqref{VCfveqsoln}.

If we assume~\hbox{$\Kfunc\le 1$} and~\hbox{$\Fgvort\ge 0$}, then to obtain
part of the sufficient conditions for stability we require that~$\modVpot$ be
positive-definite.  But when~$\modVpot$ becomes singular we cannot guarantee
this.  This was the case with CRMHD.

Note that~\hbox{$\det\modVpot=0$} does \emph{not} imply that the system will
be unstable beyond the resonance.\index{resonance} It is, however, a strong
indication that it might be.

\chapter{Conclusions}
\chlabel{conclusion}

Using the tools of Lie algebra cohomology, we have classified low-order
extensions.  We found that there were only a few normal forms for the
extensions, and that they involved no free parameters.  This is not expected
to carry over to higher orders~($n>4$).  The classification includes the
Leibniz extension, which we have shown is the maximal extension.  One of the
normal forms is the bracket appropriate to compressible reduced MHD.

We then developed techniques for finding the Casimir invariants of
Lie--Poisson brackets formed from Lie algebra extensions.  We introduced the
concept of coextension, which allowed us to explicitly write down the solution
of the Casimirs.  The coextension for the Leibniz extension can be found for
arbitrary order, so we were able obtain the corresponding Casimirs in general.

By using the method of dynamical accessibility, we derived general conditions
for the formal stability of Lie--Poisson systems.  In particular, for
compressible reduced MHD, we found the presence of a cocycle could only make a
certain class of solutions more unstable.  In general, cocycles were shown to
lead to resonances, such as the acoustic resonance for CRMHD.

The dynamical accessibility approach also allowed us to get a clearer picture
of the role of cocycles: in a pure semidirect extension, the absence of a
cocycle means that the system necessarily describes an advective system, and
the dynamically accessible variations are not independent.  In contrast, for
the nonsingular cocycle case all of the perturbations are independent.  The
form of the stability condition is thus much more complex.

It would be interesting to generalize the classification scheme presented here
to a completely general form of extension
bracket~\cite{Morrison1980a,Nore1997}. Certainly the type of coordinate
transformations allowed would be more limited, and perhaps one cannot go any
further than cohomology theory allows.

Though we have gone a long way in this respect, the interpretation of the
Casimir invariants has yet to be fully explored, both in a mathematical and a
physical sense. Mathematically, we could give a precise geometrical relation
between cocycles and the form of the Casimirs. The cocycle and Casimirs should
yield information about the holonomy of the system. For this one must study
extensions in the framework of their principal bundle
description~\cite{Azcarraga}. Physically, we would like to attach a more
precise physical meaning to these conserved quantities. The invariants
associated with simultaneous eigenvectors can be regarded as constraining the
associated field variable to move with the fluid elements~\cite{Morrison1987}.
The field variable can also be interpreted as partially labeling a fluid
element. Some attempt has been made in formulating the Casimir invariants of
brackets in such a manner~\cite{Kuznetsov1980,Thiffeault1998}, but for the
more complicated invariants a general treatment is still not yet available.



\appendices

\chapter{Proof of the Jabobi Identity}
\apxlabel{lpjacobi}

\index{Jacobi identity}
We want to show that the Lie--Poisson bracket
\begin{equation}
	{\lPB F\com G \rPB}_\pm(\fv) = \pm\lang\fv\com
		{\lpb \frac{\fd F}{\fd\fv}\com\frac{\fd G}{\fd\fv}
		\rpb}\rang ,
	\tag{\ref{eq:LPB}}
\end{equation}
where~\hbox{$\fv\in\LieA^*$}, and~\hbox{$F:\LieA^*\rightarrow\reals$}
and~\hbox{$G:\LieA^*\rightarrow\reals$} are functionals, satisfies the Jacobi
identity
\[
	{\lPB{\lPB F\com G \rPB}_\pm\com H\rPB}_\pm
	+ {\lPB{\lPB G\com H \rPB}_\pm\com F\rPB}_\pm
	+ {\lPB{\lPB H\com F \rPB}_\pm\com G\rPB}_\pm = 0.
\]
The inner bracket~$\lpb\com\rpb$ is the bracket of the Lie algebra~$\LieA$, so
it satisfies the Jacobi identity.  The overall sign of the bracket is
inconsequential, so we choose the~$+$ bracket.  We first compute the
variation of~\hbox{${\lPB F\com G \rPB}$},
\[
\begin{split}
	\fd{\lPB F\com G \rPB} &= \lang\!\fd\fv\com {\lpb \frac{\fd F}{\fd\fv}
			\com\frac{\fd G}{\fd\fv}\rpb}\rang
		+ \lang\!\fv\com
			{\lpb \frac{\fd^2 F}{\fd\fv\fd\fv}\,\fd\fv\com
			\frac{\fd G}{\fd\fv}\rpb}\rang
		+ \lang\!\fv\com
			{\lpb \frac{\fd F}{\fd\fv}\com
			\frac{\fd^2 G}{\fd\fv\fd\fv}\,\fd\fv\rpb}\rang\\[4pt]
	&= \lang\!\fd\fv\com {\lpb \frac{\fd F}{\fd\fv} \com
			\frac{\fd G}{\fd\fv}\rpb}\rang
		- \lang\!{\lpb\frac{\fd G}{\fd\fv}\com\fv\rpb}^\dagger\!\!\!
			\com\frac{\fd^2 F}{\fd\fv\fd\fv}\,\fd\fv\rang
		+ \lang\!{\lpb\frac{\fd F}{\fd\fv}\com\fv\rpb}^\dagger\!\!\!
			\com\frac{\fd^2 G}{\fd\fv\fd\fv}\,\fd\fv\rang\\[4pt]
	&= \lang\!\fd\fv\com
		{\lpb \frac{\fd F}{\fd\fv}\com\frac{\fd G}{\fd\fv}\rpb}
		- \frac{\fd^2 F}{\fd\fv\fd\fv}
			\lpb\frac{\fd G}{\fd\fv}\com\fv\rpb^\dagger
		+ \frac{\fd^2 G}{\fd\fv\fd\fv}
			\lpb\frac{\fd F}{\fd\fv}\com\fv\rpb^\dagger\rang,
\end{split}
\]
where we have used the definition of the coadjoint bracket~\eqref{coadj} and
the self-adjoint property of the second derivative operator.  Thus, we have
\[
	\frac{\fd{\lPB F\com G \rPB}}{\fd\fv} = 
		{\lpb \frac{\fd F}{\fd\fv}\com\frac{\fd G}{\fd\fv}\rpb}
		- \frac{\fd^2 F}{\fd\fv\fd\fv}
			\lpb\frac{\fd G}{\fd\fv}\com\fv\rpb^\dagger
		+ \frac{\fd^2 G}{\fd\fv\fd\fv}
			\lpb\frac{\fd F}{\fd\fv}\com\fv\rpb^\dagger.
\]
We can now evaluate the first term of the Jacobi identity,
\begin{align*}
	{\lPB{\lPB F\com G \rPB}\com H\rPB} &= \lang\fv\com
		\lpb\frac{\fd{\lPB F\com G \rPB}}{\fd\fv}\com
		\frac{\fd H}{\fd\fv}\rpb\rang\\[4pt]
	&= \lang\fv\com\!\lpb
		{\lpb \!\frac{\fd F}{\fd\fv}\com\frac{\fd G}{\fd\fv}\rpb}
		- \frac{\fd^2 F}{\fd\fv\fd\fv}
			\lpb\frac{\fd G}{\fd\fv}\com\fv\rpb^\dagger
		\! + \frac{\fd^2 G}{\fd\fv\fd\fv}
			\lpb\frac{\fd F}{\fd\fv}\com\fv\rpb^\dagger\!\!\com
		\frac{\fd H}{\fd\fv}\rpb\rang\\
	&= \lang\fv\com\lpb
		{\lpb\frac{\fd F}{\fd\fv}\com\frac{\fd G}{\fd\fv}\rpb}
		\com\frac{\fd H}{\fd\fv}\rpb\rang
	+ \lang{\lpb\frac{\fd H}{\fd\fv}\com\fv\rpb}^\dagger\com
		\frac{\fd^2 F}{\fd\fv\fd\fv}
		{\lpb\frac{\fd G}{\fd\fv}\com\fv\rpb}^\dagger\rang\\
	&\phantom{=} - \lang{\lpb\frac{\fd H}{\fd\fv}\com\fv\rpb}^\dagger\com
		\frac{\fd^2 G}{\fd\fv\fd\fv}
		{\lpb\frac{\fd F}{\fd\fv}\com\fv\rpb}^\dagger\rang.
\end{align*}
Upon adding permutations of~$F$,~$G$, and~$H$, the second-derivative terms
cancel and we are left with
\[
	\lang\fv\com
		\lpb{\lpb\frac{\fd F}{\fd\fv}\com\frac{\fd G}{\fd\fv}\rpb}
		\com\frac{\fd H}{\fd\fv}\rpb
		+ \lpb{\lpb\frac{\fd G}{\fd\fv}\com\frac{\fd H}{\fd\fv}\rpb}
		\com\frac{\fd F}{\fd\fv}\rpb
		+ \lpb{\lpb\frac{\fd H}{\fd\fv}\com\frac{\fd F}{\fd\fv}\rpb}
		\com\frac{\fd G}{\fd\fv}\rpb
	\rang,
\]
which vanishes by the Jacobi identity in~$\LieA$.

\chapter{Proof of~$\W^{(1)}=I$}
\apxlabel{woneident}

Out goal is to demonstrate that through a series of lower-triangular
coordinate transformations we can make~$\W^{(1)}$ equal to the identity
matrix, while preserving the lower-triangular nilpotent form
of~\hbox{$\W^{(2)},\dots,\W^{(n)}$}.

We first show that we can always make a series of coordinate transformations
to make~\hbox{${\W_\lambda}^{11} = {\delta_\lambda}^1$}.  First note that if
the coordinate transformation~$\M$ is of the form~\hbox{$\M = I + L$},
where~$I$ is the identity and~$L$ is lower-triangular nilpotent,
then~\hbox{$\Wt^{(1)}=\M^{-1}\,\W^{(1)}\,\M$} still has eigenvalue~$1$, and
the matrices
\[
	\Wt^{(\mu)}=\M^{-1}\,\W^{(\mu)}\,\M, \qquad \mu>1
\]
are still nilpotent.

For~$\lambda>1$ we have
\begin{equation}
	{\Wb_\lambda}^{11} = {\Wt_\lambda}{}^{11}
		+ {\Wt_\lambda}{}^{1\nu}\,{L_\nu}^1
	= {\Wt_\lambda}{}^{11}
		+ \sum_{\nu=2}^{\lambda-1}{\Wt_\lambda}{}^{1\nu}\,{L_\nu}^1
		+ {L_\lambda}^1,
	\eqlabel{apxcoordtrans}
\end{equation}
where we used~\hbox{${\Wt_\lambda}{}^{1\lambda}=1$}.  Owing to the triangular
structure of the set of equations~\eqref{apxcoordtrans} we can always solve
for the~\hbox{${L_\lambda}^1$} to make~${\Wb_\lambda}^{11}$ vanish.
This proves the first part.

We now show by induction that if~\hbox{${\W_\lambda}^{11} =
{\delta_\lambda}^1$}, as proved above, then the matrix~$\W^{(1)}$ is the
identity. For~\hbox{$\lambda = 1$} the result is trivial. Assume
that~\hbox{${\W_\mu}^{1\nu} = {\delta_\mu}^\nu$}, for~\hbox{$\mu <
\lambda$}. Setting two of the free indices to one, Eq.~\eqref{Wjacob} can be
written
\[
\begin{split}
	{\W_\lambda}^{\mu 1}\,{\W_\mu}^{1\sigma}
	&= {\W_\lambda}^{\mu\sigma}\,{\W_\mu}^{11}\\
	&= {\W_\lambda}^{\mu\sigma}\,{\delta_\mu}^{1}
	= {\W_\lambda}^{1\sigma}\,.
\end{split}
\]
Since~$\W^{(1)}$ is lower-triangular the index~$\mu$ runs from~$2$
to~$\lambda$ (since we are assuming~\hbox{$\lambda > 1$}):
\[
	\sum_{\mu = 2}^{\lambda}
	{\W_\lambda}^{\mu 1}\,{\W_\mu}^{1\sigma}
	= {\W_\lambda}^{1\sigma}\,,
\]
and this can be rewritten, for~$\sigma<\lambda$,
\[
	\sum_{\mu = 2}^{\lambda-1}
	{\W_\lambda}^{\mu 1}\,{\W_\mu}^{1\sigma}
	= 0\,.
\]
Finally, we use the inductive hypothesis
\[
	\sum_{\mu = 2}^{\lambda-1}
	{\W_\lambda}^{\mu 1}\,{\delta_\mu}^{\sigma}
	= {\W_\lambda}^{\sigma 1}= 0\,,
\]
which is valid for~$\sigma < \lambda$.  Hence,~\hbox{${\W_\lambda}^{\sigma 1}
= {\delta_\lambda}^{\sigma}$} and we have proved the result.
(${\W_\lambda}^{\lambda 1}$ must be equal to one since it lies on the diagonal
and we have already assumed degeneracy of eigenvalues.)


\index{qbert@\qbert{} extension|see{extension, Leibniz}}
\index{Leibniz extension|see{extension, Leibniz}}
\index{canonical bracket|see{bracket, canonical}}
\index{Lie--Poisson bracket|see{bracket, Lie--Poisson}}
\index{Lie bracket|see{bracket, Lie}}
\index{Lie algebra extension|see{extension}}
\index{exact sequence|see{sequence, exact}}
\index{manifold!Poisson|see{Poisson manifold}}
\index{rotation group|see{\SOthree}}
\index{ntuples@$n$-tuples!bracket|see{bracket for~$n$-tuples}}
\index{Euler's equation!for the rigid body|see{equations of motion}}
\index{Euler's equation!for the 2--D ideal fluid|see{equations of motion}}
\index{Euclidean group|see{semidirect sum}}
\index{splitting|see{semidirect sum}}
\index{delta W@$\delta W$ criterion|see{stability, $\delta W$}}
\index{semidirect extension|see{semidirect sum}}
\index{coadjoint action|see{action, coadjoint}}
\index{heavy top|see{rigid body}}
\index{Casimir invariant|see{Casimir}}
\index{coadjoint bracket|see{bracket, coadjoint}}
\index{order of extension|see{extension, order of}}
\index{direct sum|see{extension, direct sum}}
\index{adjoint action|see{action, adjoint}}
\index{MHD|see{reduced MHD and compressible
					reduced MHD}}
\index{Moore--Penrose inverse|see{pseudoinverse}}
\index{Kelvin--Stuart cat's eye|seealso{magnetic islands}}
\printindex

\end{document}